\documentclass[fleqn,usenatbib]{mnras}

\usepackage{newtxtext,newtxmath}
\usepackage{tabu}
\usepackage{longtable}

\usepackage[T1]{fontenc}

\DeclareRobustCommand{\VAN}[3]{#2}
\let\VANthebibliography\thebibliography
\def\thebibliography{\DeclareRobustCommand{\VAN}[3]{##3}\VANthebibliography}

\usepackage{graphicx}   
\usepackage{amsmath}    

\newcommand{\kms}{$\textrm{km} {\textrm{s}}^{-1}$}
\newcommand{\ergcmcms}{erg cm$^{-2}$ s$^{-1}$ }

\title[Gaia DR2 lenses]{Gravitationally lensed quasars in \textit{Gaia} - IV. 150 new lenses, quasar pairs, and projected quasars}

\author[C. Lemon et al.]{C. Lemon$^{1}$\thanks{E-mail: cameron.lemon@epfl.ch},
T. Anguita$^{2,3}$,
M. Auger$^{4,5}$,
F. Courbin$^{1}$,
A. Galan$^{1}$,
R. McMahon$^{4,5}$,
\newauthor
F. Neira$^{2,3}$,
M. Oguri$^{6,7}$,
P. Schechter $^{8,9}$,
A. Shajib$^{10,11\thanks{NHFP Einstein Fellow}}$,
and T. Treu$^{11}$
\\
$^{1}$Institute of Physics, Laboratoire d'Astrophysique, Ecole Polytechnique  F\'ed\'erale de Lausanne (EPFL), Observatoire de Sauverny, CH-1290 Versoix, Switzerland\\
$^{2}$Departamento de Ciencias Fisicas, Universidad Andres Bello, Fernandez Concha 700, Las Condes, Santiago, Chile\\
$^{3}$Millennium Institute of Astrophysics, Monse\~{n}or Nuncio Sotero Sanz 100, Oficina 104, 7500011 Providencia, Santiago, Chile\\
$^{4}$Institute of Astronomy, University of Cambridge, Madingley Road, Cambridge CB3 0HA, UK\\
$^{5}$Kavli  Institute  for  Cosmology,  University  of  Cambridge,  Madingley Road, Cambridge CB3 0HA, UK\\
$^{6}$Center for Frontier Science, Chiba University, 1-33 Yayoi-cho, Inage-ku, Chiba 263-8522, Japan\\
$^{7}$Department of Physics, Graduate School of Science, Chiba University, 1-33 Yayoi-Cho, Inage-Ku, Chiba 263-8522, Japan\\
$^{8}$MIT Kavli Institute for Astrophysics and Space Research, Cambridge, MA 02139 USA\\
$^{9}$MIT Department of Physics, Cambridge, MA 02139 USA\\
$^{10}$Department of Astronomy and Astrophysics, University of Chicago, Chicago, IL 60637, USA\\
$^{11}$Department of Physics and Astronomy, University of California, Los Angeles, CA 90095, USA\\
}

\date{Accepted XXX. Received YYY; in original form ZZZ}

\pubyear{2022}

\begin{document}
\label{firstpage}
\pagerange{\pageref{firstpage}--\pageref{lastpage}}
\maketitle

\begin{abstract}
We report the spectroscopic follow-up of 175 lensed quasar candidates selected using \textit{Gaia} Data Release 2 observations following \citet{lemon2019}. Systems include 86 confirmed lensed quasars and a further 17 likely lensed quasars based on imaging and/or similar spectra. We also confirm 11 projected quasar pairs and 11 physical quasar pairs, while 25 systems are left as unclassified quasar pairs -- pairs of quasars at the same redshift, which could be either distinct quasars or potential lensed quasars. Especially interesting objects include 8 quadruply imaged quasars of which two have BAL sources, an apparent triple, and a doubly lensed LoBaL quasar. The source redshifts and image separations of these new lenses range between 0.65--3.59 and 0.78--6.23\arcsec respectively. We compare the known population of lensed quasars to an updated mock catalogue at image separations between 1 and 4 arcseconds, showing a very good match at $z<1.5$. At $z>1.5$, only 47\% of the predicted number are known, with 56\% of these missing lenses at image separations below 1.5\arcsec. The missing higher-redshift, small-separation systems will have fainter lensing galaxies, and are partially explained by the unclassified quasar pairs and likely lenses presented in this work, which require deeper imaging. Of the 11 new reported projected quasar pairs, 5 have impact parameters below 10 kpc, almost tripling the number of such systems, which can probe the innermost regions of quasar host galaxies through absorption studies. We also report four new lensed galaxies discovered through our searches, with source redshifts ranging from 0.62 to 2.79.
\end{abstract}




\section{Introduction}

Gravitationally lensed quasars offer several unique probes of astrophysics and cosmology. Their variable sources allow measurements of the time delays between images, which can be used to constrain the Hubble constant through time-delay cosmography \citep{shajib2020, harvey2020}. The compact nature of the source allows a probe of the mass at compact scales in the lensing galaxy through microlensing; the brightness variations caused by stars as they pass in front of each quasar image can be used to constrain the Initial Mass Function \citep{jimenez2019, schechter2014} and place limits on the abundance of primordial black holes \citep{mediavilla2017}. Simultaneously, microlensing can constrain the innermost structure of quasars at high-redshift \citep{hutsemekers2021, paic2022}. The particular brightness of quasar sources provides an excellent tool for studying the geometry and kinematics of intervening matter \citep{rubin2018, okoshi2021, lemon2022}. This brightness also means that lensed quasars are often source-selected \citep{myers2003, oguri2006}, in contrast with lensed galaxy samples which are selected through their bright lensing galaxies. This offers a chance to create a well-defined statistical sample down to small separations, which can be used to probe cosmology and galaxy evolution parameters \citep{chae2003, oguri2012}. These studies are often limited by the small number of lenses that suit their particular purpose, for example requiring four well-separated images with time delays above 10 days for time-delay cosmography.

Recent optical to near-infrared surveys now provide an efficient way to select high-confidence quasar candidates across the whole sky \citep[e.g.][]{stern2012}, and have the resolution and depth to identify whether such systems are lensed, i.e. whether there are multiple images and a lensing galaxy present \citep[e.g.][]{chan2022}. However, lensed quasars are particularly rare, and outnumbered by convincing contaminant systems, mainly composed of compact star-forming galaxies and chance quasar/star projections \citep{treu2018}. The all-sky space-based catalogues from \textit{Gaia} offer a particularly effective way of removing these contaminants. In the previous papers of this series, we have shown that combining the \textit{Gaia} catalogue detections and parameters with WISE and ground-based optical imaging can efficiently remove the common contaminants of lensed quasar searches \citep{lemon2017, lemon2018, lemon2019}. In this paper, we continue our spectroscopic follow-up campaign of lensed quasar candidates selected with the techniques developed in these papers. 

In Section \ref{selection}, we describe the selection techniques and datasets used for candidate selection. Follow-up spectroscopy and imaging are described in Section \ref{observations}. Section \ref{indiv} details the characteristics of individual systems, and a general discussion on the samples and sub-samples are discussed in Section \ref{discussion}. We conclude in Section \ref{conclusion}.

\section{Candidate Selection} \label{selection}
Candidates were selected following \citet{lemon2019}, and represent the follow-up of different Right Ascension and Declination ranges, as well as untargeted systems in previously covered sky, due to lack of spectroscopic follow-up time. Two searches begin from a catalogue of quasars and quasar candidates. For these searches we used the Milliquas catalogue \citep{milliquas}, which consists of confirmed quasars and highly likely quasar candidates based on X-ray, radio, and optical and infrared colour selection. We supplement this catalogue with objects using colours from the Wide-field Infrared Survey Explorer (WISE). We take objects with W1--W2 > 0.4 (in Vega), which is less strict than most quasar selections in order to include systems with possibly large contamination from the lensing galaxy, or sources with non-standard quasar colours. Our final search begins from a morphological galaxy catalogue. Below, we briefly describe the selection techniques for lensed quasars within these quasar and galaxy samples, making use of \textit{Gaia} data release 2 \citep[GDR2, ][]{gaiadr2}. Further details on selection numbers and completeness of recovering existing samples can be found in \citet{lemon2019}.

\subsection{Multiple \textit{Gaia} detections around quasar catalogues}
This search is refined from that originally performed in \citet{lemon2018}, namely searching for multiple \textit{Gaia} detections around red ALLWISE detections \citep{mainzer2011}. The search consists of visually inspecting ground-based colour images of systems satisfying varying W1--W2 colour cuts, local stellar densities and thresholds on the significance of proper motion for the different numbers of \textit{Gaia} detections associated to the system. The Milliquas catalogue is also used as a starting catalogue for this selection. 

\subsection{Modelling unWISE pixels}
The main contaminants of the search detailed above are quasar+star projections, since they will have multiple \textit{Gaia} detections, often lacking the necessary proper motion precision to remove the star as a moving object, and having quasar-like colours in the infrared, radio, or X-ray due to blending issues in WISE (PSF FWHM of $\sim$5\arcsec). Ground-based imaging can often rule out such pairs through optical colour differences or a lack of a lensing galaxy, however contaminants still remain and the number of objects is too large to visually inspect. \citet{lemon2019} introduced a forced model fitting of WISE data based on \textit{Gaia} positions to extract the infrared colours of nearby point sources and thus classify the multiple components within such systems. We used the unWISE W1 and W2 data \citep{lang2016}, an unblurred coaddition of all WISE data to-date. Following local \textit{Gaia} density cuts, proper motion significance cuts, and finally cuts in the G--W1 vs. W1--W2 colour space, we obtained a classification procedure to retain nearly 99 per cent of known lensed quasars, while removing $\sim$80 per cent of previously followed-up contaminants.

\subsection{\textit{Gaia} detections offset from LRGs} \label{offset}
Since both search techniques presented above rely heavily on source selection, we also investigate using galaxies as a starting catalogue. We use spectroscopic galaxies from SDSS DR12, and keep all systems with one \textit{Gaia} detection within 1 and 3.5\arcsec, or two \textit{Gaia} detections within 4.5\arcsec (and separated by less than 4.5\arcsec). The need for the single detection to be further than 1\arcsec is based on known lensed quasars satisfying such a constraint, and reduces contamination. We also use morphological galaxies from Pan-STARRS, requiring at least one \textit{Gaia} detection between 1 and 3\arcsec\ away. Proper motion and astrometric excess noise cuts are performed on all matching \textit{Gaia} detections to further remove contamination from stars and star-forming galaxies.

\section{Observations} \label{observations}
\subsection{Spectroscopy}

\subsubsection{William Herschel Telescope: ISIS}
55 systems were observed with the Intermediate-dispersion Spectrograph and Imaging System (ISIS) on the WHT on the nights of 11-12 Feb. 2019. The R158R and R300B gratings were used for the red and blue arms respectively, with the standard 5300\AA\ dichroic and GG495 second order cut filter in the red arm. This provided dispersions of 1.81\AA\ pixel$^{-1}$ and 0.86\AA\ pixel$^{-1}$ for the red and blue arms. 

\subsubsection{Nordic Optical Telescope: ALFOSC}
43 systems were observed with grism \#4 and the Alhambra Faint Object Spectrograph and Camera (ALFOSC) on the NOT on 1-2 October 2019, providing a dispersion of 3.3\AA\ pixel$^{-1}$. All spectra were reliably deblended, except J2017+6204, for which high-resolution imaging confirms it as a lensed quasar.

\subsubsection{New Technology Telescope: EFOSC2}
79 systems were observed with grism \#13 of the ESO Faint Object Spectrograph and Camera version 2 (EFOSC2) over four runs (NTT1, 0104.A-0194(A): 22-24 Oct. 2019; NTT2, 0104.A-0194(B): 18-20 Jan. 2020; NTT3, 106.218K.001: 19-21 Oct. 2020; and NTT4, 106.218K.002: 15-17 Jan. 2021), providing a dispersion of 2.77\AA\ pixel$^{-1}$.

\subsubsection{General reduction procedure}
Since several spectrographs were used for long-slit spectroscopic follow-up of our candidates, we give a brief description of the common steps. The extraction follows that outlined in \citet{lemon2022}, and fits multiple Moffat components, the parameters for which are determined from the binned data themselves. In particular, the following steps are taken:

\begin{itemize}
    \item After bias-subtracting each image, cosmic rays are found and masked as all pixels above a certain threshold in the absolute Laplacian of the data. For the brightest objects this value is increased depending upon visual inspection of the cosmic ray mask.
    \item The sky background is subtracted by determining the median value within the pixels either side of the trace of the object (between 20 and 80 pixels away). We later use this same background to correct the wavelength calibration to bright sky lines.  The region is also used to determine the sky background noise at each spectral pixel (i.e., wavelength). 
    \item A Poisson noise map is created using the detector gain, and added in quadrature with the sky noise to provide a pixel noise map.
    \item A wavelength solution is determined from fitting the lines of an arc exposure (HeNe or CuAr) taken the afternoon before observations. An absolute shift is calculated for each exposure by measuring the positions of the Na 5892\AA, \textsc{[Oi]}-5577\AA\ and 6300\AA\ sky lines. This is often a significant shift ($\gtrapprox$ 1 pixel) and shows variation with wavelength, and so absolute wavelength measurements should not be trusted to more than the approximate pixel dispersion value of each instrument/grism combination.
\end{itemize}

\subsection{Imaging}
Deep imaging is key to identifying the lensing galaxy or counterimage in lensed quasar candidates, and thus we use several imaging datasets throughout this work. Principally we use \textit{grizY} imaging from Pan-STARRS \citep{chambers2016}, and the \textit{grz} imaging from the DESI Legacy Imaging Surveys \citep{dey2019}. When higher-resolution imaging is required or a system is investigated for other catalogue detections, we check the Canadian Astronomy Data Centre archive, Vizier catalogues within 5 arcseconds \citep{vizier}, and nearby NED detections and associated references. For some quadruply-imaged systems \textit{HST} multi-band imaging and models are presented in \citet{schmidt2022}.



\subsubsection{General imaging analysis}
When analysing individual systems, we often want to fit the pixels with a combination of analytic profiles. Most commonly, we want to identify if a lensing galaxy is present between two bright quasar PSFs. For each image available, we use a nearby star in the field and fit this as a Moffat profile to determine the PSF of the system \citep{moffat1969}. We then fit a square 10$\times$10\arcsec\ cutout around the system, using the relevant number of point sources. This process is performed simultaneously in all bands (or possibly a single band), and if a galaxy profile is required we use a S{\'e}rsic profile \citep{sersic1963}. The free parameters are the position(s) of the point sources, the position(s) of the galaxies, and their associated effective radii, S{\'e}rsic indices, and ellipiticity parameters. To find the best-fit parameters we use the Markov chain Monte Carlo (MCMC) sampler \textsc{emcee} Python package \citep{foreman-mackey2013}.

\subsection{Results} 
The resulting classifications from analysing both the available imaging and spectroscopy of each system are listed in Table \ref{tab:observations2}, with the selection method, maximum image separation, and \textit{Gaia} magnitudes and proper motion significances.

We broadly classify the systems into the following categories: \textit{definite lensed quasars}, \textit{likely lensed quasars} (based on the subjective opinion and experience of the authors), \textit{unclassified quasar pairs} (systems in which two quasars are spectroscopically resolved and are at the same redshift however no convincing lensing galaxy is seen but also for which there is no clear evidence for them being physically distinct quasars), \textit{binary quasars} (Section \ref{binaries}), \textit{projected quasars} (Section \ref{projected}), and \textit{contaminant systems} (definitely not lenses). We note that our use of unclassified quasar pairs is akin to Nearly Identical Quasars (NIQs) which has been used in recent literature \citep[e.g.][]{anguita2018}, however the change is simply to include those systems which have low signal-to-noise data and could readily be classified as a likely lensed quasar or binary quasar given deeper spectra.

\begin{figure*}
\centering
\includegraphics[width=\textwidth]{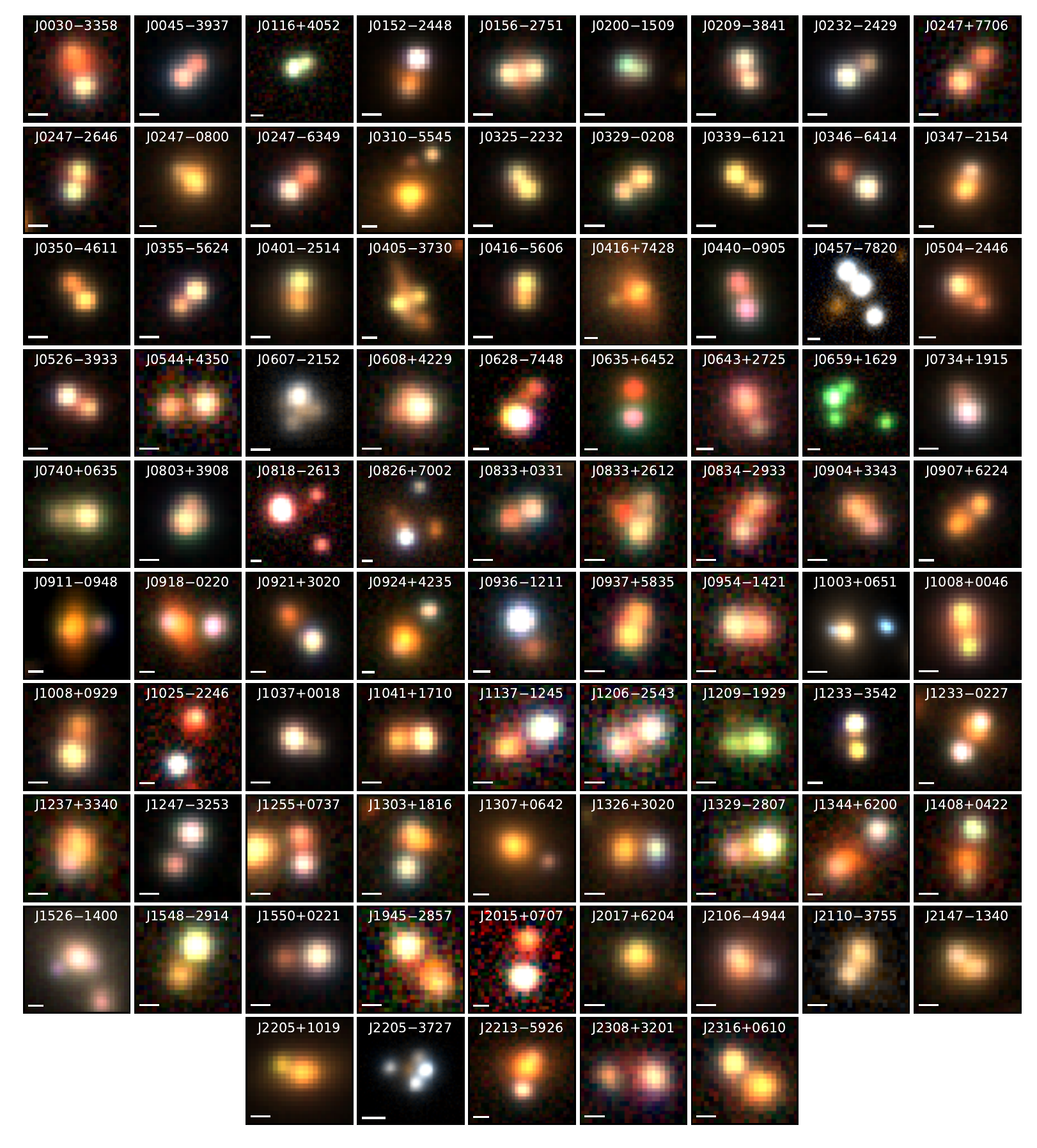}
\vspace{-0.5cm}
\caption{Colour images of the confirmed lensed quasars. HSC \textit{gri}, Legacy Survey \textit{grz}, and Pan-STARRS \textit{gri} are used for the majority of the cutouts (with preference in the given order). Other imaging is shown as explained in the relevant system subsection of Section \ref{indiv}. White scale bars are 1\arcsec. North is up, East is left.}
          \label{fig:lenses}%
\end{figure*}

\begin{figure*}
\centering
\includegraphics[width=\textwidth]{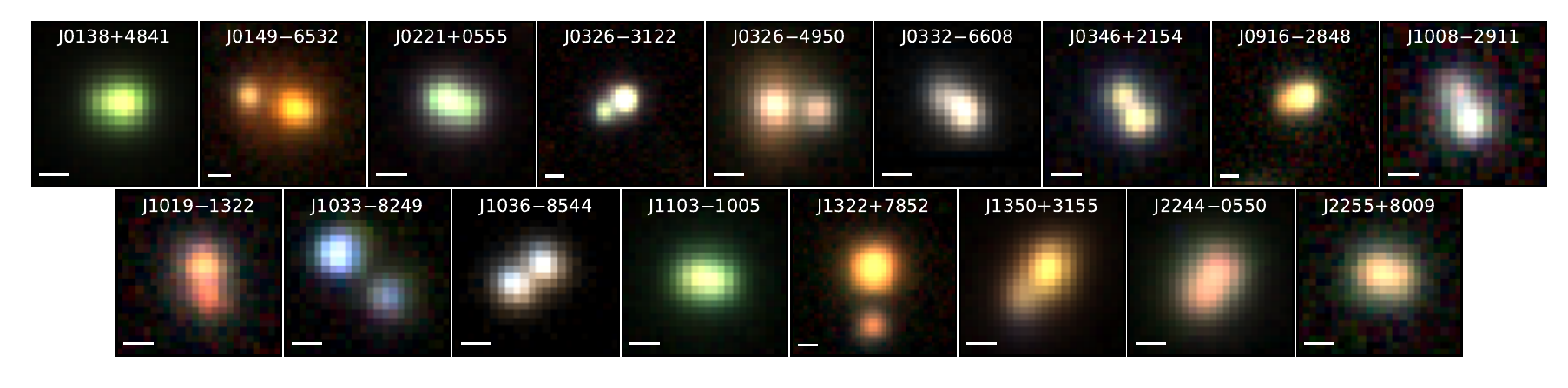}
\vspace{-0.5cm}
\caption{Colour images of the probable lensed quasars, requiring deeper imaging or spectra. Scalebars are 1 arcsecond.}
          \label{probable}%
\end{figure*}

\begin{figure*}
\centering
\includegraphics[width=\textwidth]{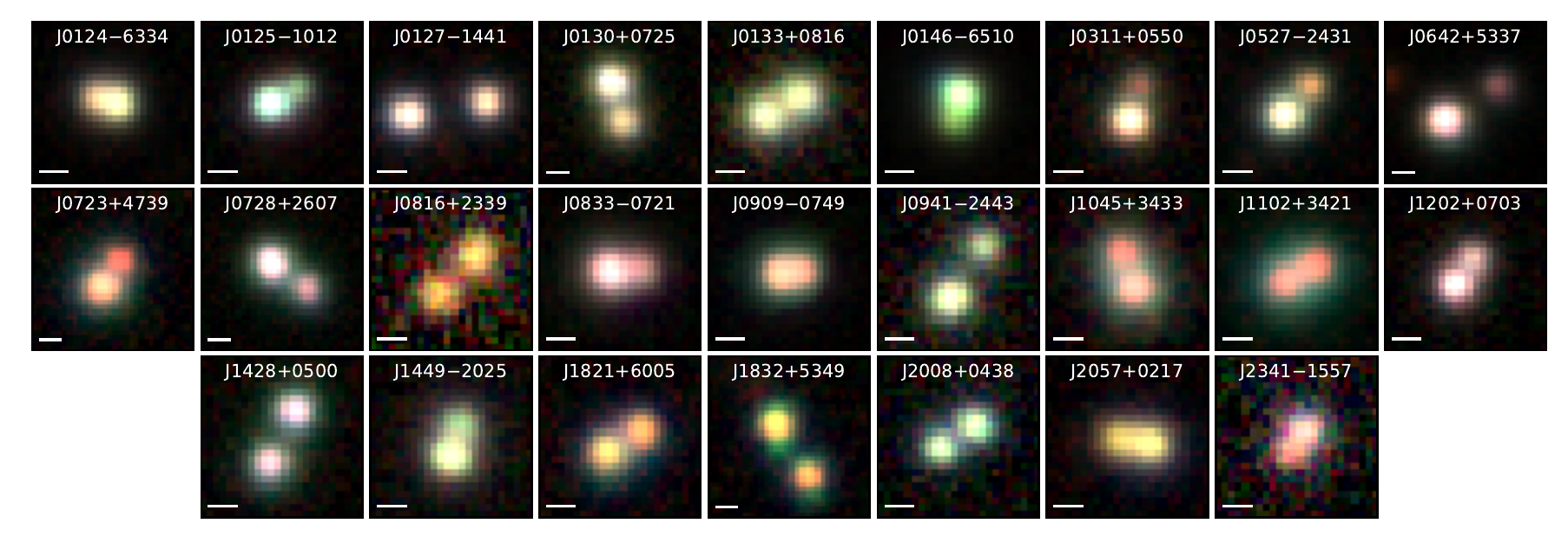}
\caption{Colour images of the unclassified quasar pairs.}
          \label{niqs}%
\end{figure*}

Extracted imaging and spectra of systems are respectively shown in Figures \ref{fig:lenses} and \ref{lens_spectra1} for confirmed lensed, \ref{probable} and \ref{probable_spectra} for likely lensed quasars, and \ref{niqs} and \ref{niqs_spectra} for unclassified quasar pairs. We also show the spectra for binary quasars and projected quasars in Figures  \ref{binaries_spectra} and \ref{projected_spectra} respectively.


\newcolumntype{P}{>{\centering\arraybackslash}p{20mm}}
\begin{table*}
    \centering
    \caption{Summary of observed targets and results based on spectroscopy and imaging. Selections are abbreviated as follows: W$=$WISE catalogue, M$=$Milliquas catalogue with multiple \textit{Gaia} detections (D, T, Q for double, triple, and quartet); GP$=$\textit{Gaia} singlet offset from a galaxy. Separations are the largest of the \textit{Gaia} separations if more than two \textit{Gaia} DR2 detections are associated to the system, or based on pixel modelling. PMSIG and runs are described in the text. }
    \label{tab:observations2}
    \begin{tabular}{lcccPcccl}
        \hline
        Name & R.A. & Dec. & Selection & \textit{Gaia} G & Sep. (\arcsec) & PMSIG & Run & Classification \\
        \hline
J0021+1927 & 5.4936 & 19.4646 & WD, MD & 20.77, 19.96 & 2.89 & 1.05, 2.74 & NOT & projected QSOs, \textit{z}$=$1.045, 1.09 \\
J0027+0438 & 6.9580 & 4.6443 & WD, MD & 17.65, 20.18 & 1.92 & 0.56, --- & NOT & projected QSOs, \textit{z}$=$0.1935, 1.972 \\
J0029--0414 & 7.3861 & -4.2472 & WD, MD & 19.27, 20.43 & 1.11 & 1.13, --- & NTT3 & \textit{z}$=$0.518 QSO + star \\
J0030--3358 & 7.6740 & -33.9767 & GP & 20.89 & 2.03 & --- & NTT1 & lens, \textit{z}$=$1.58, $z_{\textrm{lens}}$=0.715 \\
J0032--4523 & 8.1130 & -45.3884 & WD, MD & 20.57, 20.21 & 2.33 & 0.12, 0.56 & NTT3 & projected QSOs, \textit{z}$=$1.667, 1.74 \\
J0041--5350 & 10.4496 & -53.8460 & WD, MD & 20.79, 20.85 & 1.06 & 1.29, --- & NTT1 & QSO pair, \textit{z}$=$0.55 \\
J0045--3937 & 11.3665 & -39.6262 & WD & 19.70, 18.98 & 1.14 & 1.39, 0.60 & NTT2 & lens, \textit{z}$=$1.85 \\
J0116+4052 & 19.1635 & 40.8811 & WD, MD & 19.84, 18.96 & 1.28 & ---, 1.07 & NOT & lens, \textit{z}$=$1.85 \\
J0124--6334 & 21.0089 & -63.5788 & WD, MD & 20.50, 19.60 & 0.87 & 2.63, 5.19 & NTT2 & NIQ, \textit{z}$=$1.30 \\
J0125--1012 & 21.3174 & -10.2082 & WD, MD & 19.26, 20.63 & 1.12 & 1.86, --- & NTT2 & NIQ, \textit{z}$=$1.22 \\
J0127--1441 & 21.7854 & -14.6886 & WD, MD & 20.14, 20.53 & 3.01 & 0.14, 1.32 & NTT3 & NIQ, \textit{z}$=$1.754 \\
J0130+0725 & 22.5859 & 7.4212 & WD, MD & 18.98, 19.65 & 2.06 & 1.33, --- & NOT & NIQ, \textit{z}$=$1.54 \\
J0133+0816 & 23.2695 & 8.2788 & WD, MD & 20.24, 20.57 & 1.55 & ---, --- & NOT & NIQ, \textit{z}$=$1.27 \\
J0138+4841 & 24.5914 & 48.6963 & WD & 17.45, 16.42 & 0.69 & ---, 1.00 & NOT & NIQ, \textit{z}$=$1.172 \\
J0146--6510 & 26.6467 & -65.1799 & WD & 17.48, 18.71 & 1.00 & 3.49, 0.76 & NTT1 & NIQ, \textit{z}$=$1.255 \\
J0149--6532 & 27.2906 & -65.5404 & GP & 20.58 & 2.74 & --- & NTT1 & lens(?), \textit{z}$=$0.944?, \textit{z}$=$0.395 \\
J0152--2448 & 28.0797 & -24.8105 & WD, MD & 20.12, 18.66 & 1.91 & 3.81, 2.05 & NOT & lens, \textit{z}$=$1.69 \\
J0156--2751 & 29.1039 & -27.8562 & WD, MD & 20.40, 20.46 & 1.50 & 0.68, 1.81 & NTT1 & lens, \textit{z}$=$2.97 \\
J0200--1509 & 30.0872 & -15.1609 & WD, MD & 20.43, 19.72 & 0.78 & ---, --- & NTT1 & lens, \textit{z}$=$1.185 \\
J0209--3841 & 32.3047 & -38.6961 & WD, MD & 20.60, 20.21 & 1.23 & 1.06, 1.35 & NTT1 & lens, \textit{z}$=$2.87 \\
J0221+0555 & 35.4474 & 5.9192 & WD, MD & 19.96, 19.39 & 0.86 & 1.08, 1.19 & NOT & NIQ, \textit{z}$=$1.52 \\
J0232--2429 & 38.0655 & -24.4942 & WD, MD & 18.00, 19.44 & 1.44 & 2.01, 2.97 & NTT1 & lens, \textit{z}$=$1.578 \\
J0246--0131 & 41.6793 & -1.5305 & WD, MD & 19.05, 20.10 & 1.09 & 4.06, --- & NTT2 & \textit{z}$=$0.503 QSO + star \\
J0247+7706 & 41.8664 & 77.1014 & WD & 19.49, 20.20 & 1.94 & 1.30, 1.10 & NOT & lens, \textit{z}$=$2.73 \\
J0247--2646 & 41.8764 & -26.7729 & WD & 20.49, 20.78 & 1.22 & ---, --- & NTT1 & lens, \textit{z}$=$1.47 \\
J0247--0800 & 41.9561 & -8.0150 & GP & 20.55, 20.70 & 1.68 & ---, --- & NTT2 & lens, \textit{z}$=$3.28, $z_{\textrm{lens}}$=0.198 \\
J0247--6349 & 41.9780 & -63.8232 & WD, MD & 19.78, 20.94 & 1.49 & 0.29, --- & NTT1 & lens, \textit{z}$=$2.295 \\
J0249+2606 & 42.4595 & 26.1141 & WD, MD & 18.81, 20.15 & 3.21 & 2.18, 1.99 & NOT & projected QSOs, \textit{z}$=$1.514, 2.68 \\
J0310--5545 & 47.7029 & -55.7534 & GP &  & 3.57 &  & NTT3 & lens, \textit{z}$=$2.31, $z_{\textrm{lens}}$=0.298 \\
J0311+0550 & 47.7708 & 5.8367 & WD, MD & 18.36, 20.01 & 1.36 & 0.49, 1.21 & NOT & NIQ, \textit{z}$=$0.777 \\
J0315--3522 & 48.9099 & -35.3827 & WD, MD & 20.22, 17.92 & 1.30 & 1.06, 2.18 & NTT2 & QSO pair, \textit{z}$=$0.457 \\
J0325--2232 & 51.4511 & -22.5409 & WD, MD & 19.74, 19.52 & 1.01 & 0.28, 1.29 & NTT1 & lens, \textit{z}$=$1.35 \\
J0326--3122 & 51.5284 & -31.3816 & WD, MD & 20.68, 19.47 & 1.43 & 2.05, 2.16 & NTT3 & NIQ, \textit{z}$=$1.345 \\
J0326--4950 & 51.7287 & -49.8337 & WD, MD & 19.68, 20.36 & 1.63 & 1.25, 0.64 & NTT2 & \textit{z}$=$0.229 GAL + \textit{z}$=$1.72 QSO \\
J0329--0208 & 52.4280 & -2.1379 & WD & 19.99, 19.80 & 1.32 & 0.69, 0.25 & NTT3 & lens, \textit{z}$=$1.352 \\
J0332--6608 & 53.0841 & -66.1400 & WD, MD & 18.30, 19.46 & 0.94 & 7.98, 1.90 & NTT2 & NIQ, \textit{z}$=$1.97 \\
J0336--3244 & 54.0816 & -32.7410 & WD & 20.78, 18.13 & 1.29 & ---, 5.21 & NTT2 & \textit{z}$=$0.565 QSO + other \\
J0339--6121 & 54.7870 & -61.3625 & WD & 18.62, 19.52 & 1.26 & 0.30, 0.83 & NTT1 & lens, \textit{z}$=$2.89  \\
J0346+2154 & 56.5458 & 21.9124 & WD, MD & 18.98, 18.76 & 0.99 & ---, 4.55 & NOT & NIQ, \textit{z}$=$2.365 \\
J0346--6414 & 56.5579 & -64.2417 & GP & 19.41 & 1.91 & --- & NTT1 & lens, \textit{z}$=$2.96 \\
J0347--2154 & 56.7690 & -21.9095 & GP & 20.01, 19.46 & 1.87 & ---, 2.38 & WHT/NTT1 & lens, \textit{z}$=$0.81, $z_{\textrm{lens}}$=0.187 \\
J0350--4611 & 57.7146 & -46.1858 & WD, MD & 20.66, 19.68 & 1.30 & 1.52, 2.86 & NTT1 & lens, \textit{z}$=$1.505 \\
J0355--5624 & 58.8040 & -56.4147 & WD, MD & 20.93, 20.03 & 1.29 & ---, 2.14 & NTT1 & lens, \textit{z}$=$1.86 \\
J0401--2514 & 60.4504 & -25.2439 & WD, MD & 19.03, 20.22 & 1.45 & 1.67, 1.60 & WHT & lens, \textit{z}$=$1.32 \\
J0405--3730 & 61.2721 & -37.5128 & WD & 20.25, 20.57 & 1.70 & 0.58, 0.35 & NTT2 & lens, \textit{z}$=$3.585, $z_{\textrm{lens}}$=0.2295 \\
J0408+6333 & 62.2485 & 63.5522 & WD & 14.77, 16.89 & 0.99 & 3.19, 2.13 & NOT & stars \\
J0416--5606 & 64.1783 & -56.1073 & WD, MD & 20.08, 20.81 & 1.04 & 1.66, --- & NTT1 & lens, \textit{z}$=$1.45 \\
J0416+7428 & 64.1972 & 74.4827 & WD & 19.74, 19.02 & 2.64 & 1.96, 3.57 & WHT & lens, \textit{z}$=$0.900 $z_{\textrm{lens}}$=0.097 \\
J0429--2246 & 67.4161 & -22.7676 & WD & 17.42, 19.98 & 1.80 & 1.08, --- & NTT2 & galaxies, \textit{z}$=$0.206 \\
J0436+7851 & 69.1080 & 78.8530 & WD, MD & 19.00, 18.66 & 4.64 & 1.44, 0.82 & NOT & projected QSOs, \textit{z}$=$0.765, 1.975 \\
J0440--0905 & 70.0482 & -9.0911 & WD, MD & 19.92, 19.19 & 1.69 & 2.38, 2.15 & WHT & lens, \textit{z}$=$2.105 \\
J0457--7820 & 74.3483 & -78.3466 & WT & 19.60, 18.81, 19.07 & 4.89 & 1.12, 1.14, 1.50 & NTT1 & lens, \textit{z}$=$3.145 \\
J0504--2446 & 76.1616 & -24.7752 & GP & 20.50, 19.22 & 2.05 & 1.48, 5.44 & WHT & lens, \textit{z}$=$0.771 \\
J0526--3933 & 81.5470 & -39.5630 & WD & 19.99, 20.69 & 1.41 & 1.88, --- & NTT1 & lens, \textit{z}$=$2.41 \\
J0527--2431 & 81.7608 & -24.5328 & WD & 19.53, 20.57 & 1.46 & 2.60, --- & NTT2 & NIQ, \textit{z}$=$1.437 \\
J0544+4350 & 86.1448 & 43.8386 & WD & 19.64, 20.25 & 2.04 & 0.40, 1.09 & WHT & lens, \textit{z}$=$3.11 \\
J0553+0910 & 88.2723 & 9.1812 & WD, MD & 19.71, 19.73 & 1.11 & 2.47, 2.63 & NOT & projected QSOs, \textit{z}$=$1.08, 1.85 \\
J0601--2220 & 90.2664 & -22.3464 & WD, MD & 20.65, 20.43 & 0.98 & 0.44, 2.12 & NTT2 & projected QSOs, \textit{z}$=$0.809, 0.836 \\
J0607--2152 & 91.7954 & -21.8716 & WT & 20.86, 19.32, 20.90 & 1.57 & ---, 2.05, --- & WHT & quad, \textit{z}$=$1.305 \\
J0608+4229 & 92.1725 & 42.4935 & WT, MT & 20.13, 18.23, 19.86 & 1.27 & ---, 3.94, --- & WHT & quad, \textit{z}$=$2.346 \\
J0628--7448 & 97.2299 & -74.8010 & WD, MD & 17.89, 20.50 & 2.61 & 2.27, 1.96 & NTT2 & lens, \textit{z}$=$2.692 \\

        \hline
    \end{tabular}
\end{table*}

\begin{table*}
     \contcaption{}
     \label{tab:continued1}
    \centering
    \begin{tabular}{lcccPcccl}
        \hline
        Name & R.A. & Dec. & Selection & \textit{Gaia} G & Sep. (\arcsec) & \textit{Gaia} P.M. sig. & Run & Classification \\
        \hline
J0635+6452 & 98.9864 & 64.8715 & WD, MD & 20.86, 18.96 & 3.07 & ---, 0.55 & WHT & lens, \textit{z}$=$1.845, $z_{\textrm{lens}}$=0.427 \\
J0642+5337 & 100.7079 & 53.6284 & WD, MD & 19.74, 17.79 & 3.17 & 1.29, 1.31 & NOT & NIQ, \textit{z}$=$1.86 \\
J0643+2725 & 100.9259 & 27.4276 & WD & 19.48, 19.82 & 2.44 & 0.88, 1.64 & WHT & lens, \textit{z}$=$1.562, $z_{\textrm{lens}}$=0.185 \\
J0659+1629 & 104.7667 & 16.4859 & WT & 18.59, 19.94, 20.05 & 5.25 & 1.73, 1.96, 0.78 & WHT & quad, \textit{z}$=$3.09 \\
J0707+4109 & 106.9469 & 41.1640 & WD, MD & 20.53, 20.47 & 2.96 & 1.57, 1.30 & WHT & QSO pair, \textit{z}$=$0.51, 0.516 \\
J0723+4739 & 110.9364 & 47.6529 & WD, MD & 20.36, 19.72 & 1.58 & ---, 2.42 & NOT & NIQ, \textit{z}$=$0.842 \\
J0728+2607 & 112.1160 & 26.1173 & WD, MD & 18.92, 19.89 & 2.15 & 1.89, 0.38 & NOT & NIQ, \textit{z}$=$1.025 \\
J0734+1915 & 113.6936 & 19.2501 & WD, MD & 18.10, 19.96 & 1.20 & 2.10, 0.49 & NOT & lens, \textit{z}$=$0.745 \\
J0737+0925 & 114.3240 & 9.4298 & WD, MD & 19.90, 20.51 & 0.97 & 2.12, --- & NTT4 & \textit{z}$=$0.308 QSO + star \\
J0740+0635 & 115.0907 & 6.5938 & WD & 19.56, 18.32 & 1.61 & 1.55, 1.84 & WHT & lens, \textit{z}$=$1.78 \\
J0803+3908 & 120.9906 & 39.1398 & WD, MD & 18.88, 18.14 & 0.91 & 0.76, --- & WHT & quad, \textit{z}$=$2.97 \\
J0805+3550 & 121.2733 & 35.8472 & WD, MD & 19.40, 19.93 & 1.13 & 3.62, 1.34 & NOT & QSO pair, \textit{z}$=$1.66, 1.67 \\
J0816+2339 & 124.1428 & 23.6630 & WD, MD & 20.94, 20.32 & 1.99 & ---, 1.33 & WHT & NIQ, \textit{z}$=$1.22 \\
J0818--2613 & 124.6179 & -26.2237 & WQ & 19.74, 17.58, 19.94, 17.52 & 6.23 & 2.07, ---, 2.01, --- & WHT & quad, \textit{z}$=$2.155 \\
J0819+0457 & 124.8686 & 4.9522 & WD & 20.85, 20.58 & 1.84 & ---, 0.74 & WHT & \textit{z}$=$0.456 QSO + other \\
J0826+7002 & 126.5341 & 70.0456 & GP & 17.46, 19.30 & 5.83 & 1.06, 1.64 & WHT & lens, \textit{z}$=$1.618 \\
J0833+0331 & 128.3369 & 3.5247 & WD & 20.08, 21.10 & 1.55 & 1.16, --- & WHT & lens, \textit{z}$=$1.845 \\
J0833--0721 & 128.4731 & -7.35185 & WD, MD & 19.45, 18.67 & 1.22 & 2.03, 1.65 & NTT2 & NIQ, \textit{z}$=$0.828 \\
J0833+2612 & 128.4767 & 26.2029 & GP & 20.93 & 1.79 & --- & NTT4 & quad, \textit{z}$=$3.26 \\
J0834--2933 & 128.6411 & -29.5505 & WD & 19.91, 20.17 & 1.80 & 2.96, 2.11 & WHT & lens, \textit{z}$=$1.922 \\
J0839--0056 & 129.9448 & -0.9371 & GP & 18.71 & 1.66 & --- & NTT4 & \textit{z}$=$0.17 galaxy + star \\
J0904+3343 & 136.1714 & 33.7291 & WD & 20.13, 19.93 & 1.50 & 4.06, 1.19 & WHT & lens, \textit{z}$=$2.49 \\
J0907+6224 & 136.9660 & 62.4116 & WD, MD & 20.06, 19.27 & 2.49 & 1.45, 0.54 & WHT & lens, \textit{z}$=$1.86 \\
J0909--0749 & 137.4946 & -7.8179 & WD, MD & 18.24, 19.11 & 0.81 & 0.71, 1.13 & NTT4 & NIQ, \textit{z}$=$1.075 \\
J0911--0948 & 137.7845 & -9.8054 & GP & 20.84 & 2.52 & --- & NTT3 & lens, \textit{z}$=$1.47, $z_{\textrm{lens}}$=0.251 \\
J0916--2848 & 139.1646 & -28.8143 & WD, MD & 18.46, 19.47 & 1.01 & 1.88, 0.84 & NTT2 & NIQ, \textit{z}$=$1.925 \\
J0918--0220 & 139.6806 & -2.3354 & GP & 20.43 & 2.26 & --- & WHT & lens, \textit{z}$=$0.803, $z_{\textrm{lens}}$=0.460 \\
J0921+3020 & 140.2685 & 30.3421 & MD & 18.66, 20.85 & 2.93 & 1.40, --- & WHT & lens, \textit{z}$=$3.335, $z_{\textrm{lens}}$=0.428 \\
J0924+4235 & 141.1243 & 42.5947 & MD & 19.77, 20.70 & 4.64 & 0.90, --- & WHT & lens, \textit{z}$=$3.17, $z_{\textrm{lens}}$=0.415 \\
J0936--1211 & 144.2494 & -12.1836 & WD & 17.49, 20.49 & 2.11 & 0.65, --- & NTT2 & lens, \textit{z}$=$2.00 $z_{\textrm{lens}}$=0.260 \\
J0937+5835 & 144.3832 & 58.5906 & WD & 20.93, 20.45 & 1.44 & ---, 1.42 & WHT & lens, \textit{z}$=$2.115 \\
J0938+0629 & 144.6502 & 6.4973 & GP & 19.21 & 1.15 & --- & NTT4 & QSO+galaxy, \textit{z}$=$0.363 \\
J0939--0109 & 144.8548 & -1.1636 & GP & 19.05 & 1.90 & --- & NTT4 & QSO pair, \textit{z}$=$0.225 \\
J0941--2443 & 145.49455 & -24.7308 & WD, MD & 20.48, 19.51 & 2.25 & 1.21, 1.69 & NTT2 & NIQ, \textit{z}$=$0.92 \\
J0947+0247 & 146.8048 & 2.7955 & WD, MD & 19.60, 20.54 & 1.49 & 1.04, --- & NTT4 & \textit{z}$=$0.643 QSO + star \\
J0954--1421 & 148.7079 & -14.3528 & WD, MD & 20.58, 19.80 & 1.40 & ---, 1.79 & WHT & lens, \textit{z}$=$0.973 \\
J1003+0651 & 150.7886 & 6.8501 & GP & 19.85, 19.80, 20.20 & 2.62 & 0.84, ---, --- & WHT & lens, \textit{z}$=$2.565, $z_{\textrm{lens}}$=0.225 \\
J1008+0046 & 152.1932 & 0.7724 & WD, MD & 20.60, 20.60 & 1.31 & ---, --- & WHT & lens, \textit{z}$=$1.51 \\
J1008+0929 & 152.2041 & 9.4878 & GP & 19.01 & 1.76 & --- & WHT & lens, \textit{z}$=$3.08 \\
J1008--2911 & 152.2474 & -29.1887 & WD & 20.41, 19.56 & 1.14 & 1.39, 1.12 & NTT2 & NIQ, \textit{z}$=$2.49 \\
J1019--1322 & 154.8066 & -13.3692 & WD & 19.50, 20.44 & 1.14 & 0.18, --- & NTT2 & NIQ, \textit{z}$=$2.325 \\
J1019--3516 & 154.9803 & -35.2742 & WD & 17.84, 19.99 & 1.76 & 1.44, --- & NTT2 & \textit{z}$=$0.136 QSO + galaxy \\
J1025--2246 & 156.3567 & -22.7690 & WD & 19.94, 21.31 & 3.92 & 0.20, --- & WHT & lens, \textit{z}$=$1.97 \\
J1033--8249 & 158.4076 & -82.8297 & WD, MD & 18.91, 20.27 & 2.33 & 0.57, 2.37 & NTT2 & NIQ, \textit{z}$=$1.69 \\
J1036--8544 & 159.1077 & -85.7471 & WD, MD & 19.86, 19.73 & 1.22 & 1.56, 1.68 & NTT2 & NIQ, \textit{z}$=$1.09 \\
J1037+0018 & 159.3665 & 0.3057 & WD, MD & 19.95, 17.99 & 1.25 & ---, 2.46 & NTT4 & lens, \textit{z}$=$2.462 \\
J1041+1710 & 160.3081 & 17.1798 & WD, MD & 20.97, 19.49 & 1.62 & ---, 0.28 & NTT4 & lens, \textit{z}$=$2.01 \\
J1041--0836 & 160.3519 & -8.6134 & WD & 17.81, 19.38 & 1.01 & 2.75, 0.76 & NTT2 & galaxy, \textit{z}$=$0.056 \\
J1045+3433 & 161.4977 & 34.5659 & WD, MD & 20.93, 20.38 & 1.42 & ---, 1.86 & WHT & NIQ, \textit{z}$=$1.205 \\
J1102+3421 & 165.5267 & 34.3624 & WD, MD & 19.28, 19.63 & 1.25 & 2.28, 0.17 & WHT & NIQ, \textit{z}$=$1.405 \\
J1103--1005 & 165.8665 & -10.0940 & WD, MD & 17.79, 18.01 & 0.72 & ---, --- & NTT2 & NIQ, \textit{z}$=$1.292 \\
J1116--2122 & 169.1954 & -21.3799 & WD & 18.44, 19.33 & 3.29 & 2.01, 0.40 & NTT2 & QSO pair, \textit{z}$=$0.709 \\
J1137--1245 & 174.3128 & -12.7506 & WD & 20.78, 19.55 & 2.37 & 0.99, 0.52 & WHT & lens, \textit{z}$=$2.18 \\
J1140+2303 & 175.0308 & 23.0637 & WD, MD & 20.19, 19.46 & 3.63 & 1.11, 2.12 & NTT4 & \textit{z}$=$2.405 QSO + star(?) \\
J1144+1037 & 176.1344 & 10.6296 & WD, MD & 20.30, 19.69 & 1.00 & 3.64, 1.37 & NTT4 & \textit{z}$=$0.966 QSO + star \\
J1202+0703 & 180.6182 & 7.0572 & WD & 19.76, 20.74 & 1.21 & 1.99, --- & NTT4 & NIQ, \textit{z}$=$2.187 \\
J1206--2543 & 181.7448 & -25.7254 & WD, MD & 19.97, 20.40 & 2.08 & 1.27, 3.58 & WHT & lens, \textit{z}$=$1.765 \\
J1209--1929 & 182.3615 & -19.4879 & WD & 19.55, 20.66 & 1.46 & 0.89, --- & WHT & lens, \textit{z}$=$2.91 \\
J1233--3542 & 188.3261 & -35.7160 & WD & 19.56, 18.54 & 2.03 & 0.66, 1.26 & WHT & lens, \textit{z}$=$2.28 \\
J1233--0227 & 188.4219 & -2.4604 & WD & 19.74, 19.98 & 1.76 & 3.03, 2.91 & WHT & lens, \textit{z}$=$1.598, $z_{\textrm{lens}}$=0.345? \\
J1237+3340 & 189.2537 & 33.6820 & GP & 20.90 & 1.57 & --- & WHT & lens, \textit{z}$=$1.795, $z_{\textrm{lens}}$~0.35? \\
J1247--3253 & 191.9855 & -32.8919 & WD, MD & 19.09, 18.27 & 1.85 & 0.42, 2.01 & NTT2 & lens, \textit{z}$=$1.66 \\
J1255+0737 & 193.9324 & 7.6296 & WD & 20.40, 21.03 & 1.70 & 0.70, --- & WHT & lens, \textit{z}$=$2.145 \\

        \hline
    \end{tabular}
\end{table*}

\begin{table*}
     \contcaption{}
     \label{tab:continued2}
    \centering
    \begin{tabular}{lcccPcccl}
        \hline
        Name & R.A. & Dec. & Selection & \textit{Gaia} G & Sep. (\arcsec) & \textit{Gaia} P.M. sig. & Run & Classification \\
        \hline
J1303+1816 & 195.7765 & 18.2781 & WD & 20.20, 21.12 & 2.26 & 4.18, --- & WHT & lens, \textit{z}$=$2.95, $z_{\textrm{lens}}$=0.46 \\
J1307+0642 & 196.9290 & 6.7035 & GP & 20.16 & 3.51 & --- & WHT & lens, \textit{z}$=$2.03, $z_{\textrm{lens}}$=0.230 \\
J1322+7852 & 200.7303 & 78.8791 & WD & 20.15, 20.20 & 3.59 & 0.46, --- & WHT & \textit{z}$=$2.165 QSO + gal \textit{z}$=$0.352 \\
J1326+3020 & 201.7410 & 30.3400 & GP & 20.24 & 2.11 & --- & WHT & lens, \textit{z}$=$1.852, $z_{\textrm{lens}}$=0.339 \\
J1329--2807 & 202.4535 & -28.1279 & WD, MD & 20.57, 19.23 & 1.90 & ---, 1.26 & WHT & lens, \textit{z}$=$2.685 \\
J1344+6200 & 206.2040 & 62.0118 & MD & 20.14, 20.68 & 4.44 & 1.48, 0.65 & WHT & lens, \textit{z}$=$2.21 \\
J1348+2925 & 207.2060 & 29.4195 & WD, MD & 18.43, 20.31 & 1.96 & 2.62, --- & WHT & QSO+galaxy, \textit{z}$=$0.292 \\
J1350+3155 & 207.5956 & 31.9315 & WD & 18.89, 18.38 & 1.44 & 1.72, 2.91 & WHT & \textit{z}$=$1.895 QSO+ \textit{z}$=$0.211 galaxy \\
J1408+0422 & 212.1406 & 4.3747 & GP & 20.61 & 2.59 & --- & WHT & lens, \textit{z}$=$3.005 \\
J1428+0500 & 217.2309 & 5.0058 & WD, MD & 19.89, 19.96 & 2.23 & 1.50, 2.26 & WHT & NIQ, \textit{z}$=$1.375 \\
J1442--0857 & 220.7080 & -8.9504 & WD & 18.96, 18.85 & 1.50 & ---, 1.78 & NTT2 & galaxies, \textit{z}$=$0.164 \\
J1449--2025 & 222.2960 & -20.4276 & WD, MD & 18.65, 19.40 & 1.17 & 0.75, --- & NTT2 & NIQ, \textit{z}$=$0.594 \\
J1526--1400 & 231.6891 & -14.0030 & GP & 19.29, 18.51, 19.35 & 2.65 & 1.00, ---, --- & WHT & lens, \textit{z}$=$0.648, $z_{\textrm{lens}}$=0.096 \\
J1548--2914 & 237.1733 & -29.2351 & WD, MD & 18.17, 20.07 & 1.99 & 0.70, 1.49 & WHT & lens, \textit{z}$=$1.545, $z_{\textrm{lens}}$=0.380 \\
J1550+0221 & 237.7387 & 2.3629 & WD, MD & 20.34, 18.44 & 2.03 & 0.42, 1.83 & WHT & lens, \textit{z}$=$2.395 \\
J1652+4129 & 253.1935 & 41.4934 & WD & 20.10, 20.72 & 2.12 & 1.81, --- & WHT & Galaxies, \textit{z}$=$0.217 \\
J1740+0311 & 265.1550 & 3.1966 & WD, MD & 19.50, 19.13 & 0.80 & 6.48, 0.94 & NOT & \textit{z}$=$1.21 QSO + star \\
J1752+0826 & 268.1779 & 8.4339 & GP & 20.54, 20.44 & 2.04 & 3.80, 4.55 & WHT & \textit{z}$=$0.3195 galaxy + stars \\
J1820+3747 & 275.1899 & 37.7879 & WD & 20.26, 17.93 & 3.10 & 1.44, 2.33 & NOT & projected QSOs, \textit{z}$=$0.614, 0.754 \\
J1821+6005 & 275.3760 & 60.0908 & WD, MD & 20.33, 20.25 & 1.51 & 1.61, 2.27 & NOT & NIQ, \textit{z}$=$2.052 \\
J1832+5349 & 278.1236 & 53.8206 & WD, MD & 19.58, 20.15 & 3.03 & 0.68, 2.06 & NOT & NIQ, \textit{z}$=$1.160 \\
J1945--2857 & 296.3899 & -28.9548 & WD & 19.94, 20.87 & 2.69 & 2.49, --- & NTT1 & lens, \textit{z}$=$2.56  \\
J2008+0438 & 302.0095 & 4.6433 & WD, MD & 20.01, 20.00 & 1.50 & 1.73, 2.34 & NOT & NIQ(?), \textit{z}$=$1.70 \\
J2015+0707 & 303.8037 & 7.1171 & WD, MD & 20.72, 19.18 & 2.93 & ---, 0.18 & NOT & lens, \textit{z}$=$2.59 \\
J2017+6204 & 304.4544 & 62.0787 & WT, MT & 19.15, 19.71, 20.14 & 0.92 & 4.05, ---, --- & NOT & quad, \textit{z}$=$1.73 \\
J2057+0217 & 314.4672 & 2.2967 & WD, MD & 20.14, 20.47 & 1.16 & 0.19, 0.62 & NOT & NIQ, \textit{z}$=$1.52 \\
J2106--4944 & 316.5070 & -49.7482 & WD & 20.86, 20.49 & 2.09 & ---, --- & NTT1 & lens, \textit{z}$=$1.296, $z_{\textrm{lens}}$=0.290 \\
J2110--3755 & 317.5016 & -37.9183 & WD & 20.97, 20.78 & 1.52 & ---, 1.79 & NTT1 & lens, \textit{z}$=$1.50 \\
J2121+1713 & 320.3801 & 17.2322 & WD & 17.99, 20.46 & 2.89 & 2.72, 1.57 & NOT & QSO pair, \textit{z}$=$0.63 \\
J2124+2227 & 321.2344 & 22.4631 & WD, MD & 20.03, 20.01 & 0.90 & ---, --- & NOT & \textit{z}$=$2.69 QSO + star \\
J2132+3635 & 323.0301 & 36.5982 & WD & 19.95, 20.39 & 1.33 & 1.27, 1.51 & NOT & NIQ, \textit{z}$=$1.215 \\
J2145+0039 & 326.4013 & 0.6661 & GP & 20.67, 19.96 & 2.08 & ---, 2.64 & NTT1 & \textit{z}$=$0.585 QSO + star \\
J2147--1340 & 326.9957 & -13.6772 & WD & 19.85, 20.23 & 1.33 & 1.87, --- & NOT & lens, \textit{z}$=$1.382 \\
J2205+1019 & 331.4161 & 10.3307 & WD, MD & 18.53, 18.65 & 1.34 & ---, 1.69 & NOT & lens, \textit{z}$=$1.78 \\
J2205--3727 & 331.4343 & -37.4504 & WD, MD & 20.89, 20.57 & 0.76 & ---, --- & NTT1 & quad, \textit{z}$=$1.848 \\
J2212+0350 & 333.0713 & 3.8446 & WD, MD & 18.27, 20.57 & 0.94 & 1.77, --- & NOT & \textit{z}$=$0.215 QSO + gal or qso? \\
J2213--5926 & 333.3363 & -59.4376 & GP & 20.83 & 2.59 & --- & NTT3 (/NTT1) & lens, \textit{z}$=$1.72 \\
J2244--0550 & 341.0473 & -5.8471 & WD & 18.23, 17.81 & 1.02 & 1.47, 1.09 & NTT3 & NIQ, \textit{z}$=$2.84 \\
J2255+8009 & 343.9088 & 80.1662 & WD, MD & 19.15, 18.86 & 0.74 & 2.02, 2.00 & NOT & NIQ, \textit{z}$=$2.8(??) \\
J2256+2223 & 344.0512 & 22.3943 & WD, MD & 20.68, 19.96 & 1.63 & 1.91, 0.67 & NOT & QSO pair, \textit{z}$=$0.7515, 0.754 \\
J2302--4154 & 345.6308 & -41.9157 & WD, MD & 20.27, 19.45 & 1.22 & 1.01, 2.24 & NTT3 & projected QSOs, \textit{z}$=$0.98, 1.22 \\
J2308+3201 & 347.0777 & 32.0294 & WD & 20.31, 19.54 & 2.63 & 1.41, 2.00 & NOT & lens, \textit{z}$=$2.30 \\
J2311--1038 & 347.8207 & -10.6472 & WD, MD & 20.56, 18.92 & 1.72 & ---, 2.09 & NTT3 & \textit{z}$=$1.53 QSO + star \\
J2314+0323 & 348.5537 & 3.3929 & WD, MD & 19.30, 19.87 & 1.87 & 1.21, --- & NOT & projected QSOs, \textit{z}$=$0.315, 0.606 \\
J2315+4857 & 348.7680 & 48.9532 & WD, MD & 19.92, 18.24 & 1.62 & 2.89, 1.64 & NOT & \textit{z}$=$0.296 QSO + star \\
J2316+0610 & 349.1335 & 6.1804 & WD, MD & 20.05, 20.75 & 2.32 & 1.18, --- & NOT & lens, \textit{z}$=$1.955 \\
J2318+0250 & 349.5541 & 2.8411 & WD, MD & 19.59, 19.43 & 3.23 & 2.88, 0.78 & NOT & projected QSOs, \textit{z}$=$1.83, 1.945 \\
J2322+0916 & 350.6413 & 9.2796 & WD, MD & 20.51, 18.18 & 4.51 & 1.50, 1.37 & NOT & QSO pair, \textit{z}$=$1.20 \\
J2334+5036 & 353.5277 & 50.6114 & WD & 19.88, 19.18 & 2.13 & 1.99, 5.49 & NOT & Stars \\
J2341--1557 & 355.4664 & -15.9501 & WD & 20.71, 20.06 & 1.02 & ---, 0.85 & NTT3 & NIQ, \textit{z}$=$1.535 \\
J2355--4553 & 358.7761 & -45.8931 & WD & 20.01, 20.69 & 2.20 & 1.67, 0.67 & NTT3 & QSO pair, \textit{z}$=$2.85 \\

        \hline
    \end{tabular}
\end{table*}


\begin{figure}
\centering
\includegraphics[width=\columnwidth]{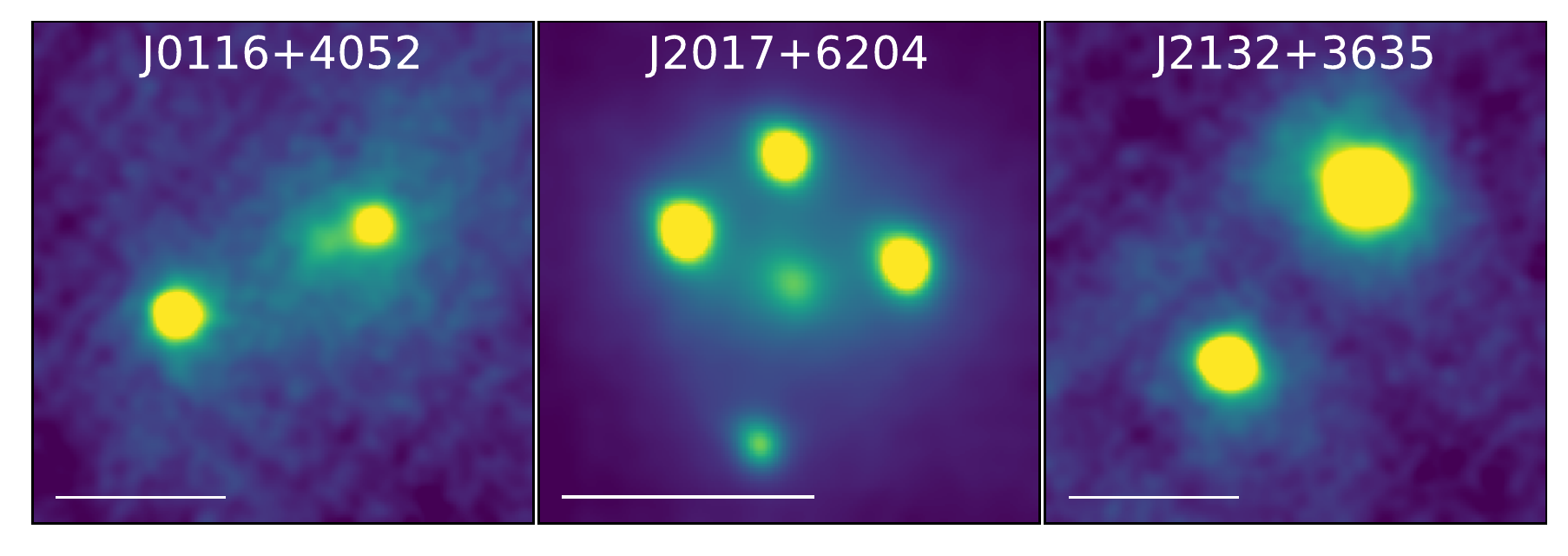}
\vspace{-0.5cm}
\caption{OSIRIS imaging of three systems. White scale bars represent 1 arcsecond. North is up, East is left. Galaxies can be seen between the quasar images for J0116+4052 ($z=1.85$) and J2017+6204 ($z=1.73$), however nothing is seen for J2132+3635 ($z=1.215$), therefore it is likely a physical binary.}
          \label{fig:osiris}%
\end{figure}

\begin{figure*}
\centering
\includegraphics[width=\textwidth]{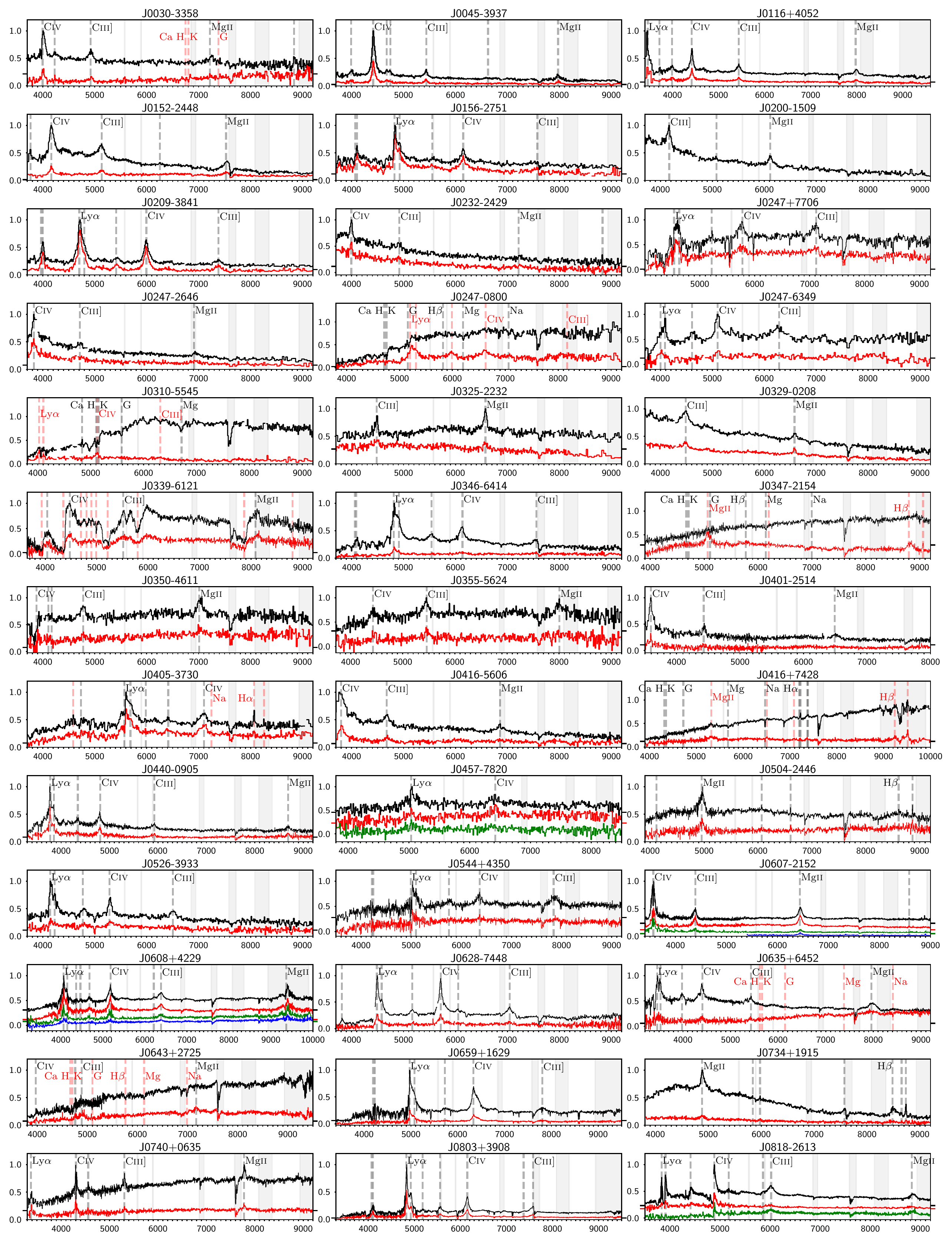}
\vspace{-0.5cm}
\caption{Spectra of the confirmed lenses in the observed frame. The most prominent quasar emission lines are marked with vertical lines, and details on certain spectra can be found in Section \ref{indiv}. In some cases one or several spectra have been offset for visualisation, and the offset is marked on the y-axis. }
          \label{lens_spectra1}%
\end{figure*}

\begin{figure*}
\centering
\includegraphics[width=\textwidth]{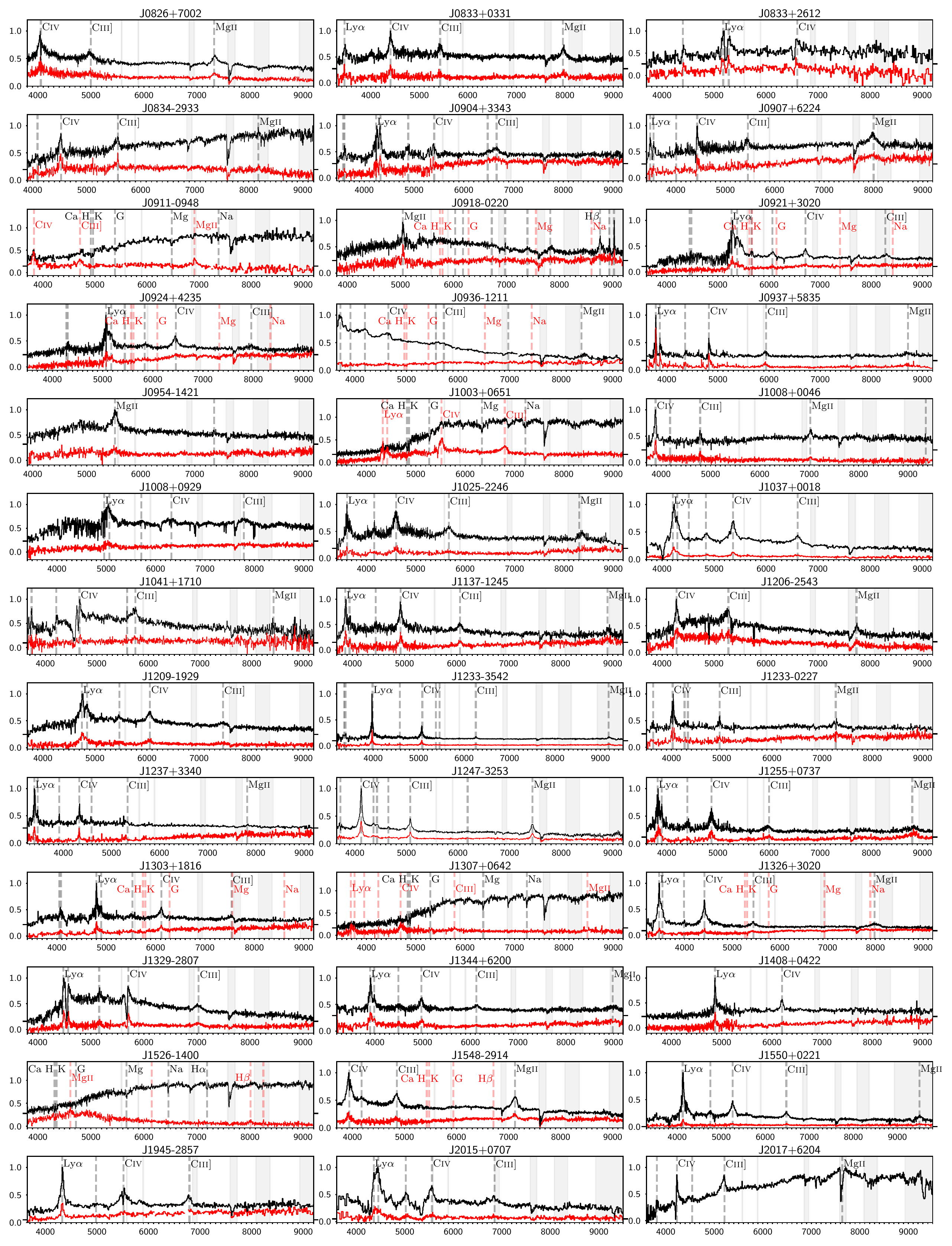}
\vspace{-0.5cm}
\contcaption{}
          \label{lens_spectra2}%
\end{figure*}

\begin{figure*}
\centering
\includegraphics[width=\textwidth]{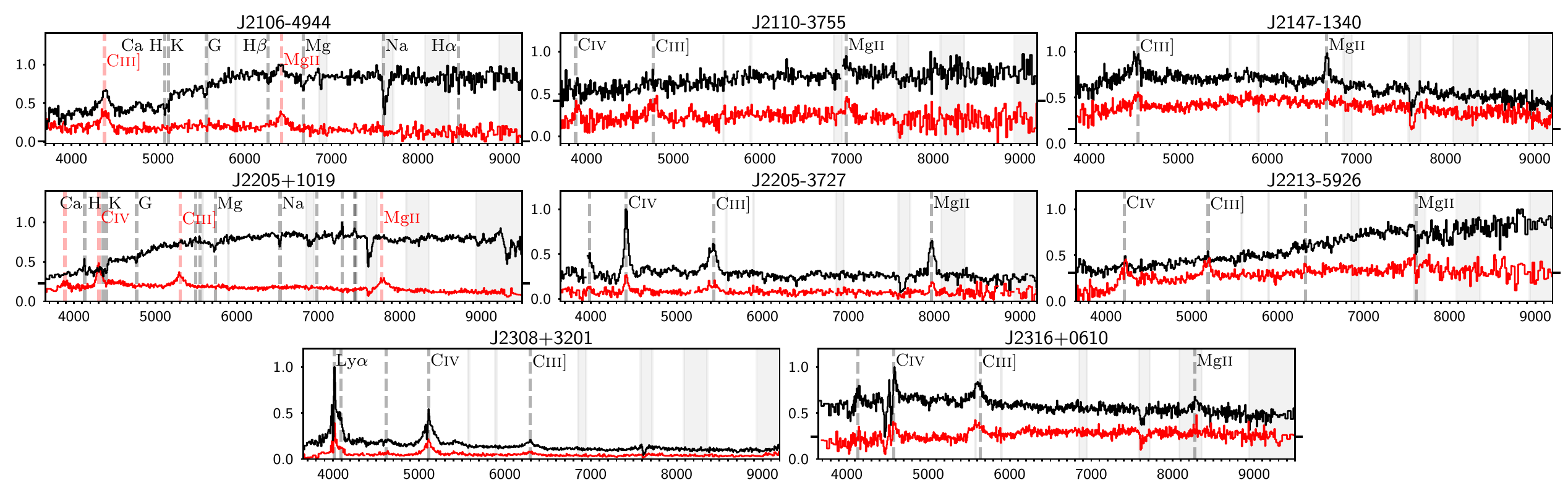}
\vspace{-0.5cm}
\contcaption{}
          \label{lens_spectra3}%
\end{figure*}

\begin{figure*}
\centering
\includegraphics[width=\textwidth]{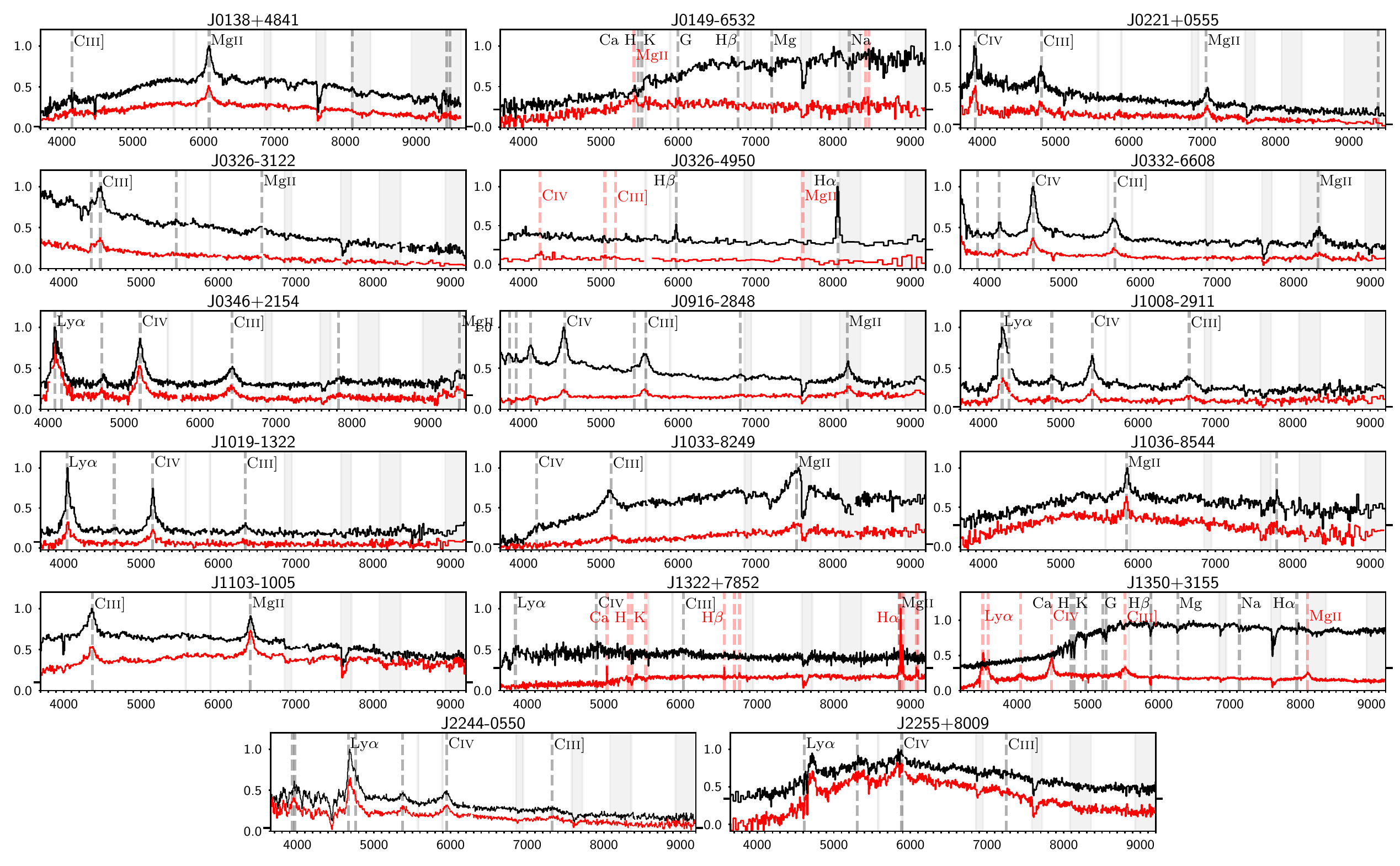}
\vspace{-0.5cm}
\caption{Spectra of very likely lensed quasars. Details of the spectra are explained in Figure \ref{lens_spectra1}.}
          \label{probable_spectra}%
\end{figure*}

\begin{figure*}
\centering
\includegraphics[width=\textwidth]{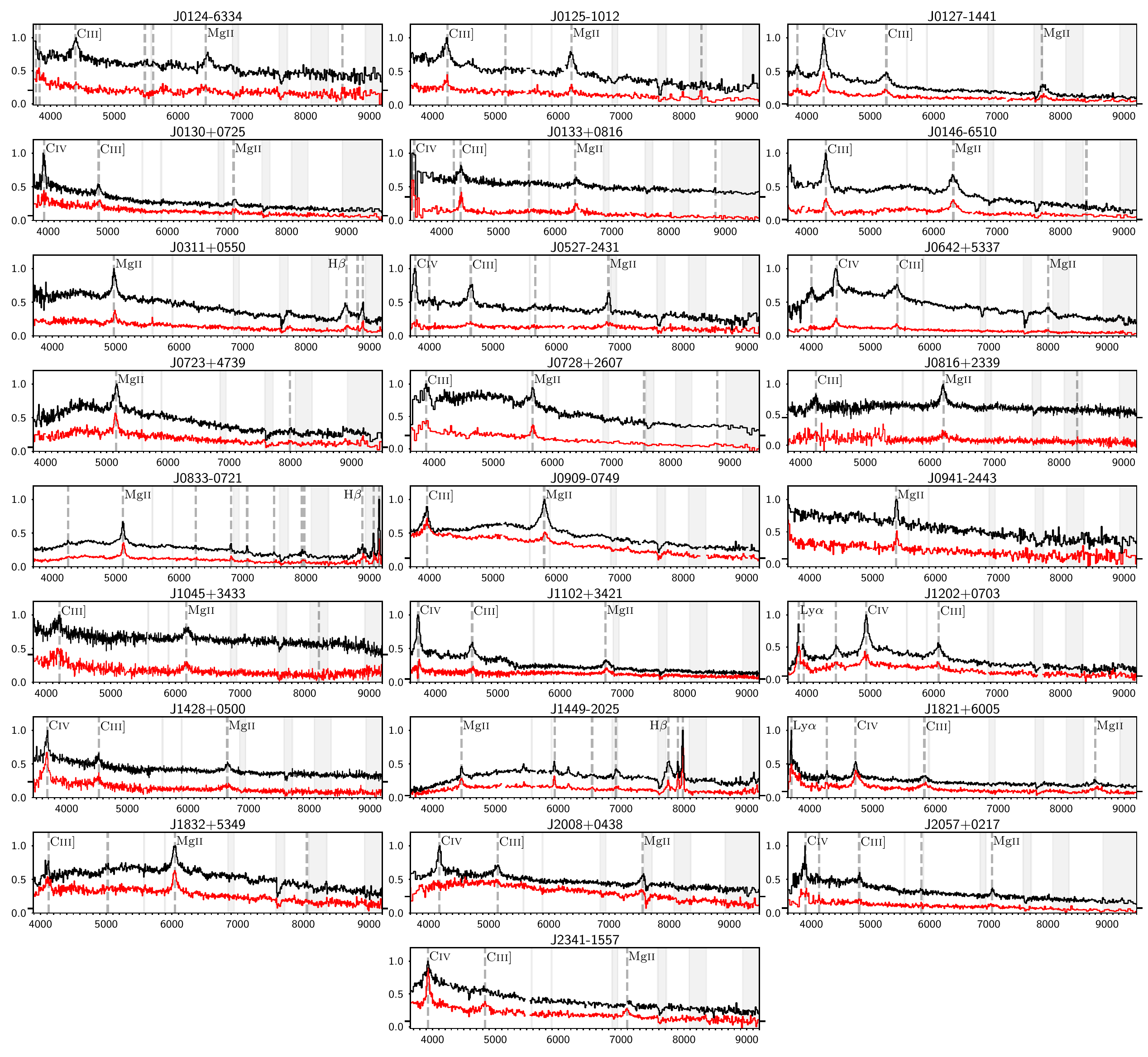}
\vspace{-0.5cm}
\caption{Spectra of unclassified quasar pairs. Details of the spectra are explained in Figure \ref{lens_spectra1}. }
          \label{niqs_spectra}%
\end{figure*}

\begin{figure*}
\centering
\includegraphics[width=\textwidth]{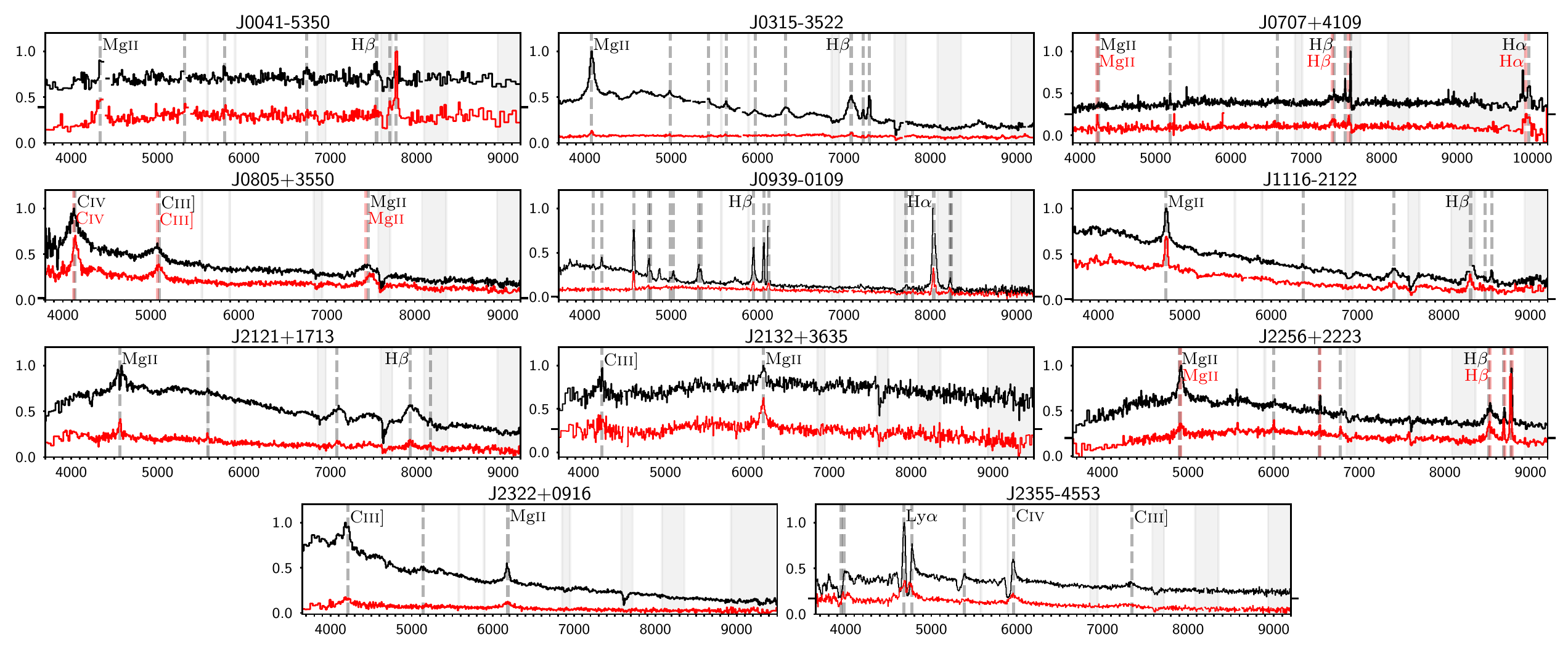}
\vspace{-0.5cm}
\caption{Spectra of binary quasars, with strong evidence for being distinct, and thus not graviationally lensed. Details of the spectra are explained in Figure \ref{lens_spectra1}. }
          \label{binaries_spectra}%
\end{figure*}

\begin{figure*}
\centering
\includegraphics[width=\textwidth]{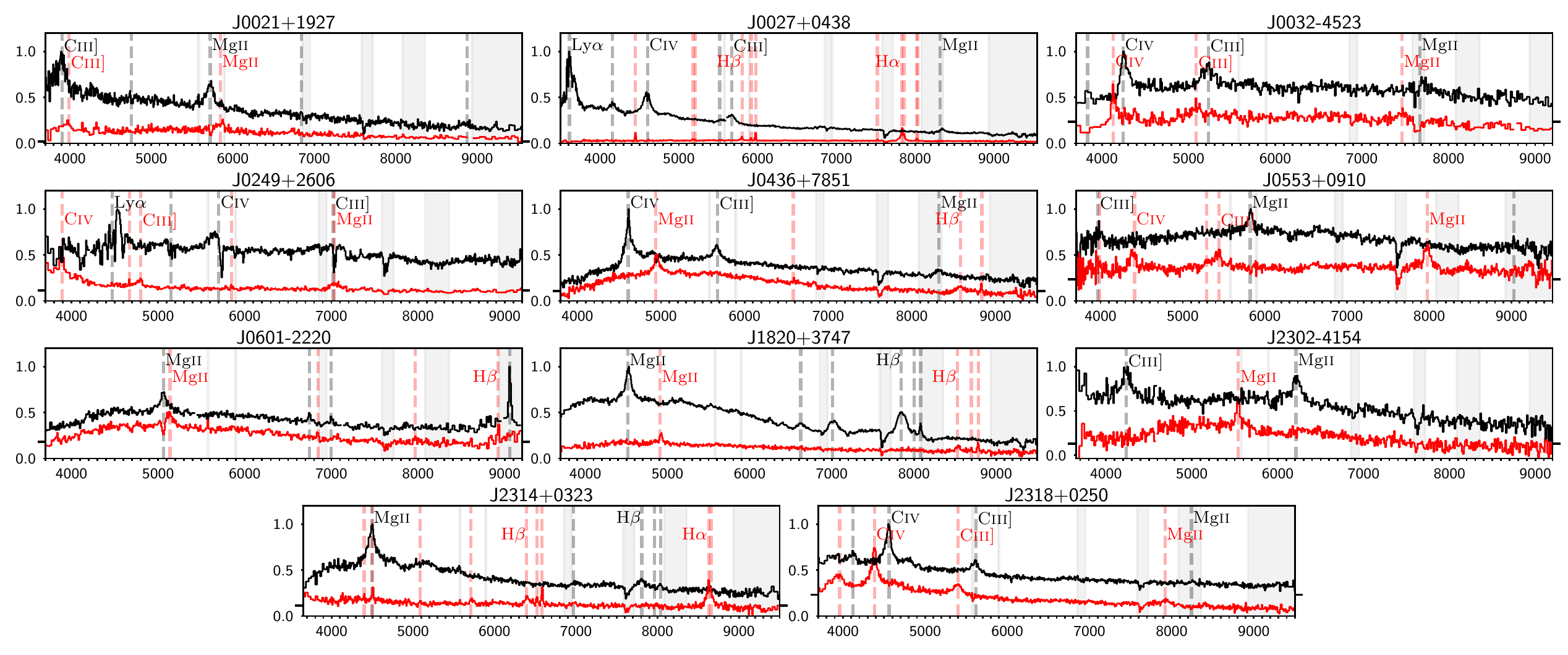}
\vspace{-0.5cm}
\caption{Spectra of projected quasar pairs. Details of the spectra are explained in Figure \ref{lens_spectra1}. }
          \label{projected_spectra}%
\end{figure*}

\section{Notes on Individual Systems} \label{indiv}
In this section we comment on any interesting aspects of the individual systems, including their photometry, spectra, and mass modelling. We divide the systems into sections based on their final classification.

\subsection{Confirmed Lensed Quasars}
\subsubsection{J0030-3358}
A jump in the continuum of the fainter image of this double is likely associated to the 4000\AA\ break of the lensing galaxy, as associated Ca H, K and G-band absorption features are also seen, placing the lens at $\textit{z}$=0.715. This is supported by the Legacy Survey photometric redshift estimate of 0.84$\pm$0.06 \citep{zhou2021}, with a small discrepancy likely due to blending with the quasar light. There is absorption on the blue side of the \textsc{Civ} line seen in both images, however this could also be attributed to Milky Way absorption since it is coincident with $\textit{z}$=0 Ca H and K absorption. 
\subsubsection{J0116+4052}
The spectra show two quasars at $\textit{z}$=1.85 with very similar continuum and broad emission line profiles, suggesting the system is a likely lens. Only shallow Pan-STARRS imaging exists for this system and shows no significant residuals upon subtraction of two PSFs. Follow-up OSIRIS imaging reveals a third source closest to the Western image, which we take as a detection of the lensing galaxy (Figure \ref{fig:osiris}).
\subsubsection{J0152-2448}
This double system at $z=1.69$ was also independently discovered by \citet{spiniello2019}.
\subsubsection{J0200-1509}
The EFOSC2 data shows a blended quasar spectrum at \textit{z}$=$1.185. Given the two obvious \textit{Gaia} detections and red residual flux upon subtraction of two PSFs in Legacy Survey imaging, which is well-fit by simultaneously including a Sersic which falls between the two PSFs, we designate this system as a lens.
\subsubsection{J0232-2429}
This system was originally selected by \citet{lemon2018}, who classified it as a likely quasar+star projection, despite poor seeing during observations and the small separation of 1.44\arcsec. However, we obtained further spectroscopy to confirm this classification, and resolve two quasars at $z=1.58$ with similar emission line profiles and continuum. There is residual red flux in the PSF-subtracted Legacy Survey image, consistent with a lensing galaxy. We therefore upgrade the classification to a secure lens. The system is in the outskirts of a cluster ($\sim1$\arcmin\ from the centre) which has galaxies with photometric redshifts of $\sim$0.85. The quasar is radio loud with detections in NVSS (4.7$\pm$0.7 mJy at 1.4GHz) and VLASS (2.7$\pm$0.4 mJy at 2--4GHz).
\subsubsection{J0247+7706}
The ALFOSC-NOT spectra clearly resolve two quasar traces at \textit{z}$=$2.73 and many narrow absorption lines common to both components. Faint hints of an extended galaxy are seen in the PSF subtracted Pan-STARRS imaging. When adding a Sersic component to the fit, flux is allocated in all bands to this component, and its best-fit position is between the two PSFs closer to the fainter image. Given this and the very similar spectra, we classify this system as a lens, however deeper imaging should be taken to verify the detection of the lensing galaxy and characterise its  properties.
\subsubsection{J0247-6349}
The seeing for this spectrum was particularly poor, leading to low signal-to-noise and a blended spectrum. The 2D spectral modelling allocates emission line flux to both components, however cannot be taken as reliable. Nevertheless, the emission lines are clearly from a \textit{z}$=$2.295 quasar, and the Legacy Survey images show two point sources either side of a red galaxy. Therefore we classify this system as a lens. This is supported by the independent spectra and discovery of this system by \citet{spiniello2019}.
\subsubsection{J0310-5545}
This double was selected as a single \textit{Gaia} detection near a possible LRG, however the \textit{Gaia} detection is not associated to one of the quasar images but to a nearby star. The possibility of two further point sources around the nearby galaxy was noticed serendipitously during visual inspection. Deep EFOSC2-NTT spectra aligned to capture the galaxy and two faint sources clearly show three traces, two of which contain emission lines of a quasar at \textit{z}$=$2.31, and the middle trace being an LRG at \textit{z}$=$0.298, therefore confirming this system as a lensed quasar.
\subsubsection{J0325-2232}
The spectra show emission lines of a \textit{z}$=$1.35 quasar in both components, however the continua and relative emission line amplitudes are clearly discrepant, with stronger broad emission lines in the redder image. Legacy Survey imaging shows a galaxy between the two components, which is well fit by a Sersic positioned between the PSFs and slightly nearer the South-Western image, B. In the $g$-band of both Pan-STARRS and Legacy Survey imaging, A is brighter than B, however B is brighter in both \textit{Gaia} (by 0.21 mag in the broad \textit{G} filter) and in the ground-based red filters, even when including the galaxy in the fit. Absorption is present around 4785\AA\ and 4890\AA\ in only the redder image. This system is likely a lens with contamination and reddening by the lensing galaxy coupled with microlensing. Deeper spectra will be useful for understanding the nature of the discrepant spectra.
\subsubsection{J0329-0208}
This double lensed quasar lies 3.5\arcmin\ from the centre of the cluster MACS J0329-0211. The \textit{HST} coverage of the cluster narrowly misses this lens system, however   \textit{Chandra} ACIS-I imaging does cover the system, revealing an extended X-ray source with a measured flux of 4.11$\times$10$^{-14}$ \ergcmcms in the 0.3--8 keV range \citep{wang2016}. The lensing galaxy may be a member of this \textit{z}$=$0.45 galaxy cluster, however there are no obvious absorption lines in our spectra around this redshift.
\subsubsection{J0339-6121}
This double shows signs of a LoBAL quasar source at $z\approx$1.89, with the troughs blueshifted by around 0.028c, typical of many BAL quasars \citep{rankine2020}. While the absorption around 5762\AA\ could be attributed to Fe\textsc{iii}-UV48 triplet, it is more likely to be due to the Cr\textsc{II}-2062\AA\ triplet since we do not see strong evidence for the defining FeLoBAL Fe\textsc{II} UV1 and UV2 lines. The flux ratio of the images across wavelength is approximately constant (2.4:1) but shows clear differences around several absorption lines, suggesting microlensing is present in the system, a particularly useful tool for constraining the quasar source geometry \citep{hutsemekers2010, sluse2015}. 
\subsubsection{J0350-4611}
The NTT-EFOSC2 spectra of this double are low signal-to-noise but each component shows emission lines of a \textit{z}$=$1.505 quasar. The \textsc{Civ} profile is not apparent, and there is a significant drop in flux at the expected wavelength and blueward of \textsc{Civ} in both components. The lens is seen in the Legacy Survey PSF-subtracted residuals.  Deeper spectra should be taken to understand the nature of this system, as it is possibly a useful lensed BAL quasar.
\subsubsection{J0405-3730}
The 2D spectra of this double show the resolved traces of the images of a quasar at \textit{z}$=$3.585, but also three narrow emission lines located between the two quasar traces, but much closer to the brighter trace. These lines are consistent with \textsc{Oii}, H$\alpha$ and \textsc{Sii} at \textit{z}$=$0.2295, and a narrow absorption line in the brighter trace can also be attributed to Na at this redshift. The imaging of this system shows that at least two galaxies are responsible for the lensing, one between the two images but offset and highly elliptical and another offset. We postulate that the lens is a merger between two galaxies, triggering star formation explaining the aforementioned narrow lines. High-resolution imaging will be needed to better understand this system.
\subsubsection{J0416+7428}
This double has a very low redshift lens (\textit{z}$=$0.098), which appears to be part of a group with two nearby galaxies apparent in the Pan-STARRS imaging. In the 2D spectra, narrow emission lines are seen both 4.6 arcseconds from the main lensing galaxy in the direction of the more distant image, and also 13 arcseconds in the other direction coincident with diffuse extended continuum emission for the latter. These lines are consistent with H$\alpha$ and \textsc{Nii} emission at the lens galaxy redshift, therefore we attribute the former to a star-forming region in the outskirts of the main lensing galaxy, and the latter to star formation in the outskirts of the galaxy group member to the West. The lensing galaxy has an \textit{i}-band brightness of 16.22. The low redshift of the lensing galaxy is rare amongst lensed quasars, with the only known lensed quasar with a redshift lower than this being the Einstein cross \citep{huchra1985}. Another double in our sample, J1526-1400, also has a low-redshift, bright lensing galaxy, so we investigate the expected number of lenses as a function of lensing galaxy brightness after describing J1526-1400 in Section \ref{J1526}.
\subsubsection{J0457-7820}
\begin{figure}
\centering
\includegraphics[width=\columnwidth]{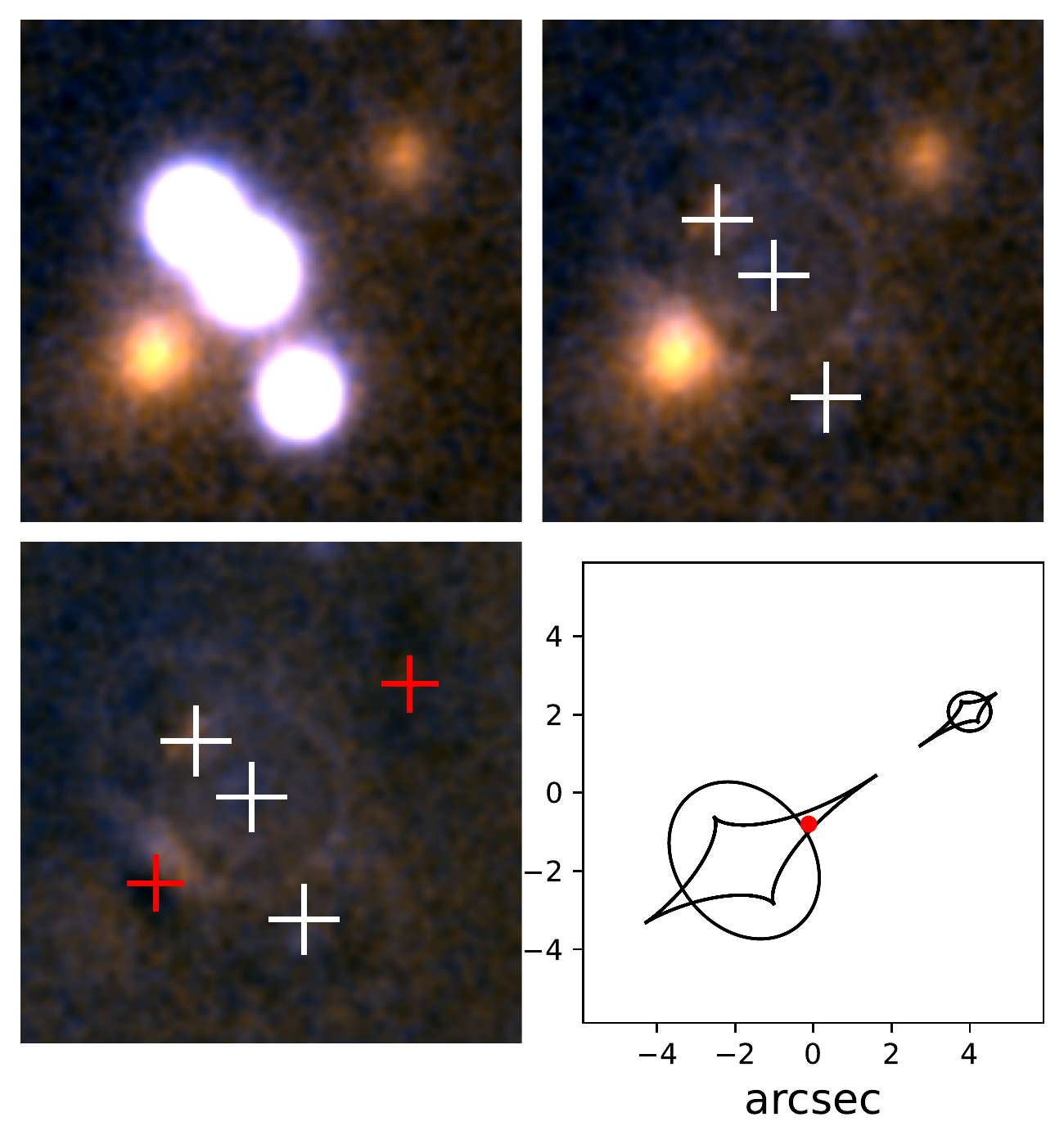}
\vspace{-0.5cm}
\caption{(a) MagIC g and i-band imaging of J0457-7820; (b) with a 3 PSF model subtracted, (c) with galaxies and PSFs subtracted, (d) fiducial mass model caustics with source position overlaid (see text for details).}
    \label{fig:J0457}
\end{figure}

This system was selected through the \textit{Gaia} multiplet WISE search, with three \textit{Gaia} detections ($G$=18.81, 19.07, and 19.60) and W1--W2=0.95. Archival NOAO DeCAM imaging shows three bright blue point sources around a galaxy. The follow-up NTT-EFOSC2 spectra, positioned to capture light from all three images simultaneously, are low signal-to-noise but clearly show three traces of a \textit{z}$=$3.145 quasar. Three $g$-band and three $i$-band exposures, each of 120s were taken with the Magellan Instant Camera, MagIC. A second galaxy is apparent to the North-West of the system. All six exposures were fit simultaneously as two galaxies and three point spread functions, with the PSF being fit as a grid of pixels, and constrained directly from the data \citep[following][]{ostrovski2018}. Figure \ref{fig:J0457} shows the colour-image stack (where the green channel is a median of the $g$ and $i$ bands), and residuals. No counterimage is detected. We repeat the image subtraction but artificially include a point source of increasing brightness until it is obvious in the residuals. This provides an upper limit on any counterimage brightness being 1\% as bright as image B, i.e. a $G$-band magnitude of $\sim$24.0. One possibility is that this is a rare three image system caused by the straddling of two galaxies. \citet{shin2008} provide a comprehensive catalogue of image configurations caused by lensing of two separated galaxies, modelled as Singular Isothermal Spheres. Several configurations allow for three image regions within the resulting caustics, where the third image is not necessarily strongly demagnified. Given that all images lie between the two galaxies, the most likely configuration is 3B-1, following the labelling of \citet{shin2008}. This naked catastrophe occurs when the two astroid caustics merge in a beak-to-beak calamity \citep[see, e.g., ][]{kassiola1992, orban2009}, but the pseudo-caustics do not overlap, creating a naked merged cusp region. Introducing external shear and mass ellipticity can create naked cusps which are not necessarily merged, again creating source plane regions corresponding to three bright images. 

To test which is the relevant scenario for J0457-7820, we perform mass models based on the astrometry from fitting the MagIC imaging. 
We use only the positions to constrain the mass model, since microlensing, extinction, and variability over the time delay can cause optical image fluxes to vary significantly from their macromodel values. This gives us only 6 free parameters. A single SIE model fixed to the main lensing galaxy, predicts a fourth image, only 2.5\% the brightness of B. Such an image could be consistent with our 1\% detection threshold coupled with reddening and microlensing demagnifying such an image, however, the model is unphysical with an axis ratio of 0.2, and the image positions are not well recovered. Including shear in the model recovers the image positions exactly, but also predicts a fourth image, about 1\% as bright as B. The lensing galaxy mass is also less elliptical (0.14) but a large shear of 0.35 is required, in the direction of the second galaxy. We note that this model has more free parameters than constraints, however some parameters are still well-constrained. This model can be forced to have three images with very little change in the mass model parameters and well-fit images ($\chi^{2}\sim0.4$). Given the shear direction, we investigate two component mass models. SIS+SIS or SIS+SIE models do not recover the three image positions. An SIE+SIE model fits the images and does not predict a fourth image, however the galaxies are predicted to have large ellipticities and lie in the same direction ($\sim$25 degrees North of West). Adding shear to this model reduces the ellipticities of the mass components and still recovers only three images, however, the lens parameters are not as well constrained. We show the caustics of this best-fit model in Figure \ref{fig:J0457}. In this instance the cusps do not merge, but since the model is still under-constrained further constraints from the lensed source host galaxy (and limits on a possible fourth image) will be needed to determine more reliable caustic configurations.

\subsubsection{J0607-2152}
\begin{figure}
\centering
\includegraphics[width=\columnwidth]{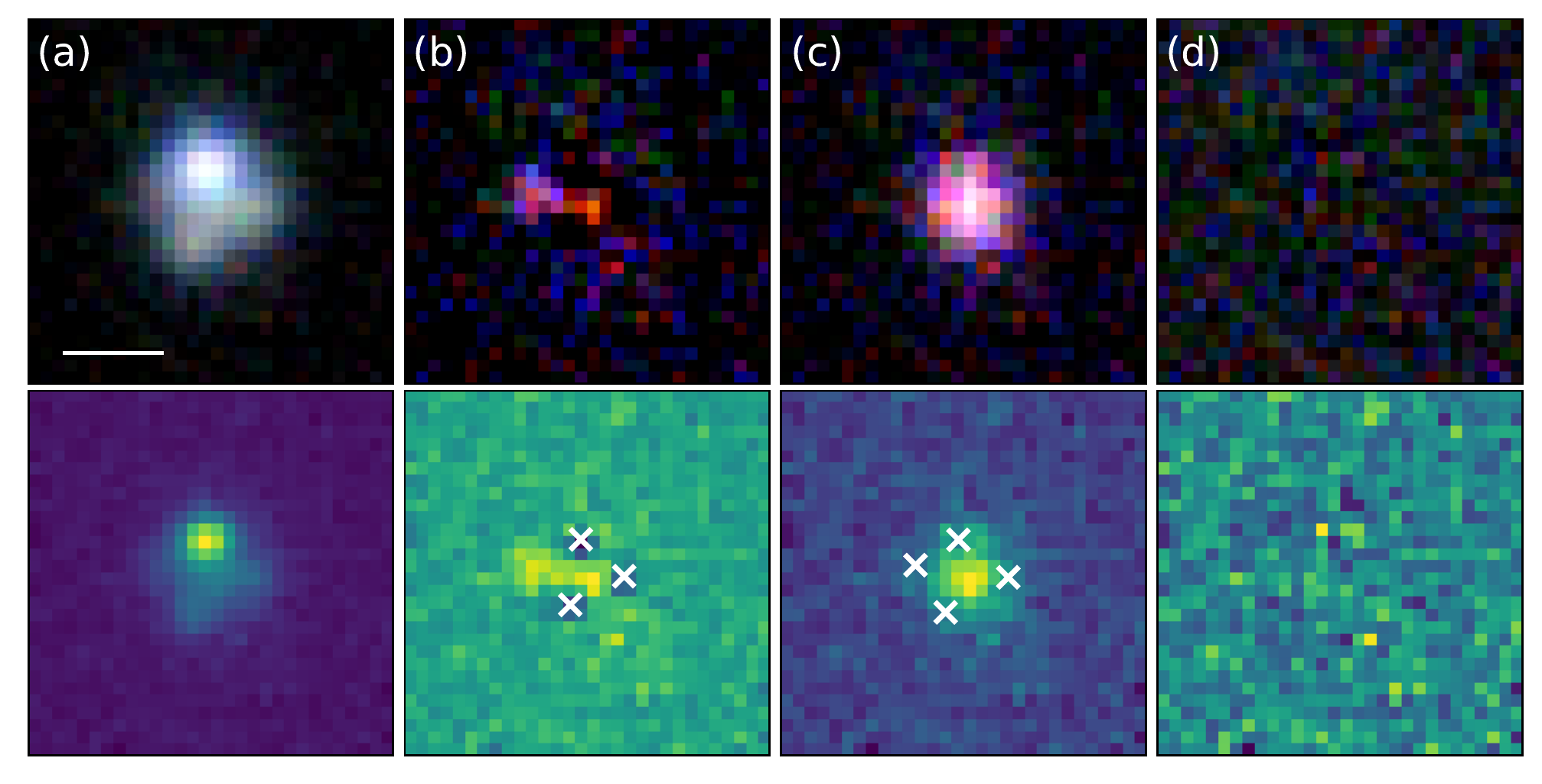}
\vspace{-0.5cm}
\caption{Analysis of Pan-STARRS imaging of J0607-2152. Top row is for \textit{gri}, while the bottom is just \textit{r}-band. (a) original data, (b) with a 3 PSF model subtracted, (c) with 4 PSFs subtracted from a 4 PSF + galaxy model, (d) residuals of the 4 PSF + galaxy model.}
    \label{fig:J0607}
\end{figure}

This quad was confirmed during the 2019 WHT observing run, in which the system was targetted at two position angles: through the brightest image and each of the other two \textit{Gaia} detections. The traces are resolved in the red arm, however in the blue arm of one observation, the images are not resolved, hence we show only the red arm extraction for the fainter image in Figure \ref{lens_spectra1}. The extracted fluxes for the brightest image between the two exposures are in good agreement in the red arm, however in the blue arm they are in disagreement likely due to slit losses and blending. Figure \ref{fig:J0607} shows the presence of two further objects between and to the East of the 3 \textit{Gaia} detections. Including a galaxy and a further PSF fits the data to the noise, so we expect that these are the lensing galaxy and fourth quasar image. Using these 5 positions, we are not able to fit an SIE+shear model to the data, suggesting a more complex lensing model is required, or the fourth image is much closer to A. This system was also independently confirmed by \citet{stern2021}. They suggest that the fourth image is between the Northern components, however this is not supported by the aforementioned residuals.

\subsubsection{J0608+4229}
\begin{figure}
\centering
\includegraphics[width=\columnwidth]{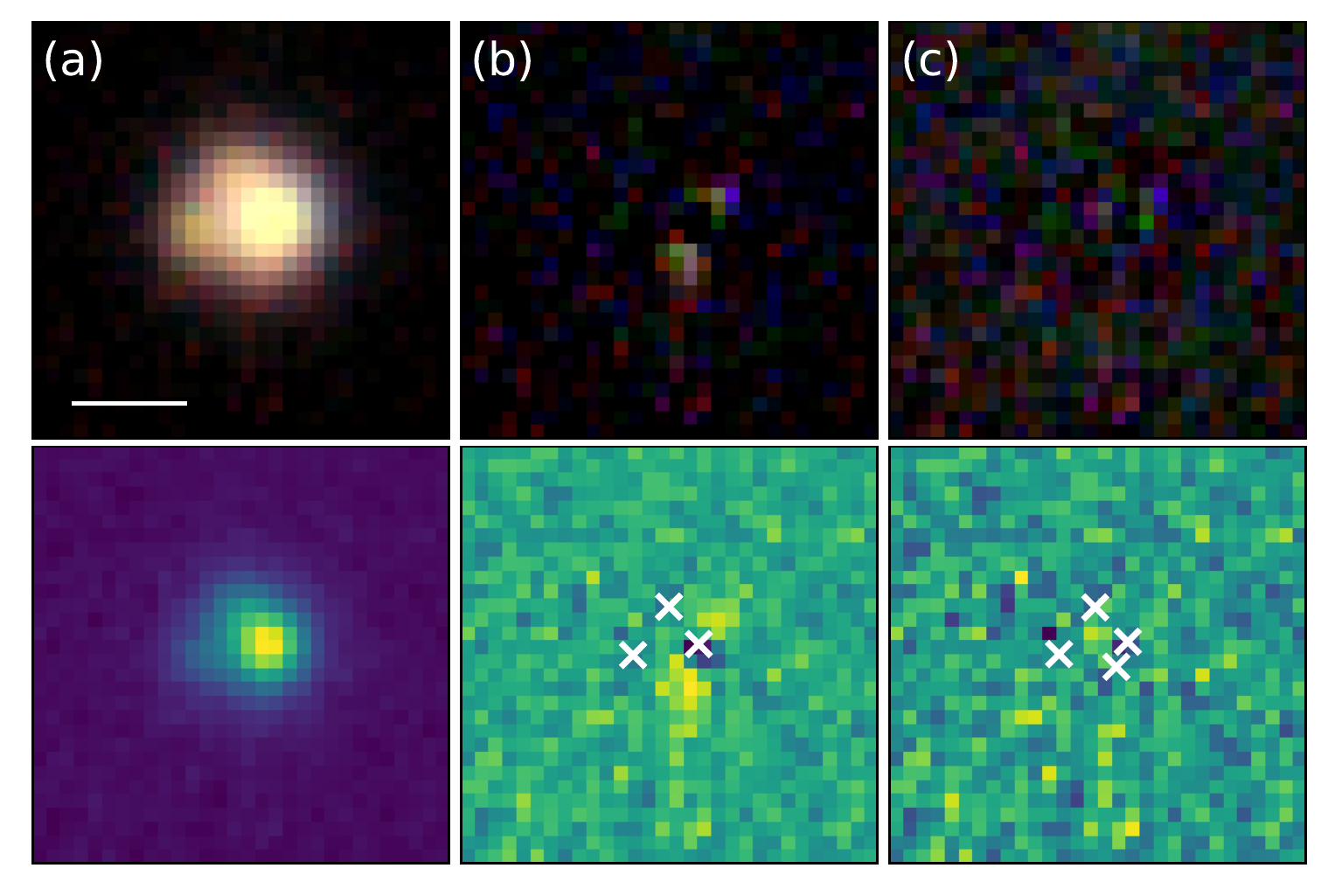}
\vspace{-0.5cm}
\caption{Analysis of Pan-STARRS imaging of J0608+4229. Top row is for \textit{gri}, while the bottom is just \textit{r}-band. (a) original data, (b) with a 3 PSF model subtracted, (c) with a 4 PSF model subtracted.}
    \label{fig:J0608}
\end{figure}

This system has three \textit{Gaia} detections with $G$=18.22, 19.86, 20.12. It was confirmed during the Feb 2019 WHT observing run, in which the system was targeted at two position angles: both through the brightest \textit{Gaia} detection and each of the other two detections. The forced fitting extracts very similar spectra for each component. The multiple extractions of the brightest image are in good agreement, with only a slight mismatch in the relative amplitudes in the red arm, likely due to the flux-matching procedure of the blue and red arm spectra. Modelling the system as three PSFs reveals residual flux just south of the brightest image, and modelling the systems as four PSFs fit the Pan-STARRS data to the noise, as shown in Figure \ref{fig:J0608}. We note, however, that no lensing galaxy is seen upon PSF subtraction, but this is often expected to be the case for higher redshift sources (here, \textit{z}$=$2.34). An SIE model fit to the four image positions shows A is observed brighter than the model, and D is observed fainter. Since D is a saddle point in this model, microlensing could be the cause of its relative demagnification \citep{schechter2002}. This system was also independently confirmed by \citet{stern2021}.

\subsubsection{J0628-7448}
Archival DECam data of this system reveal significant residuals after subtracting two PSFs, and flux is seen in the residual 2D NTT-EFOSC2 spectra towards redder wavelengths. Given also the similarity of the spectra, we classify this system as a lens. The system is associated with an XMM Slew Survey catalogue detection with a flux of 1.6$\pm$0.8$\times$10$^{-12}$ \ergcmcms in the 0.2--12keV band.
\subsubsection{J0635+6452}
This is a large flux-ratio, 3.06\arcsec-separation double. The Pan-STARRS \textit{gri} data appear consistent with only a PSF+galaxy, however a \textit{Gaia} detection associated with the galaxy suggested the presence of a counterimage, which our WHT spectrum confirms. The fainter image has dropped from $G$=20.57 to $G$=20.86 between GDR1 and GDR2. 
\subsubsection{J0643+2725}
This double has a bright lensing galaxy at \textit{z}$=$0.185. In the Pan-STARRS imaging, the quasar image closest to the lensing galaxy is the brightest (and is likely resolved since the \textit{Gaia} astrometric excess noise --- a catalogue parameter correlating with extendedness --- is 2.57, a small value compared to most galaxies).
\subsubsection{J0659+1629}
This lens is a short axis quad, with a large Einstein radius, $\sim$2.3\arcsec. We obtained spectra of the two brightest images, confirming the source to be a quasar at \textit{z}$=$3.09. The best fit SIE+shear model has a magnification of 25, and a modest shear of 0.06. The model predicts image D to be 60\% fainter than observed. Since this system has a large time delay between image D, we expect that variability over the time delay can explain this discrepancy. This system was also independently selected by \citet{delchambre2019} as a lens candidate, and independently confirmed by \citet{stern2021}.
\subsubsection{J0734+1915}
The 2D spectrum shows the traces of two quasars at \textit{z}$=$0.745 however the signal to noise of the fainter quasar is low, making comparison difficult. Both Legacy Survey and Pan-STARRS imaging reveal significant residuals in the PSF-subtracted image, which are well fit when including a galaxy which falls betweeen the two PSFs but closer to the fainter one. We classify this system as a lens, however we suggest prioritising high-resolution imaging or deeper spectroscopy of this system to verify the lensing hypothesis.
\subsubsection{J0803+3908}
\begin{figure}
\centering
\includegraphics[width=\columnwidth]{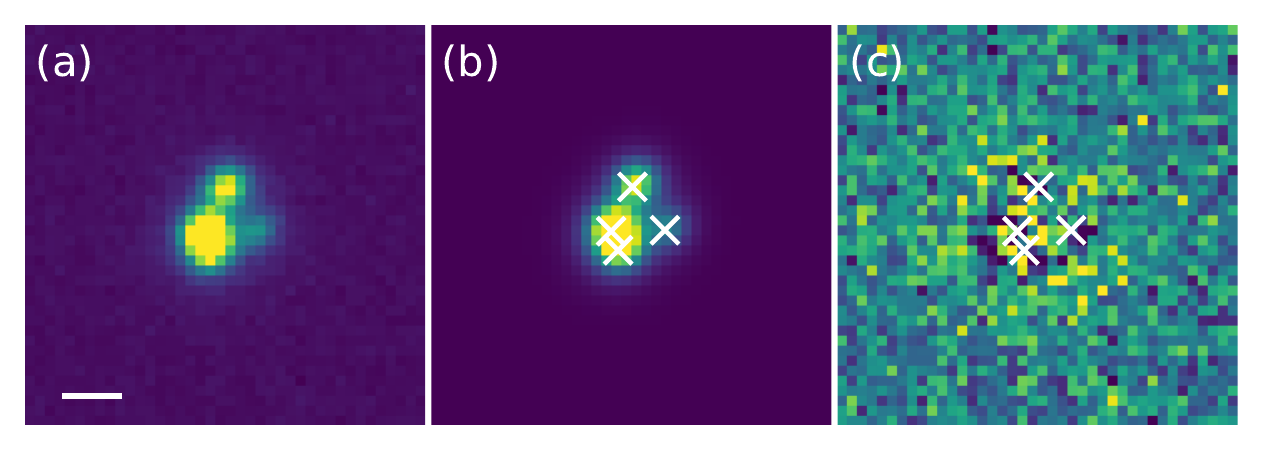}
\vspace{-0.5cm}
\caption{(a) CFHT \textit{r}-band image of J0803+3908 with a 1\arcsec\ scalebar, (b) model with PSF positions overlaid, (c) residual image possibly showing extended emission from the lensing galaxy.}
    \label{fig:J0803}
\end{figure}

This system was originally selected with the unWISE modelling technique, given it has two \textit{Gaia} detections separated by 0.91\arcsec. Good-seeing CFHT $r$-band data reveal a fold configuration quad. Figure \ref{fig:J0803} shows these data, with a 4 PSF fit and subtraction, showing faint extended residuals possibly due to a lensing galaxy. Very good seeing conditions allowed spatially resolved spectra between the bright pair and the Northern image, confirming the source to be at \textit{z}$=$2.97. \citet{schindler2018} confirmed this source as a bright quasar at \textit{z}$=$2.975 in their North Galactic Cap sample of the Extremely Luminous Quasar Survey (ELQS-N), however they did not comment on any possibility of it being gravitationally lensed. Such catalogues present a good opportunity for finding compact lensed quasars thanks to the magnification bias of lensing.

\subsubsection{J0818-2613}
This system was confirmed during the 2019 WHT observations. It was discovered as four \textit{Gaia} detections associated with a red ALLWISE detection.  Spectra were taken at two position angles, each through the bright pair, and clearly show that the source is a \textit{z}$=$2.155 BAL quasar with multiple velocity outflows \citep[e.g.,][]{korista1993}. The system is in a high stellar density environment, since it is at a galactic latitude of +5.4$^{\circ}$, an area which has been purposefully omitted from previous searches due to high contamination rates. The strong constraints from requiring four \textit{Gaia} detections around photometric quasars allows us to now probe such high-density regions. The \textit{Gaia} magnitudes are 17.51, 17.58, 19.73, and 19.94, and the system has 3 ROSAT counts within 25\arcsec. The system was targeted for spectroscopic follow-up despite the high stellar density and odd configuration, since the SEDs from optical to infrared wavelengths for all images are similar, and a faint lensing galaxy is seen upon subtraction of the PSFs. The image configuration is peculiar for a quad. When fitting an SIE+shear model to the 4 image positions and relative fluxes, the lensing galaxy position is recovered to within 0.5\arcsec. A model using the galaxy position provides a total magnification of 58, a galaxy mass ellipticity of 0.61, and position angle 34 degrees East of North, with a strong shear of 0.38 at a position angle 126 degrees East of North. This model seems highly unphysical with such a large shear and orthogonal mass and shear axis directions, and is perhaps indicative of the lensing potential not being simple. Indeed, the large Einstein radius of 2.96\arcsec\ is suggestive of a galaxy group or cluster acting as the lens. The lens was also independently confirmed by \citet{stern2021}.
\subsubsection{J0826+7002}
The system was originally targeted in a GDR1 search \citep{lemon2018}, as it was selected as a \textit{Gaia} singlet near a morphological Pan-STARRS galaxy, however the WHT spectra were not deep enough to confirm the fainter object as a quasar. It was targeted again after it was recovered by the same search technique in GDR2, and better seeing conditions and lower airmass observations reveal both \textit{Gaia} detections to be quasars at \textit{z}$=$1.62. Both images are detected in GDR1 and GDR2, but the separation is 5.82\arcsec, hence the system was not discovered by the unWISE modelling or WISE-\textit{Gaia} doublet selection. Pixel modelling shows three lensing galaxies lying closer to the brighter image. There is a third \textit{Gaia} detection in \textit{Gaia} EDR3 on the Western galaxy of these three, perhaps hinting at a third image of the quasar blended with the galaxy. High-resolution imaging is needed for this system.
\subsubsection{J0833+2612}
\begin{figure}
\centering
\includegraphics[width=\columnwidth]{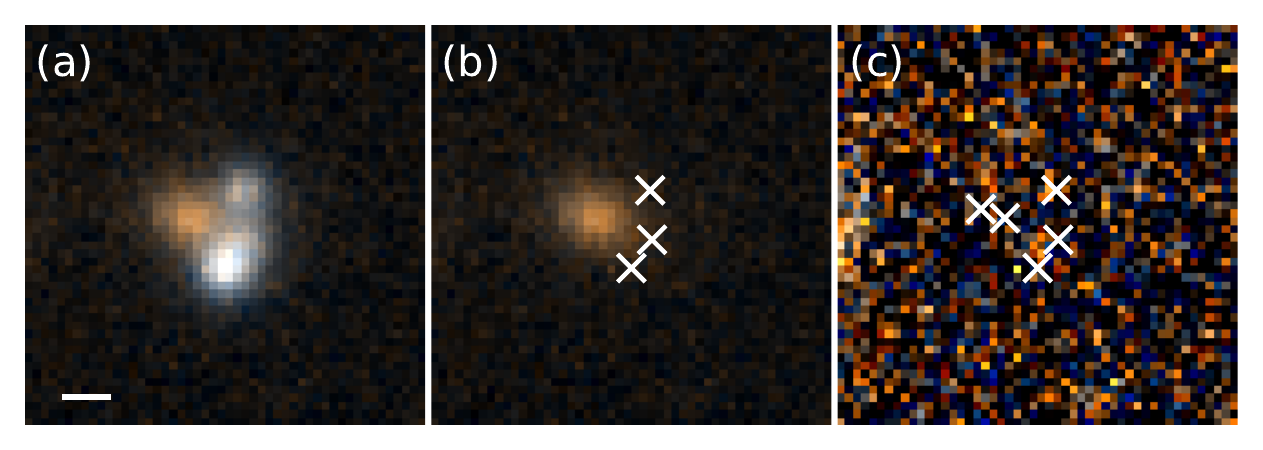}
\vspace{-0.5cm}
\caption{(a) CFHT \textit{rz} colour image of J0833+2612 with a 1\arcsec\ scalebar, (b) data with 3 PSFs subtracted showing the galaxy and bluer counterimage, (c) residuals after subtracted a 4 PSF + Sersic model.}
    \label{fig:J0833}
\end{figure}

This lens was selected as a single \textit{Gaia} detection offset from a Pan-STARRS morphological galaxy (Section \ref{offset}). The slit was positioned at 167.4 degrees East of North, and clearly resolves two traces of a BAL quasar at \textit{z}$\approx$3.26. The DECALS residual imaging shows a likely faint counterimage, confirming this system as a quad. CFHT \textit{r}- and \textit{z}-band imaging also show a counterimage, and our PSF-subtracted modelling of these data is shown in Figure \ref{fig:J0833}. There is a coincident signal in VLASS at the location of the system, suggesting the source could be radio-loud. 

\subsubsection{J0907+6224}
This double has both SDSS and BOSS spectra. It has a source redshift of \textit{z}$=$1.86 and a separation of 2.48\arcsec.  The BOSS spectrum shows two narrow absorption features in the \textsc{Civ} line. The different continua between the BOSS and SDSS spectra is likely attributed to the different fibre apertures (2 and 3\arcsec\ respectively), and contamination from the lensing galaxy. The SDSS pipeline catalogues this as two objects, with an \textit{i}-band magnitude of 18.93 for the Northern component, allowing a colour selection from SQLS to select this as a potential lens. The likely reason this was not selected was due to a colour difference in \textit{g} - \textit{r} between the two components; 0.54 and 1.01 for the quasar and companion respectively \citep{oguri2006}. 
\subsubsection{J0911-0948}
This double lens is associated with an XMM serendipitous source with a flux of 8.6$\pm$1.3$\times$10$^{-14}$ \ergcmcms in the 0.2--12keV band \citep{webb2020}.
\subsubsection{J0918-0220}
This double was confirmed during the 2019 WHT observations, revealing a source redshift of 0.803, and a lens redshift of 0.460. This system was also independently selected as a lens candidate by \citet{sonnenfeld2020} and confirmed in \citet{jaelani2021} who report consistent redshifts.
\subsubsection{J0921+3020}
This 2.93\arcsec-separation double has a source redshift of \textit{z}$=$3.33, with one bright image ($G$=18.66 with SDSS and BOSS spectra), and one faint image ($G$=20.86) blended with a galaxy. The lens redshift of this double is seen in the difference spectra between the two components, showing clear absorption lines at \textit{z}$=$0.428. There are two galaxy clusters surrounding this lens, with spectroscopic redshifts from SDSS spectra of \textit{z}$=$0.334, \textit{z}$=$0.427. Therefore, the lens belongs to the latter cluster. The system lies in the Smithsonian Hectospec Lensing Survey (SHELS) four degree field, and both objects have SHELS spectra of \textit{z}$=$3.33 quasars (Geller et al., 2014). The system has radio and X-ray counterparts in FIRST (4.33$\pm$0.14 mJy at 1.4GHz), SZA \citep[2.9m Jy at 5GHz, 1.4 mJy at 31GHz,][] {muchovej2010}, VLASS \citep[3.5$\pm$0.5 mJy at 2--4GHz, clearly resolved into two components,][]{gordon2021}, and XMM-Newton \citep[3.0$\pm$0.8$\times$10$^{-14}$ \ergcmcms in the 0.2--12keV band,][]{webb2020}.

\subsubsection{J0936-1211}
The NTT-EFOSC2 spectrum shows two clear traces, one of a \textit{z}$=$2.00 quasar with a very blue continuum, and one red trace with absorption features of a \textit{z}$=$0.260 galaxy. There are hints of broad emission features including \textsc{Civ} and the blue continuum in this latter trace. Indeed, the two \textit{Gaia} detections coupled with the alignment of the background quasar within 2.1\arcsec\ of the foreground galaxy strongly suggests the presence of a counterimage. Two PSFs and a galaxy are required for clean residuals of the Pan-STARRS imaging, with the second PSF in the location of an expected counterimage, namely 0.9\arcsec\ from the galaxy opposite the bright image. We therefore classify J0936-1211 as a lensed quasar, however, deeper imaging is needed to definitively confirm this.
\subsubsection{J1003+0651}
This double has three GDR2 detections, centred on a SDSS spectroscopic galaxy at \textit{z}$=$0.225. A feature around 4300Å is indicative of a quasar, and subtraction of the SDSS model galaxy template reveals quasar emission lines at \textit{z}$=$2.56. Our WHT data resolve two quasar traces at the same redshift, either side of the lensing galaxy, thus confirming this system as al lensed quasar.
\subsubsection{J1008+0046}
This system has a spectroscopic redshift of \textit{z}$=$1.511 from the 2df quasar redshift survey  \citep{croom2004}. Our follow-up WHT resolve two quasars at this redshift and HSC imaging clearly reveals a lensing galaxy.
\subsubsection{J1037+0018}
The NTT-EFOSC2 spectra show two quasars at \textit{z}$=$2.462, in agreement with a 2dF spectrum at \textit{z}$=$2.464. This system has HSC DR3 data, which clearly shows a lensing galaxy upon subtraction of the two PSFs. 
\subsubsection{J1041+1710}
The NTT-EFOSC2 spectra resolve two quasars at \textit{z}$=$2.00, with two broad absorption line features in both the \textsc{Civ} and Si\textsc{iv} lines. A lensing galaxy is seen in the PSF-subtracted Legacy Survey imaging.
\subsubsection{J1233-3542}
This double has two bright \textit{Gaia} detections ($G$=18.54, 19.56) separated by 2.08\arcsec. The WHT spectrum reveals particularly narrow quasar emission lines at \textit{z}$=$2.28. Archival DECam \textit{giz} imaging shows clear residuals upon subtraction of two PSFs, thus we classify this system as a lens.
\subsubsection{J1247-3253}
This double has two bright components (\textit{G}$=$18.27, 19.09) separated by 1.85\arcsec. NTT-EFOSC2 spectroscopy reveals very similarly shaped spectra of quasars at \textit{z}$=$1.66. Shallow archival \textit{griz} DECam imaging shows residual flux in the \textit{z}-band, which we take as a detection of the lensing galaxy.
\subsubsection{J1255+0737}
WHT-ISIS long-slit spectra were taken with the slit position angle at 5.6 degrees East of North. The spectra reveal two quasars at \textit{z}$=$2.145. In \textit{Gaia} EDR3 there is a third detection associated to a galaxy 2 arcseconds to the East, which could be due to a another image of the source blended with the galaxy, as in the case of SDSSJ1405+0959 \citep{rusu2014}. High-resolution imaging is needed to see if this is an interesting case of a lens with two lens planes and three images.
\subsubsection{J1303+1816}
A \textit{G}$=$12 star lies 15\arcsec\ from this lens, offering a good opportunity for natural guide star adaptive optics observations of the system. The galaxy lies in a group, with one nearby galaxy having an SDSS spectroscopic redshift of \textit{z}$=$0.410, and a further one with \textit{z}$=$0.473. Clear signs of a \textit{z}$=$0.46 lensing galaxy are seen in the spectrum of the fainter image.
\subsubsection{J1307+0642}
This system was first selected as a lens candidate by \citet{lemon2019}, however, the spectra were too shallow to detect quasar emission lines. The new 1200s WHT-ISIS exposure shows clearly two traces of a \textit{z}$=$2.03 quasar, with the fainter image blended with a \textit{z}$=$0.23 galaxy. A SDSS spectrum confirms our redshift of the lensing galaxy.  Archival Chandra data -- originally from observations of 3C281, which is situated less than 3 arcminutes from this system -- resolve two X-ray point sources coincident with two optically blue point sources either side of a bright galaxy. The fluxes of the two components are 3.7$\pm$1.9 10e-14 and 0.9$\pm$0.4 10e-14 \ergcmcms in the broad 0.5-7keV band.
\subsubsection{J1326+3020}
The WHT spectrum reveals two traces, one of a quasar at \textit{z}$=$1.852, and another of a galaxy at \textit{z}$=$0.339. The Legacy Survey imaging shows a blue point source in the model-subtracted residual image, suggesting this as a faint counterimage. Indeed there are hints of the Ly$\alpha$ line in the blue arm raw pixels of the WHT-ISIS data, thus we classify this as a lens. The lens is a member of a cluster, whose members were targeted for spectroscopy by \citet{holden1999}. They obtained a spectrum for the lensing galaxy of this system, reporting a broad emission line around 5437\AA, which they attributed to \textsc{M}g\textsc{II}, giving a tentative redshift of \textit{z}$=$0.94, however our spectra confirm this is in fact the \textsc{Ciii}] line in agreement with an archival eBOSS spectrum centred on the galaxy.
\subsubsection{J1526-1400} \label{J1526}
This double has the lowest redshift source yet discovered for lensed quasars, \textit{z}$=$0.648, and a particularly bright, low-redshift lensing galaxy at \textit{z}$=$0.096. Similar to J1003+0651, the lensing galaxy is detected by \textit{Gaia}. Given the brightness of this lens ($i=$16.02), and that of J0416+7428, we can consider how many bright lensing galaxies are expected based on the OM10 mocks. With the two doubles presented here, and Q2237+030 \citep{huchra1985}, we have 3 lenses with multiple \textit{Gaia} detections, and lensing galaxies brighter than \textit{i}$=$16.5. This seems to be consistent with the 4-7 expected across the whole sky based on OM10 but it is perhaps surprising given that our selection techniques are based mostly on source colours. Given these small-number statistics, investigating spectroscopic samples of bright galaxies up to \textit{z}$=$0.1 is needed to reveal whether there is an overabundance of low-redshift lensing galaxies compared to the mocks.

\subsubsection{J1550+0221}
This double has a BOSS spectrum, and was selected as a lens candidate independently by \citet{spiniello2018} using pixel analysis of Kilo-Degree Survey imaging. 
\subsubsection{J1945-2857}
This double has two \textit{Gaia} detections separated by 2.69\arcsec, with clearly resolved spectra of a \textit{z}$=$2.56 quasar in the NTT-EFOSC2 data. The 2D PSF-subtracted spectrum shows residuals around the Ly$\alpha$ line below both images, suggesting either a PSF not well-fit by the Moffat used (however no such residuals are seen anywhere else in the subtraction), or Ly$\alpha$ is spatially resolved, suggesting a bright lensed host galaxy or even secondary quasar source. High-resolution imaging is needed for this system. 
\subsubsection{J2017+6204}
Two NOT-ALFOSC 900s spectra were taken at position angles of 124.2 and 50.7 degrees East of North, however both showed a blended trace of a \textit{z}$=$1.72 BAL quasar. Adaptive optics imaging with OSIRIS on Keck 2 resolves the four images and lensing galaxy, and is shown in Figue \ref{fig:osiris}. The system was also independently observed by \citet{stern2021} who obtained unresolved spectroscopy, corroborating our reported redshift. 
\subsubsection{J2110-3755}
The NTT-EFOSC2 spectrum is low signal-to-noise, however, two traces are clearly visible, with at least one associated to a \textit{z}$=$1.50 quasar. Given the extended residuals in the Legacy Survey $r$-band image when fit with just two PSFs and the extended \textsc{M}g\textsc{II} emission across both traces, we classify this system as a lens.
\subsubsection{J2205+1019}
While the extracted 1D spectra only show quasar emission lines in one trace, the residual 2D spectrum clearly shows Lyman-$\alpha$ flux opposite the confirmed quasar, thus confirming the presence of a counterimage. The lens is particularly low redshift at \textit{z}$=$0.108.
\subsubsection{J2205-3727}
This quad was selected as a \textit{Gaia} double around a red WISE detection and visually inspected using archival DECam data, with subsequent spectroscopic confirmation with NTT-EFOSC2. The slit was positioned at 86.4 degrees East of North to pass through the brightest image of the cusp, and the far counterimage. \textit{Gaia} EDR3 only catalogues one image, compared to two in GDR2.  
\subsubsection{J2213-5926}
We identify a likely lens redshift of \textit{z}$=$0.545, however deeper spectra are required to confirm this.
\subsubsection{J2316+0610}
This system was originally selected by \citet{lemon2019} who suggested a possible source redshift of \textit{z}$=$1.96, however their spectra were too noisy to conclusively confirm the system as a lens. Our deeper NOT data clearly resolve two quasars at \textit{z}$=$1.955, thus confirming this system as a lens. The source is a BAL quasar with multiple components, confirmed by a recent eBOSS spectrum. We measure a lens redshift of \textit{z}$=$0.378, consistent with the redshift of a nearby galaxy with a BOSS spectrum.

\subsection{Likely Lensed Quasars}
These systems are all classified as likely lensed quasars. Most are lacking the detection of a lensing galaxy, but have particularly similar resolved spectra, and thus should be prioritised for follow-up high-resolution imaging. For some systems the spectra are too noisy to confirm the presence of a counterimage, however the imaging suggest such a counterimage to be present, and thus require deeper spectroscopy.

\subsubsection{J0138+4841}
This particularly bright, small separation pair ($G$=16.42, 17.45, and $\Delta  \theta$=0.69\arcsec) shows a clear broad emission line at 6078\AA. The 2D spectrum modelling extracts very similar spectra for the two components with a flux ratio smoothly increasing from 1.8 at 4000\AA\ to 2.4 at 9000\AA, the latter of which is in good agreement with the \textit{Gaia} flux ratio (2.6). The brighter component shows narrow absorption at the wavelength of \textsc{Ciii}]. Modelling the Pan-STARRS data reveals extended flux, which can be reasonably well-fit around the 2PSFs when including a Sersic profile which falls between the two sources, suggesting it is a lensing galaxy. Given the small-separation and the only available imaging data being low-resolution, we suggest obtaining additional high-resolution imaging before designating this system as a sure lens. This system is also associated to both a ROSAT X-ray source and a radio source in VLASS (1.65$\pm$0.35mJy at 2--4GHz).
\subsubsection{J0149-6532}
The imaging for this system clearly shows two blue PSFs either side of a luminous red galaxy (LRG). The spectrum of the brighter blue source is low signal-to-noise and shows only hints of a quasar spectrum at \textit{z}$=$0.944. 
\subsubsection{J0221+0555}
The resolved ALFOSC-NOT spectra show very similar \textit{z}$=$1.52 quasar profiles, with both components having strong absorption bluewards of the \textsc{Ciii}] line and similar narrow \textsc{Civ} profiles. Since no residual lens galaxy is seen in the imaging.
\subsubsection{J0326-3122}
This system was already reported by \citet{schechter2017} as a lensless twin at \textit{z}$=$1.34, with narrow absorption at \textit{z}$=$0.504. We obtained EFOSC-NTT spectra of the system across a larger wavelength range, revealing the \textsc{M}g\textsc{II} emission line of the components. Our results qualitatively agree with those of \citet{schechter2017}. A galaxy is seen about 1\arcsec\ North of the system in the PSF subtracted Legacy Survey images, possibly responsable for the \textit{z}$=$0.504 absorption in image A.
\subsubsection{J0326-4950}
The Legacy Survey imaging residuals for this system show an extended galaxy coincident with one of the \textit{Gaia} detections, as well as possible faint blue point source. The spectra are low signal-to-noise but the galaxy is determined to be at \textit{z}$=$0.229 through emission lines. The other blue point source is likely a \textit{z}$=$1.72 quasar, however this is not definite. The system could still be acting as a lens, so deeper imaging and spectroscopy are required. The system is associated with a ROSAT X-ray source.
\subsubsection{J0332-6608}
This system consists of two bright point sources ($G$=18.30, 19.46) separated by 0.94\arcsec. NTT-EFOSC2 spectra reveal similar emission lines of a \textit{z}$=$1.97 quasar, with a constant flux ratio of $\sim$2.5 except for a differing depth of an absorption line at 4860\AA. Only Legacy Survey $g$ and $r$-band data is available, and no obvious lens is seen upon subtracting the two PSFs. 
\subsubsection{J0346+2154}
This 0.99\arcsec-separation system was targetted for NOT-ALFOSC spectroscopy, which confirmed the presence of two quasar spectra at $z=2.365$. The system was also independently selected by \citet{kronemartins2019} as a lens candidate. They obtained resolved spectroscopy and suggest this is most likely a lensed quasar given the similar narrow absorption asssociated with the blue wing of \textsc{Civ}. We also note this absorption line, and striking similarity between the two spectra, with the only slight differences being in the height of the \textsc{Civ} and \textsc{Ciii}] lines, which is likely explained by microlensing. We model CFHT MegaCam stacked $i$-band data as 2 PSFs, however the PSF model is not good enough to clearly tell if there are residuals from a possible lensing galaxy.
\subsubsection{J0916-2848}
While the continua of the two components are quite different, the emission lines have similar profiles. If the system is a lens, the different continua can be explained by the presence of a lensing galaxy and/or microlensing, as has been seen in other systems, for example GRALJ2343+0435 \citep{kronemartins2019}. A prominent absorption line is seen in only the fainter component at 4463\AA. No significant residuals are detected upon PSF subtraction in the Pan-STARRS data. 
\subsubsection{J1008-2911}
The EFOSC2 extracted spectra show two quasars at \textit{z}$=$2.49 with very similar line profiles and continua, suggesting this is a likely lens. However, no residuals are seen in either the Pan-STARRS PSF-subtracted image or archival DECam \textit{gri} imaging.
\subsubsection{J1019-1322}
The NTT-EFOSC2 spectra reveal two quasars at \textit{z}$=$2.325 with similar continua and line profiles, however no galaxy is seen in the Pan-STARRS imaging residuals. 
\subsubsection{J1033-8249}
NTT-EFOSC2 spectra of this 2.33\arcsec-separation double reveals two quasars with similar profiles at \textit{z}$=$1.69. The spectra appear to have a red continuum, however slit losses due to atmospheric refraction and high airmass observations cannot be entirely ruled out despite our correction model. Shallow archival NOAO data exist for this system, however no obvious lens light is seen upon subtracting two point sources.
\subsubsection{J1036-8544}
NTT-EFOSC2 spectra of this 1.22\arcsec-separation double reveals two quasars with similar continua and \textsc{M}g\textsc{II} lines at \textit{z}$=$1.09. Shallow archival NOAO data clearly show the two point sources detected by \textit{Gaia}. Subtracting two PSFs from the available \textit{r} and \textit{i} band data reveal some faint residuals around the system, likely due to a lensing galaxy. We currently classify this system as a likely lens, requiring deeper imaging data.
\subsubsection{J1103-1005}
This 0.72\arcsec-separation system has two bright \textit{Gaia} detections (\textit{G}$=$17.79, 18.01), and NTT-EFOSC2 long-slit spectroscopy shows spatially resolved traces of a \textit{z}$=$1.29 quasar. The emission line profiles are very similar, with a difference in the slope of the continua in the blue. The flux ratio varies smoothly from 1.9 to 1 between 4000 and 8000\AA\ respectively. This is a very likely lens with the spectrum differences attributed to reddening by a lensing galaxy and/or microlensing. There are hints of an extended source in the Pan-STARRS \textit{i}-band PSF-subtracted image, however we suggest higher-resolution imaging is required to confirm this system as a lens.
\subsubsection{J1322+7852}
This system was selected as a \textit{Gaia} double with a red WISE source. In Pan-STARRS and Legacy Survey imaging, there is clearly one blue point source next to a galaxy, and WHT-ISIS spectroscopy shows two traces. The galaxy trace shows both an old stellar population with an obvious 4000\AA\ break and common LRG absorption lines, but also narrow star formation emission lines. The trace of the blue PSF is almost featureless and low signal-to-noise, however binning the spectrum reveals broad emission lines of a \textit{z}$=$2.165 quasar. No spectral features of this quasar appear in the trace of the galaxy as a possible counterimage, however such features would be expected to be too faint if a counterimage exists. The Legacy Survey residuals show a very faint possible counterimage. Deeper spectra and/or imaging are needed to confirm this as a counterimage.
\subsubsection{J1350+3155}
An archival SDSS spectrum of this system shows a galaxy at \textit{z}$=$0.211. Our spectrum corroborates this redshift for the galaxy, and also shows that the nearby blue PSF is a quasar at \textit{z}$=$1.895, however no quasar emission lines are seen in the combined 1500s extraction of the galaxy. We expect that the quasar is strongly lensed given the small separation between the \textit{Gaia} detections of 1.44\arcsec, and the high velocity dispersion of the galaxy (238 \kms). We estimate the Einstein radius of this lens given this dispersion, the source and lens redshifts, and assuming a singular isothermal sphere mass profile, obtaining $\theta_{E}$ $\approx$1.35\arcsec. Generally, images of such a profile are separated by twice the Einstein radius, i.e. 2.7\arcsec. The likeliest solution is that the counterimage is highly demagnified, either through microlensing, or extinction/reddening by the lensing galaxy, or some combination of both. High-resolution imaging for this system should be prioritised to search for a counterimage.
\subsubsection{J2244-0550}
NTT-EFOSC2 spectra show both components to be quasars at \textit{z}$=$2.84 with almost identical profiles. However, no lens residuals are seen in the Pan-STARRS or Legacy Survey PSF-subtracted images. 
\subsubsection{J2255+8009}
The NOT-ALFOSC spectroscopy resolves two quasars with very similar spectra and a pronounced drop in flux below 4650\AA, either due to intrinsic quasar emission or line-of-sight absorption. Two deep absorption features are seen at 4607\AA\ and 4650\AA\ consistent with absorption from Ca H and K at \textit{z}$=$0.172, however, no other clear absorption features at this redshift are seen in the spectra. The emission lines are likely associated with a \textit{z}$\approx$2.8 quasar, implying that Ly$\alpha$ is strongly absorbed (and perhaps these two absorption lines are due to proximate damped Ly$\alpha$ systems). The Pan-STARRS data are well fit by two PSFs.

\subsection{Unclassified Quasar Pairs}
The following systems have resolved spectroscopy showing two quasar components at consistent redshifts, however they lack the spectral signal-to-noise and/or lens galaxy detection to be classified as a (likely) lensed quasar. 
\subsubsection{J0124-6334}
The Legacy Survey image of this system shows two point sources of different colours, and subtracting 2 PSFs shows clear extended residuals between the PSFs. When including a Sersic profile, the data are well fit, with the position of the Sersic falling between the two PSFs closer to the fainter image, consistent with that expected of a lensing galaxy in a typical double configuration. The best-fit Sersic index is $\sim$1, typical of spiral galaxies. The spectra are low signal-to-noise and overlap due to poor seeing and small separation, however binning the spectra shows large differences between them, with different continua, broad emission line profiles, and relative intensities (e.g. \textsc{Oiii} is present in the fainter image). Thus the lensing scenario is unlikely, but not necessarily ruled out due to differential extrinsic effects like strong absorption and/or microlensing. If these are indeed distinct quasars, the projected separation of 0.87\arcsec\ is 6.52 kpc at $\textit{z}=$1.30. 
\subsubsection{J0125-1012}
The two components of this system have similar spectra of a \textit{z}=1.22 quasar, but no definite lensing galaxy is detected in the available imaging. 
\subsubsection{J0127-1441}
The NTT-EFOSC2 spectra reveal the two components to be similar quasars at \textit{z}$=$1.754, however there is only a faint red source ($i<22$) South of the system seen in the PSF subtracted image. Given the wider separation of this system, 3.00\arcsec, and assuming it is a lens, we can estimate the faintest possible lensing galaxy brightness. We use the \citet{om10} catalogue and simply search this catalogue for mock lenses with similar image separations (within 0.5\arcsec) and source redshifts (within 0.25), and consider the lens magnitudes and redshifts of the remaining systems as the possible range of lensing galaxy parameters for J0127-1441. 279 systems are recovered, with median and 1$\sigma$ uncertainties in redshift and $i$-band magnitude of 0.44 $\pm$ 0.18 and 19.08 $\pm$ 1.01 respectively. The lens properties of the system with the faintest lensing galaxy are \textit{z}$=$0.974 and $i$=20.6. It is thus unlikely that the faint source is a lensing galaxy. 
\subsubsection{J0130+0725}
The spectra show two quasars at $\textit{z}$=1.54 with different emission line profiles, as there is prominent absorption in the \textsc{Civ} and \textsc{M}g\textsc{II} lines of the fainter quasar. This is likely explained by the system being a projected pair with some small velocity separation, and line-of-sight absorption by the host galaxy of the closer (brighter) quasar. A higher resolution archival eBOSS spectrum indeed resolves both the \textsc{M}g\textsc{II} 2800\AA\ and \textsc{Civ} 1549\AA\ absorption doublets \citep{blanton2017}.  No residuals are seen in the PSF-subtracted images. Intrinsic absorption along different sightlines to the same source quasar cannot necessarily be ruled out.
\subsubsection{J0133+0816}
The spectra reveal two quasars at \textit{z}$=$1.27, with small differences in the widths of the \textsc{Ciii}] and \textsc{M}g\textsc{II} emission lines. Modelling deep CFHT \textit{riz} Megacam stacked data as two PSFs reveal no obvious lensing galaxy.
\subsubsection{J0146-6510}
The two components have very similar NTT-EFOSC2 spectra of a quasar at $\textit{z}$=1.255, however no lens is seen in the Legacy Survey $z$-band residuals after fitting 2 PSFs. The system is associated with a radio source in GLEAM \citep[0.555$\pm$0.011 Jy at 170-231MHz,][]{hurley2017}, and SUMSS \citep[214$\pm$9 mJy at 843MHz,][]{mauch2003}, and an X-ray source \citep[Slew Survey XMMSL2 Catalogue, 2.3$\pm$1.1$\times$10e--15 Wm$^2$,][]{saxton2008}.
\subsubsection{J0311+0550}
The NOT spectrum of this system reveals the two components to have similar spectra of a $z=0.777$ quasar, however modelling the Pan-STARRS data as two point sources does not show any significant residuals. 
\subsubsection{J0527-2431}
The NTT-EFOSC2 spectrum shows two resolved traces with emission lines of a \textit{z}$=$1.437 quasar, but with differences in the line profiles and relative intensities. Deeper spectra and/or imaging are required to definitively classify the two components as distinct quasars, separated by 1.46\arcsec, i.e. 12.5 kpc.
\subsubsection{J0642+5337}
The NOT-ALFOSC resolved spectrum shows two quasars at \textit{z}$=$1.86, with several narrow absorption lines in the brighter component. This is likely a distinct quasar pair due to the lack of lens in the PSF-subtracted Legacy Survey image, the different \textsc{Civ} profile, and the sky separation of 3.17\arcsec.
\subsubsection{J0723+4739}
This pair shows no residuals in PSF-subtracted Legacy Survey imaging, and the peaks of the \textsc{M}g\textsc{II} profiles are offset in velocity. Given possible extrinsic absorption in a lensing configuration causing this difference, we categorise the system as an unclassified quasar pair, but likely a binary quasar. If the components are indeed distinct quasars, their projected separation would be 10.9 kpc (1.58\arcsec). Deeper spectra and imaging are required. This system was targeted by \citet{lemon2018} but their results were inconclusive due to low signal-to-noise.
\subsubsection{J0728+2607}
The NOT-ALFOSC spectrum shows two quasars at \textit{z}$=$1.025, with large differences in their emission line profiles and continua, thus making it a likely physical quasar pair. Their projected separation is 17.6 kpc at \textit{z}$=$1.025 (2.15\arcsec).
\subsubsection{J0816+2339}
Despite the WHT spectrum being low signal-to-noise, the two components are clearly quasars at \textit{z}$=$1.22. This is likely a distinct quasar pair given the slight differences in the \textsc{M}g\textsc{II} profile.
\subsubsection{J0833-0721}
The NTT-EFOSC2 spectrum shows two similar traces of a \textit{z}$=$0.828 quasar, however there is a small offset in the \textsc{Oiii}-4960 and 5008\AA\ lines, which appears to be real in the 2D spectrum. This would immediately rule out the lensing hypothesis, however the spectra are particularly similar, and small wavelength variations in the spatial direction could explain the apparent offset. No lens is seen in the Pan-STARRS imaging residuals. Further spectra are required to confirm a velocity difference in the \textsc{Oiii} lines. If the quasars are indeed distinct, the projected separation would be 9.39 kpc.
\subsubsection{J0909-0749}
The NTT-EFOSC2 spectra resolve this 0.81\arcsec-separated pair, showing clear \textit{z}=1.075 quasar emission lines in each component, however the continua are significantly different. Modelling the Pan-STARRS data as 2 point sources shows no significant residuals in the redder bands. The system is associated with a VLASS detection with a flux of 2.6$\pm$0.3mJy.
\subsubsection{J0941-2443}
The NTT-EFOSC2 spectrum shows two traces, each with an emission line around 5400\AA. This is most likely \textsc{M}g\textsc{II}, as then other common emission lines would fall outside the wavelength range of our spectrum. \textsc{Ciii}] would then lie at the very bluest end of our spectrum, and indeed an increase in flux may hint at the red wing of this emission line. The \textsc{M}g\textsc{II} profiles are different, and there is no apparent lensing galaxy in the PSF-subtracted Legacy Survey or Pan-STARRS imaging. This is likely a distinct quasar pair, with a projected separation of 15.8 kpc (2.25\arcsec).
\subsubsection{J1045+3433}
WHT-ISIS spectroscopy reveals two components with similar spectra of a \textit{z}$=$1.205 quasar. No lens seen in Legacy Survey residuals but emission lines and continua are similar enough to warrant deeper follow-up imaging.
\subsubsection{J1102+3421}
The WHT-ISIS spectrum reveals two quasars at \textit{z}$=$1.405, however strong absorption is seen in the \textsc{Civ} line of only the fainter component. Coupled with no lens residuals in the PSF-subtracted Pan-STARRS residuals, this system is likely a physical quasar pair, with the fainter component at slightly higher redshift than the bright component, however differential line-of-sight absorption in a lensing scenario cannot necessarily be ruled out.
\subsubsection{J1202+0703}
This 1.21\arcsec-separated pair shows two spatially resolved traces of a \textit{z}$=$2.187 quasar in the NTT-EFOSC2 spectrum, with each trace having similar continua and profiles, except for absorption in the \textsc{Civ} line of the fainter trace. No lens is seen in the residuals.
\subsubsection{J1428+0500}
This system was previously followed up by \citet{lemon2019} who concluded the two components had spectra of quasars at \textit{z}$=$1.38. Our new WHT-ISIS spectra now reveal the \textsc{Civ} profile and place both quasars at \textit{z}$=$1.375, with differing \textsc{M}g\textsc{II} profiles due to absorption in the fainter component. The Legacy Survey data are well fit by two PSFs. If the quasars are distinct, the projected separation of 2.23\arcsec\ translates to 17.0 kpc at \textit{z}$=$1.375.
\subsubsection{J1449-2025}
EFOSC2-NTT spectroscopy shows that both components of this 1.17\arcsec-separation pair are quasars at \textit{z}$=$0.594 with similar spectra. A two PSF fit to the Pan-STARRS imaging shows significant residuals around the system, as expected if the system is either a lens (residuals due to a lensing galaxy) or simply a distinct quasar pair (residuals due to the quasar hosts given the low redshift). Adding a single Sersic fits these residuals and this component falls between the two PSFs. If this system is a distinct pair, the projected separation would be 7.9 kpc.
\subsubsection{J1821+6005}
The NOT-ALFOSC spectrum shows two traces of quasar at \textit{z}$=$2.052 with similar continua, but a slight difference in the shape of the \textsc{Civ} line. \citet{lemon2018} also presented WHT-ISIS follow-up of this system, concluding the system to be a distinct quasar pair. No lens is seen in the PSF-subtracted Pan-STARRS, Legacy Survey, or CFHT $r$-band imaging.
\subsubsection{J1832+5349}
The two components of this system are separated by 3.03\arcsec, and NOT-ALFOSC spectra reveal two quasars at \textit{z}$=$1.16, with similar \textsc{M}g\textsc{II} emission line profiles and continua. However, no lensing galaxy is seen in the Legacy Survey residuals after PSF subtraction. In Pan-STARRS there is a hint of residual flux between the two quasars in the \textit{Y}-band. Given the separation and relatively low redshift, we expect this is a distinct quasar pair.
\subsubsection{J2008+0438}
This system lies in a high stellar density environment, at a galactic latitude of --14$^{\circ}$. One trace of the NOT-ALFOSC spectra is clearly a quasar at \textit{z}$=$1.70, however the other does not have similarly obvious broad emission lines. The latter trace has a broad emission line at the same wavelength as the \textsc{M}g\textsc{II} line of the first trace, however the telluric absorption band obscures the red wing. There is a hint of a line at the same wavelength as \textsc{Ciii}] of the first trace suggesting this source is also a quasar at \textit{z}$=$1.70, but no \textsc{Civ} emission is seen.

\subsubsection{J2057+0217}
The NOT-ALFOSC spectra show two components with similar continua but different emission line profiles suggesting a distinct quasar pair. If true, the sky separation of 1.16\arcsec\ would correspond to a physical projected separation of 8.9 kpc at \textit{z}$=$1.52. The WHT-ISIS observations of this system by \citet{lemon2018} do not reveal a \textsc{Civ} emission line as we see in these newer spectra, which is likely due to unaccounted slit losses.
\subsubsection{J2341-1557}
NTT-EFOSC2 spectra of this small-separation system (1.02\arcsec) reveal two quasars at \textit{z}$=$1.535 with different continua and relative emission line strengths. It is likely a distinct quasar pair -- and thus would have a projected separation of 8.8 kpc.

\subsection{Physical Quasar Pairs}
\subsubsection{J0041-5350}
Resolved spectra show that each component of this system is a quasar at \textit{z}$=$0.55, however the lensing scenario is ruled out since the \textsc{Oiii} profiles are very different. The extended asymetric flux seen in the ground-based imaging is likely a sign of an ongoing merger between the two quasar host galaxies. The sky separation of 1.06\arcsec\ translates to 6.2 kpc, making it one of the smallest separation confirmed quasar binaries above $\textit{z}$=0.5 \citep{chen2022}.
\subsubsection{J0315-3522}
Legacy Survey imaging shows two point sources and a galaxy offset to one side of the fainter component. NTT-EFOSC2 spectra reveal two quasars at \textit{z}$=$0.457, with residuals around \textsc{Oiii}-5008\AA\ suggesting spatially resolved emission. Given the difference between the spectra for both the continua and emission line profiles, and lack of lensing galaxy candidate in the imaging, we designate this system as a distinct quasar pair. The separation is 1.30\arcsec\ at \textit{z}$=$0.457 -- a projected separation of 6.9 kpc.
\subsubsection{J1116-2122}
Given the 3.29\arcsec\ separation and relatively low redshift sources (\textit{z}$=$0.709), we categorise this as a distinct quasar pair. The projected separation is 23.9 kpc.
\subsubsection{J2121+1713}
The spectra reveal two quasars at \textit{z}$=$0.63, however, the emission lines and continua are markedly different. Given also the low redshift, large separation of 2.89\arcsec, and lack of lensing galaxy in the residuals, we classify this system as a distinct quasar pair. The projected physical separation is 17.9 kpc.
\subsubsection{J2132+3635}
The NOT-ALFOSC spectrum shows two quasars at \textit{z}$=$1.215 with slight differences in the \textsc{M}g\textsc{II} profile. Follow-up OSIRIS imaging reveals no lensing galaxy (Figure \ref{fig:osiris}), thus we designate this as a distinct quasar pair. The sky separation of 1.33\arcsec\ translates to a physical projected separation of 9.9 kpc. 
\subsubsection{J2256+2223}
Despite similar profiles of a \textit{z}$=$0.75 quasar, the \textsc{Oiii} narrow emission lines are slightly offset in wavelength. This cannot be a calibration drift in the spatial direction since the sky emission lines stay within the same pixel centroid across the width of the system and over several tens of arcseconds. Therefore, this is a distinct quasar pair with sky separation of 1.63\arcsec, translating to 10.7 kpc in projected separation at \textit{z}$=$0.75.
\subsubsection{J2322+0916}
ALFOSC-NOT spectroscopy shows that both components of this pair are quasars at \textit{z}$=$1.20, in agreement with the SDSS spectrum of the fainter component. The emission line shapes and continua are different, and no residuals are seen in the PSF-subtracted Legacy Survey imaging. Therefore, we classify this system as a distinct quasar pair, with projected separation of 37.9 kpc (4.51\arcsec).
\subsubsection{J2355-4553}
The NTT-EFOSC2 spectrum reveals two quasars at \textit{z}$=$2.85, however one has strong broad absorption lines. We suggest this must imply that the two quasars are distinct, and thus have a projected separation of 16.7 kpc given their 2.20\arcsec\ sky separation.

\subsection{Projected Quasar Pairs}
\subsubsection{J0027+0438}
The spectra show a bright quasar at \textit{z}$=$1.972, and an AGN (broad H$\alpha$ emission) with extended host galaxy at \textit{z}$=$0.1935. The projected separation of the system is 1.92\arcsec. While no signs of a counterimage are seen in the imaging, the lensing scenario is still plausible as the quasar host galaxy requires only a modest Einstein radius, $\sim$1\arcsec, and would explain the particularly large apparent brightness of the quasar through an associated magnification. High resolution imaging or deeper spectroscopy will be required to rule out the lensing hypothesis fully.
\subsubsection{J1820+3747}
This projected quasar pair system is associated with an X-ray source in ROSAT. The quasar redshifts are 0.614 and 0.754, and Legacy Survey imaging and photometric redshifts suggests the system is coincident with a cluster at the background quasar source redshift.
\subsubsection{J2302-4154}
One trace shows \textsc{Ciii}] and \textsc{M}g\textsc{II} emission lines of a \textit{z}$=$1.22 quasar, while the other only shows one emission line at 5540\AA, and no other clear emission lines. This is most likely \textsc{M}g\textsc{II}, since at this redshift no other common quasar emission lines are expected to fall within the wavelength range of our spectra. This would put the second component at a lower redshift of 0.98, with the background quasar passing at 8.7 kpc in projection.

\subsection{Other}
\subsubsection{J0336-3244}
One trace of this system is a $z=0.545$ quasar, however the trace of the second object is noisy, with hints of \textit{z}$=$0 absorption lines, i.e. it is possibly a star. Deeper spectra are needed to verify this.
\subsubsection{J0819+0457}
The signal-to-noise of the fainter trace is too low to determine any robust features. It is likely that this is a distinct quasar pair or quasar-star projection, however deeper spectra are needed.
\subsubsection{J0938+0629}
The spectrum shows resolved narrow line emission at \textit{z}$=$0.363, however only the brighter component has broad emission lines. We therefore classify this as a quasar+galaxy pair.
\subsubsection{J1140+2303}
The resolved spectra show a \textit{z}$=$2.405 quasar and a likely star given H$\alpha$ and H$\beta$ absorption at \textit{z}$=$0. However, there are also similarities between the two spectra, in particular around the wavelength of Lyman alpha, which is unlikely to be due to extraction problems given the wide separation of 3.63\arcsec. We suggest this is a quasar-star projection, however, deeper spectroscopy and/or imaging is needed to definitively conclude this.
\subsubsection{J1442-0857}
The extracted spectra reveal galaxies at \textit{z}$=$0.164, the redder one of which has strong narrow line emission (OII, H$\alpha$, NII, SII). The redder component also appears to be coincident with a bright radio detection in VLASS.
\subsubsection{J1752+0826}
This system appears as two blue point sources either side of a galaxy, however follow-up WHT-ISIS spectra reveal likely \textit{z}$=$0 absorption lines in both traces, alongside narrow line emission at the redshift of the galaxy, \textit{z}$=$0.3195. Such similarities to double quasar configurations are expected at high stellar densities. This system lies at a galactic latitude of \textit{b}=15$^{\circ}$.

\subsection{Lensed galaxies}

As a by-product of looking for lensed quasars, lensed galaxies may be discovered semi-serendipitously since they share unique morphological similarities with lensed quasars, namely multiple blue detections around luminous red galaxies. We present four new lensed galaxies that were found in our searches, two with single \textit{Gaia} detections and two with two blue SDSS detections around a red galaxy.

\subsubsection{J0343-2828}
This lens was selected as a potential lensed quasar candidate through the single \textit{Gaia} detection offset method using DR2 ($G_{DR2}$= 21.29), and looked promising given the compact PSFs. However, spectroscopic follow-up with the NTT in October 2019 confirmed the source as a \textit{z}$=$1.655 galaxy, with a lens at \textit{z}$=$0.385. The coordinates are: R.A., Dec. = (55.7978, -28.4779). \textit{HST} imaging (\textit{HST} Proposal 15652, P.I.:Treu) shows that indeed the images are not consistent with point sources, and further show two lensing galaxies with a fifth image between them \citep{schmidt2022}. 

\subsubsection{J0500-5534}
This lens was selected as a potential lensed quasar through the single \textit{Gaia} detection offset method in \textit{Gaia} DR2 ($G_{\textrm{DR2}}$= 20.28) and a marginally red colour of W1--W2=0.53. The coordinates are: R.A., Dec. = (75.1017, -55.5796). The DES imaging is well-fit by four point sources and a Sersic galaxy profile, with the positions forming a short-axis cusp configuration, which is indeed well-fit by an SIE+shear lens model. Follow-up spectroscopy with FIRE on the 6.5-meter Baade Magellan telescope shows clearly resolved narrow emission lines of a \textit{z}$=$0.620 galaxy. Figure \ref{fig:lensedgals} shows the 1D spectrum, with the following emission lines marked: \textsc{H}$\alpha$, [\textsc{Siii}], and He\textsc{i}. In EDR3, there is an additional detection for the counterimage and is the only known lensed galaxy with two \textit{Gaia} detections to our knowledge ($G_{\textrm{DR2}}$ = 20.84, 20.98). 

\subsubsection{J0920+4521}
This system was discovered through a search for multiple blue catalogued detections in SDSS within a few arcseconds of a red photometric galaxy. Legacy Survey imaging shows a blue arc East of a group of four galaxies all with similar colours (see Figure \ref{fig:lensedgals}). Two of these galaxies have existing SDSS spectra, one an LRG at \textit{z}$=$0.546, and the other with narrow emission lines at \textit{z}$=$0.956. However, closer inspection of the latter spectrum shows also an LRG spectrum consistent with \textit{z}$=$0.546, suggesting the emission lines are from a background source, possibly a counterimage of the observed arc. Follow-up spectroscopy with ISIS on the WHT were taken on 31 March 2017 at two position angles -- along the arc, and through the brightest part of the arc and the galaxy with the spectrum with narrow emission lines. The spectra are shown in Figure \ref{fig:lensedgals}. The arc shows clear absorption lines of a \textit{z}$=$2.633 galaxy, and the galaxy spectrum is qualitatively similar to the SDSS one, showing the narrow OII emission of a \textit{z}$=$0.956 source. One possibility is that the main lensing galaxy is indeed at \textit{z}$=$0.956 and any signs of a \textit{z}$=$0.546 LRG are due to contamination or blending with a nearby galaxy. However, this is unlikely given that the ground-based imaging shows clearly similar colours between the multiple galaxies. We suggest the most likely situation is that there are two main lensing galaxies at \textit{z}$\approx$0.55 with two sources at \textit{z}$=$0.956 and \textit{z}$=$2.633. While the lensing potential may be complex, if well-constrained, the system can be a very useful tool for cosmography \citep[e.g.][]{collett2014}.

\subsubsection{J1059+4251}
\begin{figure*}
\centering
\includegraphics[width=\textwidth]{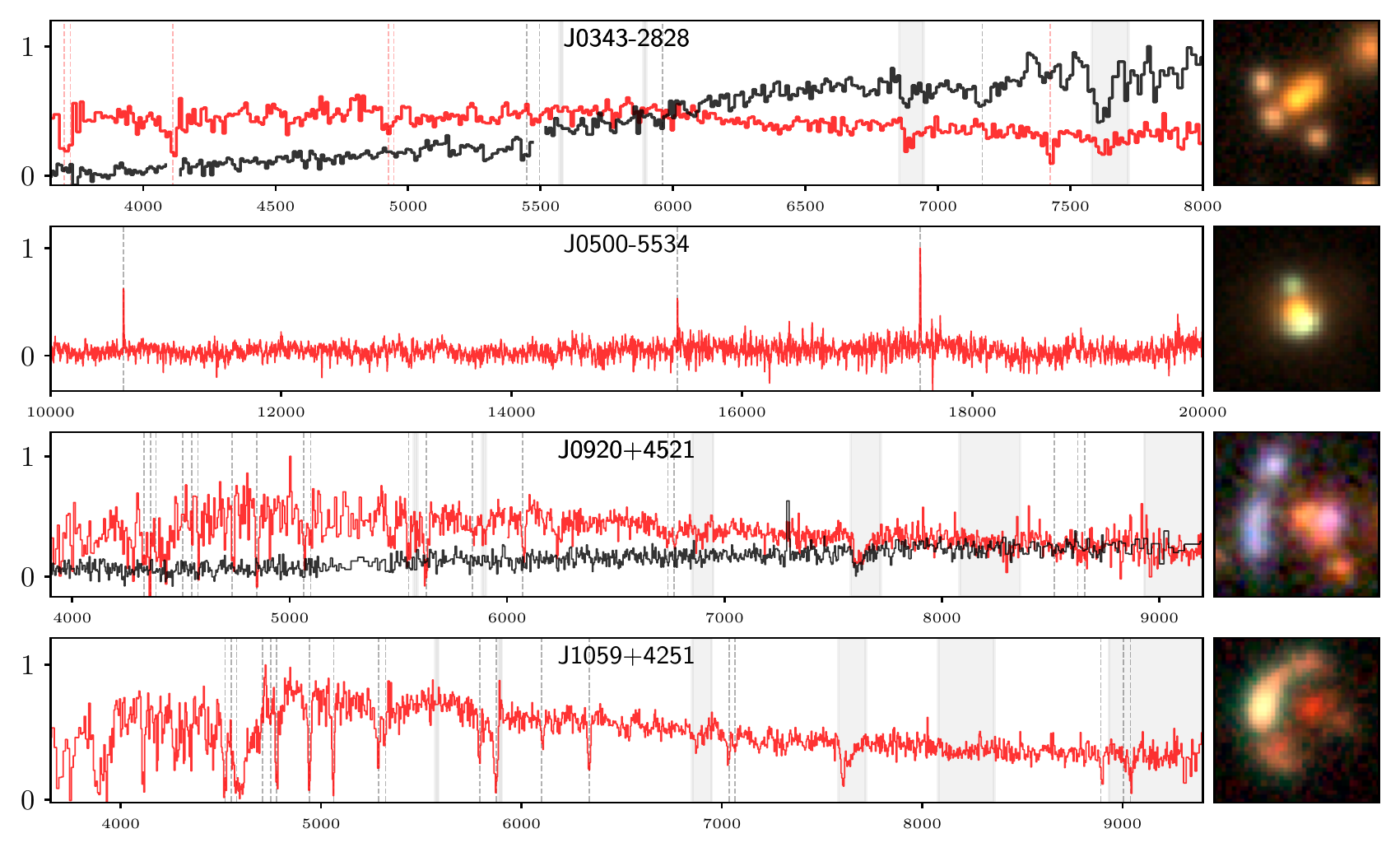}
\vspace{-0.5cm}
\caption{Spectra and \textit{grz} images of serendipitously discovered lensed galaxies. Fluxes of the spectra are normalised to have a peak value of 1, and the x-axis is wavelength in Angstroms. Cutouts are from DECaLS and are 10 arcseconds on the side.}
    \label{fig:lensedgals}
\end{figure*}

This system was discovered through a search in SDSS imaging for multiple photometric quasar candidates around photometric red galaxies. Follow-up spectroscopy with ISIS on the WHT on 31 March 2017 targeting the bright arc reveals a \textit{z}$=$2.793 galaxy. Figure \ref{fig:lensedgals} shows the 1D spectrum and the DECaLS \textit{grz} colour image of the system. \textit{HST} imaging reveals three lensing galaxies and a complex lensing configuration. This system is examined in detail in \citet{citro2021}. 

\section{Discussion} \label{discussion}
\subsection{Lenses} 
\subsubsection{Sample characteristics} \label{samplechar}
We can compare several features of the lenses we have discovered with those previously known in order to elucidate possible biases in our selection methods and/or those of previous discovery methods. For the following analysis, we compile a list of 224 lensed quasars from the literature, mainly from the database of \citet{lemon2019}, but including those from more recent publications \citep[e.g.,][]{desira2022}. Since our goal is to find lenses with optically bright images detected by \textit{Gaia}, we remove lenses with no \textit{Gaia} detections, or unmeasured source redshifts. From the known sample, this eliminates 38 lenses, mainly composed of CLASS radio-loud sources, lenses with faint optical images \citep{more2016, chan2022}, particularly reddened or high-redshift quasar sources \citep[e.g.,][]{fan2019}, and two systems which have \textit{Gaia} detections but unmeasured or insecure source redshifts: B2108+213 from \citet{mckean2005} and J2218-3322 from \citet{chen2022}. It also removes one object from the new sample, J0310-5545, which was serendipitously discovered because of a \textit{Gaia} detection from a nearby star. This leaves 85 lenses (8 quads, 1 triple, 76 doubles) in our new sample, and 186 (2 five-image lenses, 41 quads, 3 triples, 140 doubles) in the previously known sample. We can immediately compare the multiplicity ratio of the known sample against those of this paper: the ratio of lenses with more than two images to those with just two is 46 to 140 (1 to 3.0), and 9 to 76 (1 to 8.4) respectively. These very different ratios are likely due to the ease of finding quads (generally easily recognisable even without the presence of a bright lensing galaxy) compared to doubles coupled with the bias of following up quad candidates and reporting them in earlier papers or by others \citep[e.g.,][]{schechter2017, lucey2018}. It cannot be taken as a bias in our selection since our selection criteria recover all optically bright quads.

\begin{figure}
\centering
\includegraphics[width=\columnwidth]{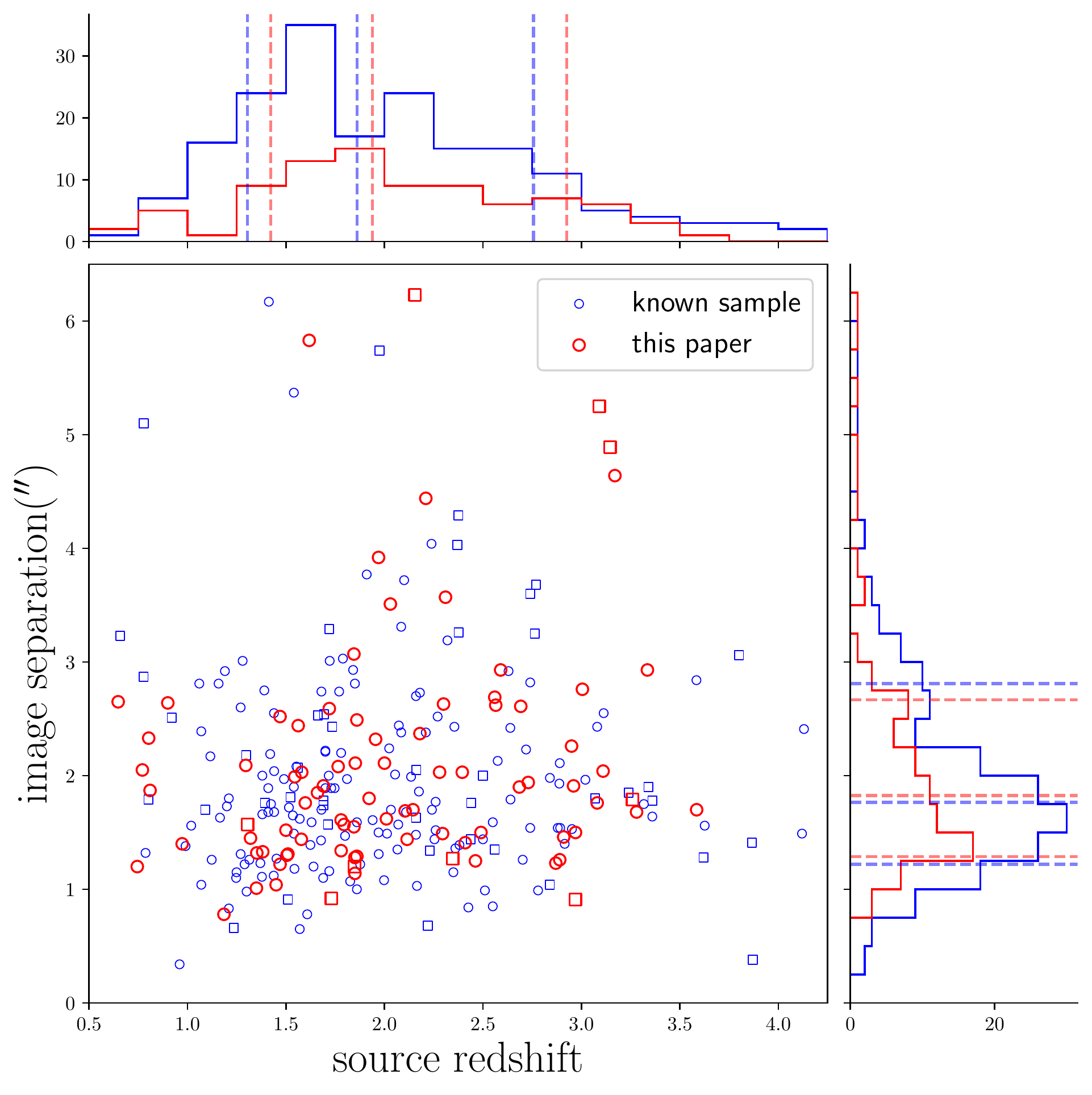}
\vspace{-0.5cm}
\caption{Image separation against source redshift for previously known lensed quasars (blue) and those presented in this work (red). Circles are doubles, squares are lenses with three or more images. Only those lenses falling within the image separation and source redshift limits of the plot are included in the histograms. Lines are overlaid for the 16th, 50th, and 84th percentiles of each distribution. }
    \label{fig:imsep_sourcez_known_new}
\end{figure}

In Figure \ref{fig:imsep_sourcez_known_new}, we show the source redshift and maximum image separation of these two samples, marking the 16th, 50th, and 84th percentiles, which highlights two marginal differences: (i) the sources of the lensed quasars from this work have a higher median redshift than the known lens sample (1.94 and 1.86 respectively), likely due to redshift constraints of previous searches relying on UV-excess for pre-selection \citep{oguri2006}; and (ii) the image separation distribution is marginally more peaked than that of the existing sample (medians of 1.83 and 1.77, and standard deviations of 0.92 and 1.07 respectively). While not significantly different populations, these differences can be explained by the cutoff at larger separations in our original selection criteria, and also the cutoff at lower separations from our candidates which are likely lenses but do not show lensing galaxies in the shallow existing imaging. The counterparts in the existing sample have already received the necessary deeper follow-up imaging. Including these unclassified quasar pairs as lenses would remove this difference at smaller separations, but exacerbate the redshift offset between the samples. See Section \ref{mockcomparison} for a comparison of these samples to mock catalogues, to assess the relevant completeness of known lensed quasars.

\subsubsection{Comparing \textit{Gaia} DR2 to \textit{Gaia} EDR3}
We investigate the differences in detection of lensed quasar images between \textit{Gaia} DR2 and early \textit{Gaia} DR3 (EDR3), not only to elucidate the completeness and biases in our current and previous selections, but also as a potential probe of discovering new lenses through catalogue parameter comparison. We combine the known and new samples of lensed quasars from Section \ref{samplechar}, resulting in 274 lenses (now including those without known source redshifts). We inspect cutouts of the lenses with the detections from both data releases overlaid and note the following: 

\begin{itemize}
    \item 241 (88\%) have equal numbers of \textit{Gaia} detections between data releases; however in 3 cases detections have been replaced: HS0810+2554 \citep{reimers2002}, the DR2 detections coincide with one of the bright pair and the Northern image, while EDR3 replaces the Northern image with the South-Eastern image -- interestingly, the first data release catalogues the same images as EDR3; PSJ0840+3550 \citep{lemon2018}, the counterimage was catalogued in DR2 but in EDR3 it has been removed and the galaxy is catalogued; and SDSSJ1251+2935 \citep{kayo2007}, the merging pair was catalogued in DR2, but now has just the Southern image of this pair and the Western image catalogued in EDR3.
    
    \item 25 (9\%) have gained one detection. For 19 of these, the detection is associated with the lensing galaxy. For 5 (3 quads, 2 doubles), the detection is associated with another image. And for 1, PMNJ0134-0931, it did not previously have a detection in DR2.
    
    \item 7 have lost a detection. 5 (1 quad, 4 doubles) have lost an image detection, 1 double (J1003+0651) has lost a galaxy detection, and 1 quad (SDSSJ1640+1932) has lost both of its image detections and gained a detection of the lensing galaxy.
\end{itemize}

\begin{figure*}
\centering
\includegraphics[width=\textwidth]{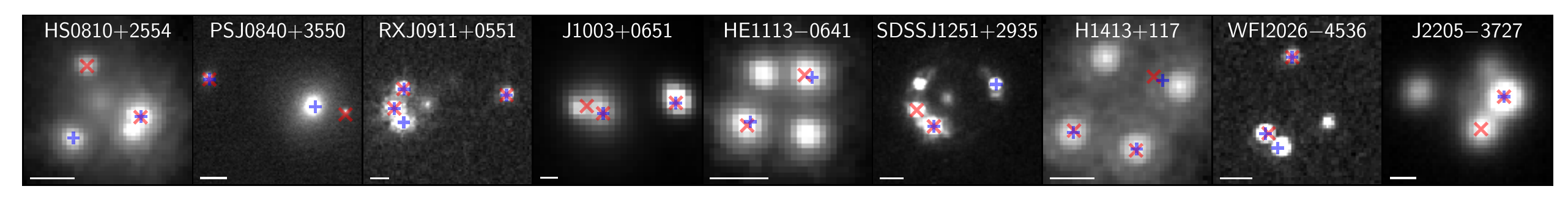}
\vspace{-0.5cm}
\caption{Overlay of DR2 and EDR3 catalogue detections on the best available imaging data for various interesting cases. Blue plus symbols are EDR3, while red crosses are DR2. The white scalebar represents 0.5 arcseconds. The filter has been chosen to match the peak of the \textit{Gaia} G bandpass when possible.}
    \label{fig:overlay}
\end{figure*}

Figure \ref{fig:overlay} shows an assortment of lens systems with the DR2 and EDR3 detections overlaid. Interestingly, for H1413+117, the \textit{Gaia} position for image C is offset from the true position, i.e. from \textit{HST} imaging, and has shifted closer to this true position in EDR3, however, it is still inconsistent with the position given the EDR3 positional uncertainty of $\sim$6 milliarcseconds. Conversely, the positions of the two EDR3 detections in HE1113-0641 seem to be further from the true positions than in DR2. 

\begin{figure}
\centering
\includegraphics[width=\columnwidth]{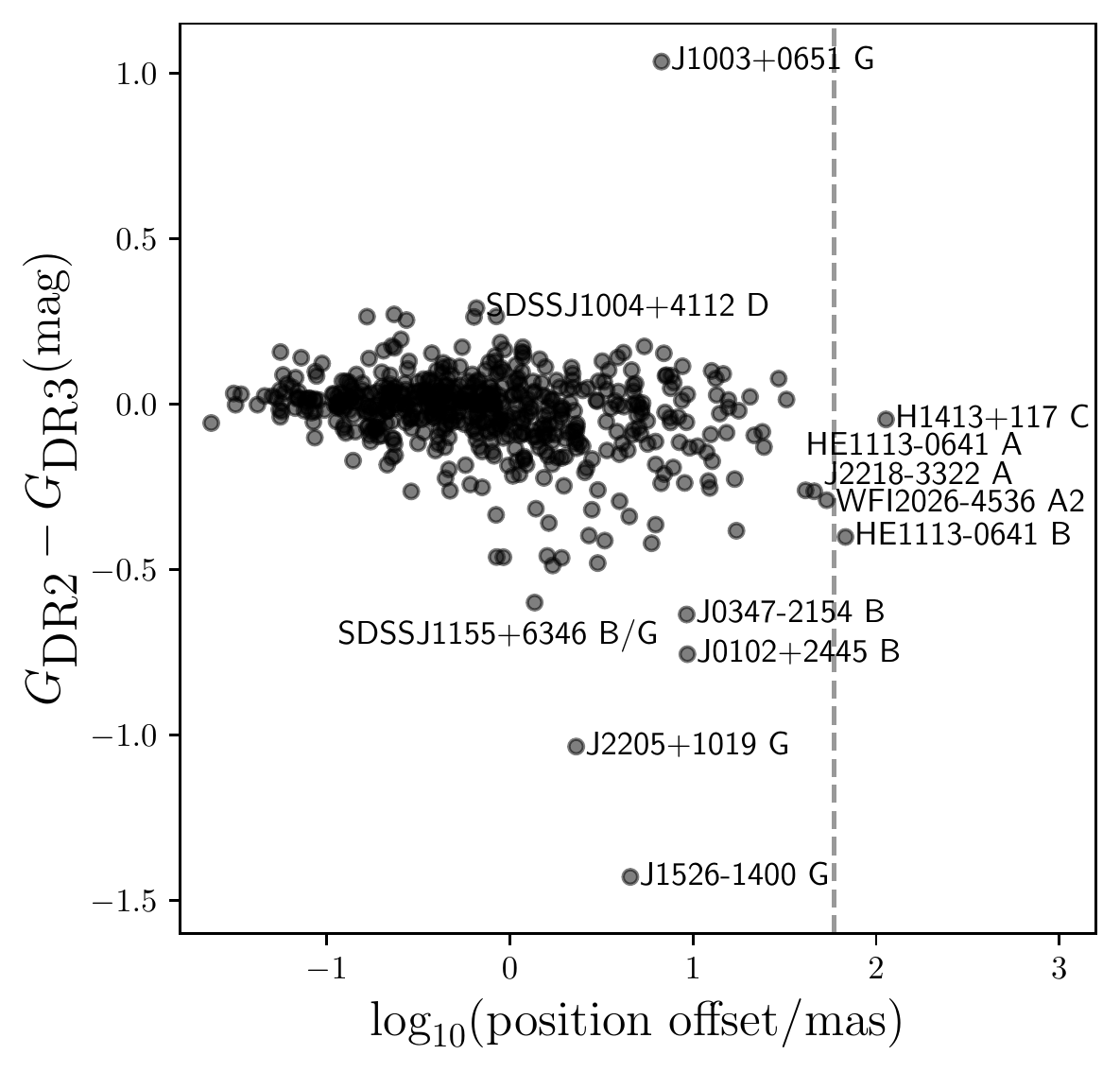}
\vspace{-0.5cm}
\caption{$G$-band magnitude difference against positional offset between DR2 and EDR3 for all \textit{Gaia} detections in lensed quasars. Several outliers have been labelled and are discussed further in the text. The pixel scale of the \textit{Gaia} detectors in the scanning direction (59 mas) is plotted as a dashed vertical line. }
    \label{fig:phot_astrom_offsets}
\end{figure}

To investigate further these positional offsets as a possible method for finding compact quads and other lenses, we match images between DR2 and EDR3. We record the $G$-band magnitude difference and positional offset for all lensed quasar images clearly belonging to the same component (i.e. galaxy or specific image). For one image of 2M1310-1714 there is no \textit{G} magnitude in EDR3, so this single image is omitted from our investigation. The distribution of these offsets in position and photometry are shown in Figure \ref{fig:phot_astrom_offsets}. The largest positional offsets are those of images in small separation quads, namely H1413+117, HE1113-0641, WFI2026-4536, and J2218-3322 (see Figure \ref{fig:overlay}). In the absence of the raw \textit{Gaia} data, these offsets suggest a promising avenue for discovering more compact quads through large astrometric offsets between \textit{Gaia} data releases. However, to properly assess the efficacy of such a method would require understanding the same metric distribution in common contaminants, which is beyond the scope of this paper.

There are also several systems with large photometric differences between data releases. These are all systems with bright, low-redshift, lensing galaxies. In the case of the lensing galaxy of J1003+0651, a large increase in flux (positive $G_{\textrm{DR2}}-G_{\textrm{DR3}}$) is seen, and can be attributed to the counterimage being removed in the most recent data release (see Figure \ref{fig:overlay}). Conversely, several galaxies have a large reduction in flux (e.g., J1526-1400 and J2205+1019), despite having the same detections in both releases. This is likely explained by a combination of deblending problems with a nearby counterimage and the known increase in variability for elongated galaxies due to average scanning direction differences \citep{riello2021}.

\subsubsection{Comparison to mock catalogues} \label{mockcomparison}
We can also compare the current sample of lenses to those expected from a mock catalogue to understand the limitations of our selection. For this we use an updated version of the mock catalogue from \citet{om10}. The updates include adopting a galaxy velocity function for all types of galaxies from \citet{bernardi2010} combined with the redshift evolution predicted by the Illustris cosmological hydrodynamical simulation \citep{torrey2015} as described in detail in \citet{oguri2018}, adopting cosmological parameters from the latest Planck satellite mission result \citep{planck2020}, and adopting an improved redshift-dependent model of the external shear. 

\begin{figure}
\centering
\includegraphics[width=\columnwidth]{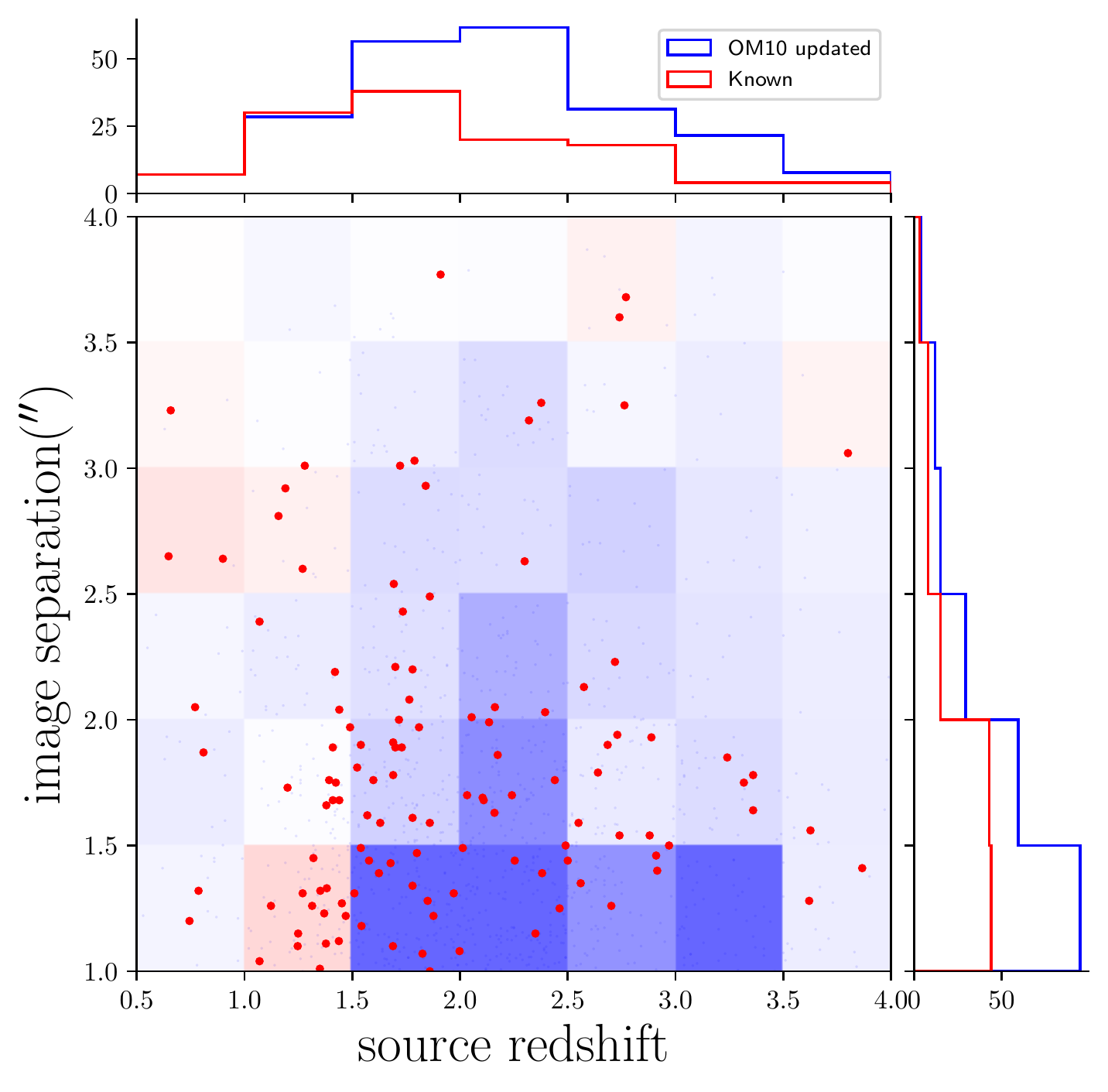}
\vspace{-0.5cm}
\caption{Distribution of image separations against source redshifts for a statistical sample of known lensed quasars (red) as described in the text. Circles are doubles, while squares are quads. Also overlaid in the histograms are the predicted numbers satisfying the same criteria from an updated version of the \citet{om10} mock catalogue; blue is an over-prediction relative to observed numbers, and red is an under-prediction.}
    \label{fig:om10_updated}
\end{figure}

We decide to compare only systems with multiple \textit{Gaia} detections, since this is where we expect the current selection to be most complete; systems with single \textit{Gaia} detections can have much fainter images that prevent spectroscopic confirmation. To compare the mock catalogue to the true sky, we limit the area of sky to the typical areas searched for lenses, namely those not too highly contaminated by stars. We define this as any sky area with local \textit{Gaia} detection densities lower than 20,000 detections per square degree (as measured by the number of detections within a 100 arcsecond radius) and within the Pan-STARRS footprint, i.e. declinations above --30 degrees. This leaves 53.9\% of the full-sky. We also include limits on image separation of 1 to 4 arcseconds, and source redshifts below 4. These criteria keep 121 known lensed quasars. To provide mock quasars in the catalogue with \textit{Gaia} magnitudes, we follow \citet{lemon2019} by using SDSS quasars with both \textit{Gaia} detections and measured \textit{i}-band magnitudes, and draw a \textit{G}-\textit{i} colour for each mock source from the known quasars within 0.025 in redshift. The number of mocks that pass, normalised to the sky area used for the real lens sample, is $\sim$214. The image separation and source redshift distributions are shown in Figure \ref{fig:om10_updated}.

The predicted numbers are in good agreement with the known lenses for larger image separations and redshifts below 1.5. This is naturally explained by prioritising lens candidates with clear signs of a possible lensing galaxy, which is favoured by lower redshift sources and larger image separations. 

The mock lens catalogue of \citet{yue2022} underpredicts the number of observed lenses by 44\%, suggesting a discrepancy within their imposed galaxy velocity dispersion function, quasar luminosity function, or shear distribution.

\subsection{Physical quasar pairs} \label{binaries}
We classify 11 systems as distinct quasar pairs at very similar redshifts. Following \citet{hennawi2010}, we take our operational definition of a binary quasar to have measured redshifts within 2000 \kms. This accounts for possible offsets of emission lines from the systemic redshift due to the dynamics of the broad line region, inflows/outflows, and absorption, while also allowing for peculiar velocities of the two quasars while still remaining in the same halo. The best-fit redshifts are found by minimising the ${\chi}^{2}$ of fitting a quasar template using \textsc{PyQSOFit} within a range of redshifts \citep{guo2018}. Uncertainties are not quoted since they depend strongly on the prior for potential velocity offsets of certain lines. A summary of key parameters is given in Table \ref{tab:binaries}. Of particular interest are the smallest-separation distinct quasar pairs, which probe the latest stage in the merger process and are the least well-characterised population. Only recently have binary quasars below sub-10kpc been discovered above redshift 1 \citep{inada2008, anguita2018, tang2021}, due to typical ground-based observing conditions and blending at such small angular separations (typically under 1 arcsecond). \citet{lemon2022} argued that the discovery of the first lensed binary quasar (with a sub-10kpc source) likely suggests a larger population of unlensed small-separation binaries. Several convincing candidate systems are already being found using \textit{Gaia} and \textit{HST} follow-up imaging \citep{shen2021, chen2022}. We therefore expect to find such systems as a byproduct of a lensed quasar search. However, systems with residual flux between the putative lensed images are preferentially observed for spectroscopy, which explains our bias towards low-redshift distinct quasar pairs relative to the lensed quasar sources (median redshifts of 0.71 and 1.94 respectively).

\begin{table}
    \centering
    \caption{Summary of confirmed binary quasars. $^{\dagger}$The brighter component of J2355-4553 is a BAL quasar, and thus is compatible with two redshifts even when masking the absorption features. We quote the more likely lower redshift (2.832) and relevant velocity difference, but a redshift of 2.8506 ($\Delta\theta$ = 60\kms) is also possible.}
    \label{tab:binaries}
    \begin{tabular}{ccccc}
\hline
Name & $z$ & $\Delta$v (\kms) & $\Delta\theta$ (\arcsec) & $d$ (kpc) \\
\hline
J0041--5350 & 0.5489, 0.5504 & 290 & 1.06 & 6.8 \\
J0315--3522 & 0.4572, 0.4573 & 20 & 1.3 & 7.6 \\
J0707+4109  & 0.5114, 0.5149 & 690 & 2.96 & 18.3 \\
J0805+3550  & 1.6604, 1.6761 & 1760 & 1.13 & 9.6 \\
J0939--0109 & 0.2240, 0.2247 & 170 & 1.9 & 6.8 \\
J1116--2122 & 0.7094, 0.7106 & 210 & 3.29 & 23.6 \\
J2121+1713  & 0.6298, 0.6307 & 170 & 2.89 & 19.8 \\
J2132+3635  & 1.2135, 1.2148 & 180 & 1.33 & 11 \\
J2256+2223  & 0.7506, 0.7534 & 480 & 1.63 & 12 \\
J2322+0916  & 1.2064, 1.2066 & 30 & 4.51 & 37.4 \\
J2355--4553$^{\dagger}$ & 2.8320, 2.8498 & 1390 & 2.20 & 17.2 \\

\hline
\end{tabular}
\end{table}

\subsection{Projected quasars} \label{projected}
We report 11 new projected quasar pairs, with angular separations ranging from 0.98 - 4.64\arcsec, and physical impact parameters from 6.2 - 34.3 kpc. We summarise the key parameters of these pairs in Table \ref{tab:projected_quasars}. Such systems are valuable for understanding quasar feedback by probing the host galaxy of the foreground quasar through absorption of the background quasar continuum \citep[see the Quasars probing quasars paper series, e.g.][]{hennawi2006}. Studies are typically limited to probing regions above 25 kpc from the foreground quasars \citep[e.g.][]{johnson2015}, since samples are built only from well-separated quasars (more likely to both have been targeted for spectroscopy). Of particular interest are the systems with small transverse proper distances (i.e. under 20 kpc) since they are both intrinsically rare (P($d$)$\ \sim d^2$), and are harder to discover due to observational constraints (e.g. systems become blended in ground-based imaging and are therefore not selected for spectroscopic follow-up as quasars). There are discrepancies between simulations and observations regarding Ly$\alpha$ absorption statistics below 100kpc in quasar host galaxies \citep{prochaska2013, sorini2020}, suggesting observational constraints on the absorption properties at small projected separations will be key to constraining more advanced feedback models.

The QSO pair Spectral Database of \citet{findlay2018} lists only 4 projected quasar pairs with angular separations under 3 arcseconds and velocity differences above 2000 \kms (after removing known lensed quasars and spurious cross-matches), compared to $\sim$3000 such pairs with projections under 30 arcseconds. Sub-10-kpc-separation projected pairs from the literature have come from lens searches: DESJ0215--4728 ($b$=6.3 kpc, $z_{\textrm{fg}}$=0.467, 1.07\arcsec) from \citet{anguita2018}, J0240--0208 ($b$=7.6 kpc, $z_{\textrm{fg}}$=1.058, 0.94\arcsec) from \citet{shalyapin2018} and \citet{anguita2018}, and J2251+0016 ($b$=10.0 kpc, $z_{\textrm{fg}}$=0.41, 1.86\arcsec) from \citet{tang2021}. The projected quasars presented in this work almost triple the number of systems probing the foreground galaxies within 10 kpc (from 3 to 8 systems).


The number of such systems expected at each separation can be predicted through well-constrained quasar luminosity functions, along with the predicted luminosity and redshift distributions. We investigate this using the quasar luminosity function from \citet{ross2013}. We recover both the redshift distribution and absolute numbers of the spectroscopically confirmed quasar sample in a 500 square degree patch of SDSS. We estimate a sky density of quasars ($0<z<4$) with $i$-band magnitude above 20.7 of $\approx$30.0 per square degree. In an area of 20,000 square degrees, this predicts 110 pairs within 5 arcseconds, and 17 pairs within 2 arcseconds. Several other studies report sub-2\arcsec\ projected pairs \citep{inada2008, more2016, schechter2017}. Including the previously mentioned systems, the total is $\sim$10, below the predicted number which was a conservative value given some components of the known systems have $i >$ 20.7. Designing searches to find these missing projected pairs should be considered in future lensed quasar and quasar pair searches.

Finally, we note that some of these systems could still be acting as strong lenses when the foreground galaxy is massive enough, and any counterimage is blended with the foreground host and is undetected in both the available low signal-to-noise imaging and spectra. High-resolution follow-up imaging would provide strong constraints on any such strong lensing, and, in the case of no detection, can still be used to obtain constraints on the mass profile \citep{smith2018}.


\begin{table}
    \centering
    \caption{Summary of confirmed projected quasar pairs.}
    \label{tab:projected_quasars}
    \begin{tabular}{ccccc}
\hline
Name & $z_{fg}$ & $z_{bg}$ & $\Delta\theta$ (\arcsec) & $d$ (kpc) \\
\hline
J0021+1927 & 1.045 & 1.09 & 2.89 & 23.4 \\
J0027+0438 & 0.194 & 1.972 & 1.92 & 6.2\\ 
J0032--4523 & 1.667 & 1.74 & 2.34 & 19.8 \\
J0249+2606 & 1.514 & 2.68 & 3.21 & 27.2 \\
J0436+7851 & 0.765 & 1.975 & 4.64 & 34.3 \\
J0553+0910 & 1.08 & 1.85 & 1.11 & 9.0 \\
J0601--2220 & 0.809 & 0.836 & 0.98 & 7.4 \\
J1820+3747 & 0.614 & 0.754 & 3.10 & 20.9 \\
J2302--4154 & 0.98 & 1.22 & 1.22 & 9.7 \\
J2314+0323 & 0.315 & 0.606 & 1.87 & 8.6\\
J2318+0250 & 1.83 & 1.945 & 3.23 & 27.2\\
\hline
\end{tabular}
\end{table}

\subsection{Contaminants}
It is prudent to reflect on the main contaminants from our search, principally to aid future discovery methods. Given the utility and rarity of small-separation projected and physical quasar pairs, we include them in our true positive sample. Of the 175 observed systems, 25 are confirmed as contaminant systems: 17 quasar+star/non-quasar, 5 star-forming galaxies, and 3 star pairs. Of the stellar pairs, two have red WISE colours similar to those of quasars, which are rare but a well-known contaminant in infrared-colour selection of quasars, and one (J1752+0826) was selected as a \textit{Gaia} detection near a galaxy and whose Pan-STARRS image shows a very convincing possible lensed quasar. Of the 5 star-forming galaxies, one was a potential quad candidate selected through the single offset method, and the rest were \textit{Gaia} multiplets around red WISE quasar candidates. Finally, of the 17 quasar+star pairs, 15 were selected as \textit{Gaia} multiplets around WISE quasars, and 2 as \textit{Gaia} singlets near galaxies. We expect that future improvements in the \textit{Gaia} catalogue -- through more precise catalogue measurements -- to continue to help remove such contaminants.

\section{Conclusion} \label{conclusion}
We have presented spectroscopic follow-up of 175 lensed quasar candidates, confirming 103 new lensed quasars, 17 of which simply lack photometric detection of the lensing galaxy, however, whose spectra are either very similar or whose imaging shows a faint counterimage. We also report 25 further unclassified quasar pairs which likely contain a mixture of lenses and distinct quasar pairs at the same redshift, however, further data is required in each case. We present 11 projected quasar pairs, and 11 binary pairs, which substantially increases the population of known systems at angular separations of a few arcseconds and below.

We used a mock catalogue of lensed quasars with an updated quasar luminosity function, galaxy velocity dispersion function, and treatment of shear to compare the current population of known lensed quasars with multiple \textit{Gaia} detections. We show that the majority of lensed quasars with sources below redshift 1.5 are known, however, almost half of lenses with sources above \textit{z}$=$1.5 are yet to be discovered, which mostly have image separations below 1.5\arcsec. The likely and possible lensed quasars discovered in this work represent already a subset of these missing lenses. 

With the goal of creating a magnitude-limited sample of lensed quasars, future \textit{Gaia} data releases should help through increasing completeness in detections at small separations, and reliable classification of quasars through increased precision of proper motions and parallaxes, as well as better variability constraints due to a longer baseline. To test for the completeness and biases of each lens search, other searches making use of spectroscopic and cadenced photometric datasets will be crucial. Such searches in LSST, making use of detecting nearby or extended variability, should not only be an efficient method for selecting lensed quasars but also for selecting small-separation physical quasar pairs and projected quasars \citep{kochanek2006, chao2020}. As targets become fainter and more numerous -- with up to 400 lensed quasars with measured time delays in LSST \citep{liao2015} -- a complete survey for spectroscopic redshifts must be carefully designed to convert these delays into a powerful probe of measuring the Hubble constant, with special consideration to minimising follow-up of common contaminants.

\section*{Acknowledgements}
We would like to thank Amanda S. Bosh and Stephen E. Levine for obtaining MagIC observations of lensed quasars.

This work is supported by the Swiss National Science Foundation (SNSF) and by the European Research Council (ERC) under the European Union's Horizon 2020 research and innovation program (COSMICLENS: grant agreement No 787886).
This work was supported by JSPS KAKENHI Grant Numbers JP22H01260, JP18K03693, JP20H05856, JP20H00181. T.A. acknowledges support from ANID-FONDECYT Regular
1190335, the Millennium Science Initiative ICN12\_009 and
the ANID BASAL project FB210003. TT acknowledges support by the National Science Foundation through grant NSF-AST-1906976. Support for this work was provided by NASA through the NASA Hubble Fellowship grant
HST-HF2-51492 awarded to AJS by the Space Telescope Science Institute, which is operated by the Association of Universities for Research in Astronomy, Inc., for NASA, under contract NAS5-26555.

This paper includes data gathered with the 6.5 meter Magellan Telescopes located at Las Campanas Observatory, Chile. This research used the facilities of the Canadian Astronomy Data Centre operated by the National Research Council of Canada with the support of the Canadian Space Agency.

This research has made use of the VizieR catalogue access tool, CDS, Strasbourg, France (DOI: 10.26093/cds/vizier). The original description of the VizieR service was published in A\&AS 143, 23.

This research has made use of the NASA/IPAC Extragalactic Database (NED), which is funded by the National Aeronautics and Space Administration and operated by the California Institute of Technology.

The Pan-STARRS1 Surveys (PS1) and the PS1 public science archive have been made possible through contributions by the Institute for Astronomy, the University of Hawaii, the Pan-STARRS Project Office, the Max-Planck Society and its participating institutes, the Max Planck Institute for Astronomy, Heidelberg and the Max Planck Institute for Extraterrestrial Physics, Garching, The Johns Hopkins University, Durham University, the University of Edinburgh, the Queen's University Belfast, the Harvard-Smithsonian Center for Astrophysics, the Las Cumbres Observatory Global Telescope Network Incorporated, the National Central University of Taiwan, the Space Telescope Science Institute, the National Aeronautics and Space Administration under Grant No. NNX08AR22G issued through the Planetary Science Division of the NASA Science Mission Directorate, the National Science Foundation Grant No. AST-1238877, the University of Maryland, Eotvos Lorand University (ELTE), the Los Alamos National Laboratory, and the Gordon and Betty Moore Foundation.

The Legacy Surveys consist of three individual and complementary projects: the Dark Energy Camera Legacy Survey (DECaLS; Proposal ID \#2014B-0404; PIs: David Schlegel and Arjun Dey), the Beijing-Arizona Sky Survey (BASS; NOAO Prop. ID \#2015A-0801; PIs: Zhou Xu and Xiaohui Fan), and the Mayall z-band Legacy Survey (MzLS; Prop. ID \#2016A-0453; PI: Arjun Dey). DECaLS, BASS and MzLS together include data obtained, respectively, at the Blanco telescope, Cerro Tololo Inter-American Observatory, NSF’s NOIRLab; the Bok telescope, Steward Observatory, University of Arizona; and the Mayall telescope, Kitt Peak National Observatory, NOIRLab. The Legacy Surveys project is honored to be permitted to conduct astronomical research on Iolkam Du’ag (Kitt Peak), a mountain with particular significance to the Tohono O’odham Nation.

NOIRLab is operated by the Association of Universities for Research in Astronomy (AURA) under a cooperative agreement with the National Science Foundation.

This project used data obtained with the Dark Energy Camera (DECam), which was constructed by the Dark Energy Survey (DES) collaboration. Funding for the DES Projects has been provided by the U.S. Department of Energy, the U.S. National Science Foundation, the Ministry of Science and Education of Spain, the Science and Technology Facilities Council of the United Kingdom, the Higher Education Funding Council for England, the National Center for Supercomputing Applications at the University of Illinois at Urbana-Champaign, the Kavli Institute of Cosmological Physics at the University of Chicago, Center for Cosmology and Astro-Particle Physics at the Ohio State University, the Mitchell Institute for Fundamental Physics and Astronomy at Texas A\&M University, Financiadora de Estudos e Projetos, Fundacao Carlos Chagas Filho de Amparo, Financiadora de Estudos e Projetos, Fundacao Carlos Chagas Filho de Amparo a Pesquisa do Estado do Rio de Janeiro, Conselho Nacional de Desenvolvimento Cientifico e Tecnologico and the Ministerio da Ciencia, Tecnologia e Inovacao, the Deutsche Forschungsgemeinschaft and the Collaborating Institutions in the Dark Energy Survey. The Collaborating Institutions are Argonne National Laboratory, the University of California at Santa Cruz, the University of Cambridge, Centro de Investigaciones Energeticas, Medioambientales y Tecnologicas-Madrid, the University of Chicago, University College London, the DES-Brazil Consortium, the University of Edinburgh, the Eidgenossische Technische Hochschule (ETH) Zurich, Fermi National Accelerator Laboratory, the University of Illinois at Urbana-Champaign, the Institut de Ciencies de l’Espai (IEEC/CSIC), the Institut de Fisica d’Altes Energies, Lawrence Berkeley National Laboratory, the Ludwig Maximilians Universitat Munchen and the associated Excellence Cluster Universe, the University of Michigan, NSF’s NOIRLab, the University of Nottingham, the Ohio State University, the University of Pennsylvania, the University of Portsmouth, SLAC National Accelerator Laboratory, Stanford University, the University of Sussex, and Texas A\&M University.

BASS is a key project of the Telescope Access Program (TAP), which has been funded by the National Astronomical Observatories of China, the Chinese Academy of Sciences (the Strategic Priority Research Program “The Emergence of Cosmological Structures” Grant \# XDB09000000), and the Special Fund for Astronomy from the Ministry of Finance. The BASS is also supported by the External Cooperation Program of Chinese Academy of Sciences (Grant \# 114A11KYSB20160057), and Chinese National Natural Science Foundation (Grant \# 11433005).

The Legacy Survey team makes use of data products from the Near-Earth Object Wide-field Infrared Survey Explorer (NEOWISE), which is a project of the Jet Propulsion Laboratory/California Institute of Technology. NEOWISE is funded by the National Aeronautics and Space Administration.

The Legacy Surveys imaging of the DESI footprint is supported by the Director, Office of Science, Office of High Energy Physics of the U.S. Department of Energy under Contract No. DE-AC02-05CH1123, by the National Energy Research Scientific Computing Center, a DOE Office of Science User Facility under the same contract; and by the U.S. National Science Foundation, Division of Astronomical Sciences under Contract No. AST-0950945 to NOAO.




\bibliographystyle{mnras}
\bibliography{example} 

\begin{thebibliography}{}
\makeatletter
\relax
\def\mn@urlcharsother{\let\do\@makeother \do\$\do\&\do\#\do\^\do\_\do\%\do\~}
\def\mn@doi{\begingroup\mn@urlcharsother \@ifnextchar [ {\mn@doi@}
  {\mn@doi@[]}}
\def\mn@doi@[#1]#2{\def\@tempa{#1}\ifx\@tempa\@empty \href
  {http://dx.doi.org/#2} {doi:#2}\else \href {http://dx.doi.org/#2} {#1}\fi
  \endgroup}
\def\mn@eprint#1#2{\mn@eprint@#1:#2::\@nil}
\def\mn@eprint@arXiv#1{\href {http://arxiv.org/abs/#1} {{\tt arXiv:#1}}}
\def\mn@eprint@dblp#1{\href {http://dblp.uni-trier.de/rec/bibtex/#1.xml}
  {dblp:#1}}
\def\mn@eprint@#1:#2:#3:#4\@nil{\def\@tempa {#1}\def\@tempb {#2}\def\@tempc
  {#3}\ifx \@tempc \@empty \let \@tempc \@tempb \let \@tempb \@tempa \fi \ifx
  \@tempb \@empty \def\@tempb {arXiv}\fi \@ifundefined
  {mn@eprint@\@tempb}{\@tempb:\@tempc}{\expandafter \expandafter \csname
  mn@eprint@\@tempb\endcsname \expandafter{\@tempc}}}

\bibitem[\protect\citeauthoryear{{Anguita} et~al.,}{{Anguita}
  et~al.}{2018}]{anguita2018}
{Anguita} T.,  et~al., 2018, \mn@doi [\mnras] {10.1093/mnras/sty2172}, \href
  {https://ui.adsabs.harvard.edu/abs/2018MNRAS.480.5017A} {480, 5017}

\bibitem[\protect\citeauthoryear{{Bernardi}, {Shankar}, {Hyde}, {Mei},
  {Marulli}  \& {Sheth}}{{Bernardi} et~al.}{2010}]{bernardi2010}
{Bernardi} M.,  {Shankar} F.,  {Hyde} J.~B.,  {Mei} S.,  {Marulli} F.,
  {Sheth} R.~K.,  2010, \mn@doi [\mnras] {10.1111/j.1365-2966.2010.16425.x},
  \href {https://ui.adsabs.harvard.edu/abs/2010MNRAS.404.2087B} {404, 2087}

\bibitem[\protect\citeauthoryear{{Blanton} et~al.,}{{Blanton}
  et~al.}{2017}]{blanton2017}
{Blanton} M.~R.,  et~al., 2017, \mn@doi [\aj] {10.3847/1538-3881/aa7567}, \href
  {https://ui.adsabs.harvard.edu/abs/2017AJ....154...28B} {154, 28}

\bibitem[\protect\citeauthoryear{{Chae}}{{Chae}}{2003}]{chae2003}
{Chae} K.-H.,  2003, \mn@doi [\mnras] {10.1111/j.1365-2966.2003.07092.x}, \href
  {https://ui.adsabs.harvard.edu/abs/2003MNRAS.346..746C} {346, 746}

\bibitem[\protect\citeauthoryear{{Chambers} et~al.,}{{Chambers}
  et~al.}{2016}]{chambers2016}
{Chambers} K.~C.,  et~al., 2016, arXiv e-prints, \href
  {https://ui.adsabs.harvard.edu/abs/2016arXiv161205560C} {p. arXiv:1612.05560}

\bibitem[\protect\citeauthoryear{{Chan} et~al.,}{{Chan}
  et~al.}{2022}]{chan2022}
{Chan} J.~H.~H.,  et~al., 2022, \mn@doi [\aap] {10.1051/0004-6361/202142389},
  \href {https://ui.adsabs.harvard.edu/abs/2022A&A...659A.140C} {659, A140}

\bibitem[\protect\citeauthoryear{{Chao}, {Chan}, {Suyu}, {Yasuda}, {More},
  {Oguri}, {Morokuma}  \& {Jaelani}}{{Chao} et~al.}{2020}]{chao2020}
{Chao} D. C.~Y.,  {Chan} J. H.~H.,  {Suyu} S.~H.,  {Yasuda} N.,  {More} A.,
  {Oguri} M.,  {Morokuma} T.,   {Jaelani} A.~T.,  2020, \mn@doi [\aap]
  {10.1051/0004-6361/201936806}, \href
  {https://ui.adsabs.harvard.edu/abs/2020A&A...640A..88C} {640, A88}

\bibitem[\protect\citeauthoryear{{Chen}, {Hwang}, {Shen}, {Liu}, {Zakamska},
  {Yang}  \& {Li}}{{Chen} et~al.}{2022}]{chen2022}
{Chen} Y.-C.,  {Hwang} H.-C.,  {Shen} Y.,  {Liu} X.,  {Zakamska} N.~L.,  {Yang}
  Q.,   {Li} J.~I.,  2022, \mn@doi [\apj] {10.3847/1538-4357/ac401b}, \href
  {https://ui.adsabs.harvard.edu/abs/2022ApJ...925..162C} {925, 162}

\bibitem[\protect\citeauthoryear{{Citro}, {Erb}, {Pettini}, {Auger}, {Becker}
  \& {James}}{{Citro} et~al.}{2021}]{citro2021}
{Citro} A.,  {Erb} D.~K.,  {Pettini} M.,  {Auger} M.~W.,  {Becker} G.~D.,
  {James} B.~L.,  2021, \mn@doi [\apj] {10.3847/1538-4357/ac24a2}, \href
  {https://ui.adsabs.harvard.edu/abs/2021ApJ...922..187C} {922, 187}

\bibitem[\protect\citeauthoryear{{Collett} \& {Auger}}{{Collett} \&
  {Auger}}{2014}]{collett2014}
{Collett} T.~E.,  {Auger} M.~W.,  2014, \mn@doi [\mnras]
  {10.1093/mnras/stu1190}, \href
  {https://ui.adsabs.harvard.edu/abs/2014MNRAS.443..969C} {443, 969}

\bibitem[\protect\citeauthoryear{{Croom}, {Smith}, {Boyle}, {Shanks}, {Miller},
  {Outram}  \& {Loaring}}{{Croom} et~al.}{2004}]{croom2004}
{Croom} S.~M.,  {Smith} R.~J.,  {Boyle} B.~J.,  {Shanks} T.,  {Miller} L.,
  {Outram} P.~J.,   {Loaring} N.~S.,  2004, \mn@doi [\mnras]
  {10.1111/j.1365-2966.2004.07619.x}, \href
  {https://ui.adsabs.harvard.edu/abs/2004MNRAS.349.1397C} {349, 1397}

\bibitem[\protect\citeauthoryear{{Delchambre} et~al.,}{{Delchambre}
  et~al.}{2019}]{delchambre2019}
{Delchambre} L.,  et~al., 2019, \mn@doi [\aap] {10.1051/0004-6361/201833802},
  \href {https://ui.adsabs.harvard.edu/abs/2019A&A...622A.165D} {622, A165}

\bibitem[\protect\citeauthoryear{{Desira}, {Shu}, {Auger}, {McMahon}, {Lemon},
  {Anguita}  \& {Neira}}{{Desira} et~al.}{2022}]{desira2022}
{Desira} C.,  {Shu} Y.,  {Auger} M.~W.,  {McMahon} R.~G.,  {Lemon} C.~A.,
  {Anguita} T.,   {Neira} F.,  2022, \mn@doi [\mnras] {10.1093/mnras/stab2960},
  \href {https://ui.adsabs.harvard.edu/abs/2022MNRAS.509..738D} {509, 738}

\bibitem[\protect\citeauthoryear{{Dey} et~al.,}{{Dey} et~al.}{2019}]{dey2019}
{Dey} A.,  et~al., 2019, \mn@doi [\aj] {10.3847/1538-3881/ab089d}, \href
  {https://ui.adsabs.harvard.edu/abs/2019AJ....157..168D} {157, 168}

\bibitem[\protect\citeauthoryear{{Fan} et~al.,}{{Fan} et~al.}{2019}]{fan2019}
{Fan} X.,  et~al., 2019, \mn@doi [\apjl] {10.3847/2041-8213/aaeffe}, \href
  {https://ui.adsabs.harvard.edu/abs/2019ApJ...870L..11F} {870, L11}

\bibitem[\protect\citeauthoryear{{Findlay} et~al.,}{{Findlay}
  et~al.}{2018}]{findlay2018}
{Findlay} J.~R.,  et~al., 2018, \mn@doi [\apjs] {10.3847/1538-4365/aabee5},
  \href {https://ui.adsabs.harvard.edu/abs/2018ApJS..236...44F} {236, 44}

\bibitem[\protect\citeauthoryear{{Flesch}}{{Flesch}}{2015}]{milliquas}
{Flesch} E.~W.,  2015, \mn@doi [\pasa] {10.1017/pasa.2015.10}, \href
  {https://ui.adsabs.harvard.edu/abs/2015PASA...32...10F} {32, e010}

\bibitem[\protect\citeauthoryear{{Foreman-Mackey}, {Hogg}, {Lang}  \&
  {Goodman}}{{Foreman-Mackey} et~al.}{2013}]{foreman-mackey2013}
{Foreman-Mackey} D.,  {Hogg} D.~W.,  {Lang} D.,   {Goodman} J.,  2013, \mn@doi
  [\pasp] {10.1086/670067}, \href
  {https://ui.adsabs.harvard.edu/abs/2013PASP..125..306F} {125, 306}

\bibitem[\protect\citeauthoryear{{Gaia Collaboration} et~al.,}{{Gaia
  Collaboration} et~al.}{2018}]{gaiadr2}
{Gaia Collaboration} et~al., 2018, \mn@doi [\aap]
  {10.1051/0004-6361/201833051}, \href
  {https://ui.adsabs.harvard.edu/abs/2018A&A...616A...1G} {616, A1}

\bibitem[\protect\citeauthoryear{{Gordon} et~al.,}{{Gordon}
  et~al.}{2021}]{gordon2021}
{Gordon} Y.~A.,  et~al., 2021, VizieR Online Data Catalog, \href
  {https://ui.adsabs.harvard.edu/abs/2021yCat..22550030G} {p. J/ApJS/255/30}

\bibitem[\protect\citeauthoryear{{Guo}, {Shen}  \& {Wang}}{{Guo}
  et~al.}{2018}]{guo2018}
{Guo} H.,  {Shen} Y.,   {Wang} S.,  2018, {PyQSOFit: Python code to fit the
  spectrum of quasars}, Astrophysics Source Code Library (\mn@eprint {ascl}
  {1809.008})

\bibitem[\protect\citeauthoryear{{Harvey}}{{Harvey}}{2020}]{harvey2020}
{Harvey} D.,  2020, \mn@doi [\mnras] {10.1093/mnras/staa2522}, \href
  {https://ui.adsabs.harvard.edu/abs/2020MNRAS.498.2871H} {498, 2871}

\bibitem[\protect\citeauthoryear{{Hennawi} et~al.,}{{Hennawi}
  et~al.}{2006}]{hennawi2006}
{Hennawi} J.~F.,  et~al., 2006, \mn@doi [\apj] {10.1086/507069}, \href
  {https://ui.adsabs.harvard.edu/abs/2006ApJ...651...61H} {651, 61}

\bibitem[\protect\citeauthoryear{{Hennawi} et~al.,}{{Hennawi}
  et~al.}{2010}]{hennawi2010}
{Hennawi} J.~F.,  et~al., 2010, \mn@doi [\apj] {10.1088/0004-637X/719/2/1672},
  \href {https://ui.adsabs.harvard.edu/abs/2010ApJ...719.1672H} {719, 1672}

\bibitem[\protect\citeauthoryear{{Holden}, {Nichol}, {Romer}, {Metevier},
  {Postman}, {Ulmer}  \& {Lubin}}{{Holden} et~al.}{1999}]{holden1999}
{Holden} B.~P.,  {Nichol} R.~C.,  {Romer} A.~K.,  {Metevier} A.,  {Postman} M.,
   {Ulmer} M.~P.,   {Lubin} L.~M.,  1999, \mn@doi [\aj] {10.1086/301069}, \href
  {https://ui.adsabs.harvard.edu/abs/1999AJ....118.2002H} {118, 2002}

\bibitem[\protect\citeauthoryear{{Huchra}, {Gorenstein}, {Kent}, {Shapiro},
  {Smith}, {Horine}  \& {Perley}}{{Huchra} et~al.}{1985}]{huchra1985}
{Huchra} J.,  {Gorenstein} M.,  {Kent} S.,  {Shapiro} I.,  {Smith} G.,
  {Horine} E.,   {Perley} R.,  1985, \mn@doi [\aj] {10.1086/113777}, \href
  {https://ui.adsabs.harvard.edu/abs/1985AJ.....90..691H} {90, 691}

\bibitem[\protect\citeauthoryear{{Hurley-Walker} et~al.,}{{Hurley-Walker}
  et~al.}{2017}]{hurley2017}
{Hurley-Walker} N.,  et~al., 2017, \mn@doi [\mnras] {10.1093/mnras/stw2337},
  \href {https://ui.adsabs.harvard.edu/abs/2017MNRAS.464.1146H} {464, 1146}

\bibitem[\protect\citeauthoryear{{Hutsem{\'e}kers} \&
  {Sluse}}{{Hutsem{\'e}kers} \& {Sluse}}{2021}]{hutsemekers2021}
{Hutsem{\'e}kers} D.,  {Sluse} D.,  2021, \mn@doi [\aap]
  {10.1051/0004-6361/202141820}, \href
  {https://ui.adsabs.harvard.edu/abs/2021A&A...654A.155H} {654, A155}

\bibitem[\protect\citeauthoryear{{Hutsem{\'e}kers}, {Borguet}, {Sluse}, {Riaud}
   \& {Anguita}}{{Hutsem{\'e}kers} et~al.}{2010}]{hutsemekers2010}
{Hutsem{\'e}kers} D.,  {Borguet} B.,  {Sluse} D.,  {Riaud} P.,   {Anguita} T.,
  2010, \mn@doi [\aap] {10.1051/0004-6361/200913247}, \href
  {https://ui.adsabs.harvard.edu/abs/2010A&A...519A.103H} {519, A103}

\bibitem[\protect\citeauthoryear{{Inada} et~al.,}{{Inada}
  et~al.}{2008}]{inada2008}
{Inada} N.,  et~al., 2008, \mn@doi [\aj] {10.1088/0004-6256/135/2/496}, \href
  {https://ui.adsabs.harvard.edu/abs/2008AJ....135..496I} {135, 496}

\bibitem[\protect\citeauthoryear{{Jaelani} et~al.,}{{Jaelani}
  et~al.}{2021}]{jaelani2021}
{Jaelani} A.~T.,  et~al., 2021, \mn@doi [\mnras] {10.1093/mnras/stab145}, \href
  {https://ui.adsabs.harvard.edu/abs/2021MNRAS.502.1487J} {502, 1487}

\bibitem[\protect\citeauthoryear{{Jim{\'e}nez-Vicente} \&
  {Mediavilla}}{{Jim{\'e}nez-Vicente} \& {Mediavilla}}{2019}]{jimenez2019}
{Jim{\'e}nez-Vicente} J.,  {Mediavilla} E.,  2019, \mn@doi [\apj]
  {10.3847/1538-4357/ab46b8}, \href
  {https://ui.adsabs.harvard.edu/abs/2019ApJ...885...75J} {885, 75}

\bibitem[\protect\citeauthoryear{{Johnson}, {Chen}  \& {Mulchaey}}{{Johnson}
  et~al.}{2015}]{johnson2015}
{Johnson} S.~D.,  {Chen} H.-W.,   {Mulchaey} J.~S.,  2015, \mn@doi [\mnras]
  {10.1093/mnras/stv1481}, \href
  {https://ui.adsabs.harvard.edu/abs/2015MNRAS.452.2553J} {452, 2553}

\bibitem[\protect\citeauthoryear{{Kassiola}, {Kovner}  \&
  {Blandford}}{{Kassiola} et~al.}{1992}]{kassiola1992}
{Kassiola} A.,  {Kovner} I.,   {Blandford} R.~D.,  1992, \mn@doi [\apj]
  {10.1086/171693}, \href
  {https://ui.adsabs.harvard.edu/abs/1992ApJ...396...10K} {396, 10}

\bibitem[\protect\citeauthoryear{{Kayo} et~al.,}{{Kayo}
  et~al.}{2007}]{kayo2007}
{Kayo} I.,  et~al., 2007, \mn@doi [\aj] {10.1086/521652}, \href
  {https://ui.adsabs.harvard.edu/abs/2007AJ....134.1515K} {134, 1515}

\bibitem[\protect\citeauthoryear{{Kochanek}, {Mochejska}, {Morgan}  \&
  {Stanek}}{{Kochanek} et~al.}{2006}]{kochanek2006}
{Kochanek} C.~S.,  {Mochejska} B.,  {Morgan} N.~D.,   {Stanek} K.~Z.,  2006,
  \mn@doi [\apjl] {10.1086/500559}, \href
  {https://ui.adsabs.harvard.edu/abs/2006ApJ...637L..73K} {637, L73}

\bibitem[\protect\citeauthoryear{{Korista}, {Voit}, {Morris}  \&
  {Weymann}}{{Korista} et~al.}{1993}]{korista1993}
{Korista} K.~T.,  {Voit} G.~M.,  {Morris} S.~L.,   {Weymann} R.~J.,  1993,
  \mn@doi [\apjs] {10.1086/191825}, \href
  {https://ui.adsabs.harvard.edu/abs/1993ApJS...88..357K} {88, 357}

\bibitem[\protect\citeauthoryear{{Krone-Martins} et~al.,}{{Krone-Martins}
  et~al.}{2019}]{kronemartins2019}
{Krone-Martins} A.,  et~al., 2019, arXiv e-prints, \href
  {https://ui.adsabs.harvard.edu/abs/2019arXiv191208977K} {p. arXiv:1912.08977}

\bibitem[\protect\citeauthoryear{{Lang}, {Hogg}  \& {Schlegel}}{{Lang}
  et~al.}{2016}]{lang2016}
{Lang} D.,  {Hogg} D.~W.,   {Schlegel} D.~J.,  2016, \mn@doi [\aj]
  {10.3847/0004-6256/151/2/36}, \href
  {https://ui.adsabs.harvard.edu/abs/2016AJ....151...36L} {151, 36}

\bibitem[\protect\citeauthoryear{{Lemon}, {Auger}, {McMahon}  \&
  {Koposov}}{{Lemon} et~al.}{2017}]{lemon2017}
{Lemon} C.~A.,  {Auger} M.~W.,  {McMahon} R.~G.,   {Koposov} S.~E.,  2017,
  \mn@doi [\mnras] {10.1093/mnras/stx2094}, \href
  {https://ui.adsabs.harvard.edu/abs/2017MNRAS.472.5023L} {472, 5023}

\bibitem[\protect\citeauthoryear{{Lemon}, {Auger}, {McMahon}  \&
  {Ostrovski}}{{Lemon} et~al.}{2018}]{lemon2018}
{Lemon} C.~A.,  {Auger} M.~W.,  {McMahon} R.~G.,   {Ostrovski} F.,  2018,
  \mn@doi [\mnras] {10.1093/mnras/sty911}, \href
  {https://ui.adsabs.harvard.edu/abs/2018MNRAS.479.5060L} {479, 5060}

\bibitem[\protect\citeauthoryear{{Lemon}, {Auger}  \& {McMahon}}{{Lemon}
  et~al.}{2019}]{lemon2019}
{Lemon} C.~A.,  {Auger} M.~W.,   {McMahon} R.~G.,  2019, \mn@doi [\mnras]
  {10.1093/mnras/sty3366}, \href
  {https://ui.adsabs.harvard.edu/abs/2019MNRAS.483.4242L} {483, 4242}

\bibitem[\protect\citeauthoryear{{Lemon}, {Millon}, {Sluse}, {Courbin},
  {Auger}, {Chan}, {Paic}  \& {Agnello}}{{Lemon} et~al.}{2022}]{lemon2022}
{Lemon} C.,  {Millon} M.,  {Sluse} D.,  {Courbin} F.,  {Auger} M.,  {Chan}
  J.~H.~H.,  {Paic} E.,   {Agnello} A.,  2022, \mn@doi [\aap]
  {10.1051/0004-6361/202142138}, \href
  {https://ui.adsabs.harvard.edu/abs/2022A&A...657A.113L} {657, A113}

\bibitem[\protect\citeauthoryear{{Liao} et~al.,}{{Liao}
  et~al.}{2015}]{liao2015}
{Liao} K.,  et~al., 2015, \mn@doi [\apj] {10.1088/0004-637X/800/1/11}, \href
  {https://ui.adsabs.harvard.edu/abs/2015ApJ...800...11L} {800, 11}

\bibitem[\protect\citeauthoryear{{Lucey}, {Schechter}, {Smith}  \&
  {Anguita}}{{Lucey} et~al.}{2018}]{lucey2018}
{Lucey} J.~R.,  {Schechter} P.~L.,  {Smith} R.~J.,   {Anguita} T.,  2018,
  \mn@doi [\mnras] {10.1093/mnras/sty243}, \href
  {https://ui.adsabs.harvard.edu/abs/2018MNRAS.476..927L} {476, 927}

\bibitem[\protect\citeauthoryear{{Mainzer} et~al.,}{{Mainzer}
  et~al.}{2011}]{mainzer2011}
{Mainzer} A.,  et~al., 2011, \mn@doi [\apj] {10.1088/0004-637X/743/2/156},
  \href {https://ui.adsabs.harvard.edu/abs/2011ApJ...743..156M} {743, 156}

\bibitem[\protect\citeauthoryear{{Mauch}, {Murphy}, {Buttery}, {Curran},
  {Hunstead}, {Piestrzynski}, {Robertson}  \& {Sadler}}{{Mauch}
  et~al.}{2003}]{mauch2003}
{Mauch} T.,  {Murphy} T.,  {Buttery} H.~J.,  {Curran} J.,  {Hunstead} R.~W.,
  {Piestrzynski} B.,  {Robertson} J.~G.,   {Sadler} E.~M.,  2003, \mn@doi
  [\mnras] {10.1046/j.1365-8711.2003.06605.x}, \href
  {https://ui.adsabs.harvard.edu/abs/2003MNRAS.342.1117M} {342, 1117}

\bibitem[\protect\citeauthoryear{{McKean} et~al.,}{{McKean}
  et~al.}{2005}]{mckean2005}
{McKean} J.~P.,  et~al., 2005, \mn@doi [\mnras]
  {10.1111/j.1365-2966.2004.08516.x}, \href
  {https://ui.adsabs.harvard.edu/abs/2005MNRAS.356.1009M} {356, 1009}

\bibitem[\protect\citeauthoryear{{Mediavilla}, {Jim{\'e}nez-Vicente},
  {Mu{\~n}oz}, {Vives-Arias}  \& {Calder{\'o}n-Infante}}{{Mediavilla}
  et~al.}{2017}]{mediavilla2017}
{Mediavilla} E.,  {Jim{\'e}nez-Vicente} J.,  {Mu{\~n}oz} J.~A.,  {Vives-Arias}
  H.,   {Calder{\'o}n-Infante} J.,  2017, \mn@doi [\apjl]
  {10.3847/2041-8213/aa5dab}, \href
  {https://ui.adsabs.harvard.edu/abs/2017ApJ...836L..18M} {836, L18}

\bibitem[\protect\citeauthoryear{{Moffat}}{{Moffat}}{1969}]{moffat1969}
{Moffat} A.~F.~J.,  1969, \aap, \href
  {https://ui.adsabs.harvard.edu/abs/1969A&A.....3..455M} {3, 455}

\bibitem[\protect\citeauthoryear{{More} et~al.,}{{More}
  et~al.}{2016}]{more2016}
{More} A.,  et~al., 2016, \mn@doi [\mnras] {10.1093/mnras/stv2813}, \href
  {https://ui.adsabs.harvard.edu/abs/2016MNRAS.456.1595M} {456, 1595}

\bibitem[\protect\citeauthoryear{{Muchovej} et~al.,}{{Muchovej}
  et~al.}{2010}]{muchovej2010}
{Muchovej} S.,  et~al., 2010, \mn@doi [\apj] {10.1088/0004-637X/716/1/521},
  \href {https://ui.adsabs.harvard.edu/abs/2010ApJ...716..521M} {716, 521}

\bibitem[\protect\citeauthoryear{{Myers} et~al.,}{{Myers}
  et~al.}{2003}]{myers2003}
{Myers} S.~T.,  et~al., 2003, \mn@doi [\mnras]
  {10.1046/j.1365-8711.2003.06256.x}, \href
  {https://ui.adsabs.harvard.edu/abs/2003MNRAS.341....1M} {341, 1}

\bibitem[\protect\citeauthoryear{{Ochsenbein}, {Bauer}  \&
  {Marcout}}{{Ochsenbein} et~al.}{2000}]{vizier}
{Ochsenbein} F.,  {Bauer} P.,   {Marcout} J.,  2000, \mn@doi [\aaps]
  {10.1051/aas:2000169}, \href
  {https://ui.adsabs.harvard.edu/abs/2000A&AS..143...23O} {143, 23}

\bibitem[\protect\citeauthoryear{{Oguri}}{{Oguri}}{2018}]{oguri2018}
{Oguri} M.,  2018, \mn@doi [\mnras] {10.1093/mnras/sty2145}, \href
  {https://ui.adsabs.harvard.edu/abs/2018MNRAS.480.3842O} {480, 3842}

\bibitem[\protect\citeauthoryear{{Oguri} \& {Marshall}}{{Oguri} \&
  {Marshall}}{2010}]{om10}
{Oguri} M.,  {Marshall} P.~J.,  2010, \mn@doi [\mnras]
  {10.1111/j.1365-2966.2010.16639.x}, \href
  {https://ui.adsabs.harvard.edu/abs/2010MNRAS.405.2579O} {405, 2579}

\bibitem[\protect\citeauthoryear{{Oguri} et~al.,}{{Oguri}
  et~al.}{2006}]{oguri2006}
{Oguri} M.,  et~al., 2006, \mn@doi [\aj] {10.1086/506019}, \href
  {https://ui.adsabs.harvard.edu/abs/2006AJ....132..999O} {132, 999}

\bibitem[\protect\citeauthoryear{{Oguri} et~al.,}{{Oguri}
  et~al.}{2012}]{oguri2012}
{Oguri} M.,  et~al., 2012, \mn@doi [\aj] {10.1088/0004-6256/143/5/120}, \href
  {https://ui.adsabs.harvard.edu/abs/2012AJ....143..120O} {143, 120}

\bibitem[\protect\citeauthoryear{{Okoshi} et~al.,}{{Okoshi}
  et~al.}{2021}]{okoshi2021}
{Okoshi} K.,  et~al., 2021, \mn@doi [\aj] {10.3847/1538-3881/ac0bbb}, \href
  {https://ui.adsabs.harvard.edu/abs/2021AJ....162..175O} {162, 175}

\bibitem[\protect\citeauthoryear{{Orban de Xivry} \& {Marshall}}{{Orban de
  Xivry} \& {Marshall}}{2009}]{orban2009}
{Orban de Xivry} G.,  {Marshall} P.,  2009, \mn@doi [\mnras]
  {10.1111/j.1365-2966.2009.14925.x}, \href
  {https://ui.adsabs.harvard.edu/abs/2009MNRAS.399....2O} {399, 2}

\bibitem[\protect\citeauthoryear{{Ostrovski} et~al.,}{{Ostrovski}
  et~al.}{2018}]{ostrovski2018}
{Ostrovski} F.,  et~al., 2018, \mn@doi [\mnras] {10.1093/mnrasl/slx173}, \href
  {https://ui.adsabs.harvard.edu/abs/2018MNRAS.473L.116O} {473, L116}

\bibitem[\protect\citeauthoryear{{Paic}, {Vernardos}, {Sluse}, {Millon},
  {Courbin}, {Chan}  \& {Bonvin}}{{Paic} et~al.}{2022}]{paic2022}
{Paic} E.,  {Vernardos} G.,  {Sluse} D.,  {Millon} M.,  {Courbin} F.,  {Chan}
  J.~H.,   {Bonvin} V.,  2022, \mn@doi [\aap] {10.1051/0004-6361/202141808},
  \href {https://ui.adsabs.harvard.edu/abs/2022A&A...659A..21P} {659, A21}

\bibitem[\protect\citeauthoryear{{Planck Collaboration} et~al.,}{{Planck
  Collaboration} et~al.}{2020}]{planck2020}
{Planck Collaboration} et~al., 2020, \mn@doi [\aap]
  {10.1051/0004-6361/201833910}, \href
  {https://ui.adsabs.harvard.edu/abs/2020A&A...641A...6P} {641, A6}

\bibitem[\protect\citeauthoryear{{Prochaska} et~al.,}{{Prochaska}
  et~al.}{2013}]{prochaska2013}
{Prochaska} J.~X.,  et~al., 2013, \mn@doi [\apj] {10.1088/0004-637X/776/2/136},
  \href {https://ui.adsabs.harvard.edu/abs/2013ApJ...776..136P} {776, 136}

\bibitem[\protect\citeauthoryear{{Rankine}, {Hewett}, {Banerji}  \&
  {Richards}}{{Rankine} et~al.}{2020}]{rankine2020}
{Rankine} A.~L.,  {Hewett} P.~C.,  {Banerji} M.,   {Richards} G.~T.,  2020,
  \mn@doi [\mnras] {10.1093/mnras/staa130}, \href
  {https://ui.adsabs.harvard.edu/abs/2020MNRAS.492.4553R} {492, 4553}

\bibitem[\protect\citeauthoryear{{Reimers}, {Hagen}, {Baade}, {Lopez}  \&
  {Tytler}}{{Reimers} et~al.}{2002}]{reimers2002}
{Reimers} D.,  {Hagen} H.~J.,  {Baade} R.,  {Lopez} S.,   {Tytler} D.,  2002,
  \mn@doi [\aap] {10.1051/0004-6361:20011798}, \href
  {https://ui.adsabs.harvard.edu/abs/2002A&A...382L..26R} {382, L26}

\bibitem[\protect\citeauthoryear{{Riello} et~al.,}{{Riello}
  et~al.}{2021}]{riello2021}
{Riello} M.,  et~al., 2021, \mn@doi [\aap] {10.1051/0004-6361/202039587}, \href
  {https://ui.adsabs.harvard.edu/abs/2021A&A...649A...3R} {649, A3}

\bibitem[\protect\citeauthoryear{{Ross} et~al.,}{{Ross}
  et~al.}{2013}]{ross2013}
{Ross} N.~P.,  et~al., 2013, \mn@doi [\apj] {10.1088/0004-637X/773/1/14}, \href
  {https://ui.adsabs.harvard.edu/abs/2013ApJ...773...14R} {773, 14}

\bibitem[\protect\citeauthoryear{{Rubin} et~al.,}{{Rubin}
  et~al.}{2018}]{rubin2018}
{Rubin} K. H.~R.,  et~al., 2018, \mn@doi [\apj] {10.3847/1538-4357/aaaeb7},
  \href {https://ui.adsabs.harvard.edu/abs/2018ApJ...859..146R} {859, 146}

\bibitem[\protect\citeauthoryear{{Rusu}, {Oguri}, {Minowa}, {Iye}, {More},
  {Inada}  \& {Oya}}{{Rusu} et~al.}{2014}]{rusu2014}
{Rusu} C.~E.,  {Oguri} M.,  {Minowa} Y.,  {Iye} M.,  {More} A.,  {Inada} N.,
  {Oya} S.,  2014, \mn@doi [\mnras] {10.1093/mnras/stu1621}, \href
  {https://ui.adsabs.harvard.edu/abs/2014MNRAS.444.2561R} {444, 2561}

\bibitem[\protect\citeauthoryear{{Saxton}, {Read}, {Esquej}, {Freyberg},
  {Altieri}  \& {Bermejo}}{{Saxton} et~al.}{2008}]{saxton2008}
{Saxton} R.~D.,  {Read} A.~M.,  {Esquej} P.,  {Freyberg} M.~J.,  {Altieri} B.,
   {Bermejo} D.,  2008, \mn@doi [\aap] {10.1051/0004-6361:20079193}, \href
  {https://ui.adsabs.harvard.edu/abs/2008A&A...480..611S} {480, 611}

\bibitem[\protect\citeauthoryear{{Schechter} \& {Wambsganss}}{{Schechter} \&
  {Wambsganss}}{2002}]{schechter2002}
{Schechter} P.~L.,  {Wambsganss} J.,  2002, \mn@doi [\apj] {10.1086/343856},
  \href {https://ui.adsabs.harvard.edu/abs/2002ApJ...580..685S} {580, 685}

\bibitem[\protect\citeauthoryear{{Schechter}, {Pooley}, {Blackburne}  \&
  {Wambsganss}}{{Schechter} et~al.}{2014}]{schechter2014}
{Schechter} P.~L.,  {Pooley} D.,  {Blackburne} J.~A.,   {Wambsganss} J.,  2014,
  \mn@doi [\apj] {10.1088/0004-637X/793/2/96}, \href
  {https://ui.adsabs.harvard.edu/abs/2014ApJ...793...96S} {793, 96}

\bibitem[\protect\citeauthoryear{{Schechter}, {Morgan}, {Chehade}, {Metcalfe},
  {Shanks}  \& {McDonald}}{{Schechter} et~al.}{2017}]{schechter2017}
{Schechter} P.~L.,  {Morgan} N.~D.,  {Chehade} B.,  {Metcalfe} N.,  {Shanks}
  T.,   {McDonald} M.,  2017, \mn@doi [\aj] {10.3847/1538-3881/aa6899}, \href
  {https://ui.adsabs.harvard.edu/abs/2017AJ....153..219S} {153, 219}

\bibitem[\protect\citeauthoryear{{Schindler} et~al.,}{{Schindler}
  et~al.}{2018}]{schindler2018}
{Schindler} J.-T.,  et~al., 2018, \mn@doi [\apj] {10.3847/1538-4357/aad2dd},
  \href {https://ui.adsabs.harvard.edu/abs/2018ApJ...863..144S} {863, 144}

\bibitem[\protect\citeauthoryear{{Schmidt} et~al.,}{{Schmidt}
  et~al.}{2022}]{schmidt2022}
{Schmidt} T.,  et~al., 2022, arXiv e-prints, \href
  {https://ui.adsabs.harvard.edu/abs/2022arXiv220604696S} {p. arXiv:2206.04696}

\bibitem[\protect\citeauthoryear{{S{\'e}rsic}}{{S{\'e}rsic}}{1963}]{sersic1963}
{S{\'e}rsic} J.~L.,  1963, Boletin de la Asociacion Argentina de Astronomia La
  Plata Argentina, \href
  {https://ui.adsabs.harvard.edu/abs/1963BAAA....6...41S} {6, 41}

\bibitem[\protect\citeauthoryear{{Shajib} et~al.,}{{Shajib}
  et~al.}{2020}]{shajib2020}
{Shajib} A.~J.,  et~al., 2020, \mn@doi [\mnras] {10.1093/mnras/staa828}, \href
  {https://ui.adsabs.harvard.edu/abs/2020MNRAS.494.6072S} {494, 6072}

\bibitem[\protect\citeauthoryear{{Shalyapin}, {Sergeyev}, {Goicoechea}  \&
  {Zheleznyak}}{{Shalyapin} et~al.}{2018}]{shalyapin2018}
{Shalyapin} V.~N.,  {Sergeyev} A.~V.,  {Goicoechea} L.~J.,   {Zheleznyak}
  A.~P.,  2018, \mn@doi [\mnras] {10.1093/mnras/sty2028}, \href
  {https://ui.adsabs.harvard.edu/abs/2018MNRAS.480.2853S} {480, 2853}

\bibitem[\protect\citeauthoryear{{Shen} et~al.,}{{Shen}
  et~al.}{2021}]{shen2021}
{Shen} Y.,  et~al., 2021, \mn@doi [Nature Astronomy]
  {10.1038/s41550-021-01323-1}, \href
  {https://ui.adsabs.harvard.edu/abs/2021NatAs...5..569S} {5, 569}

\bibitem[\protect\citeauthoryear{{Shin} \& {Evans}}{{Shin} \&
  {Evans}}{2008}]{shin2008}
{Shin} E.~M.,  {Evans} N.~W.,  2008, \mn@doi [\mnras]
  {10.1111/j.1365-2966.2008.13738.x}, \href
  {https://ui.adsabs.harvard.edu/abs/2008MNRAS.390..505S} {390, 505}

\bibitem[\protect\citeauthoryear{{Sluse}, {Hutsem{\'e}kers}, {Anguita},
  {Braibant}  \& {Riaud}}{{Sluse} et~al.}{2015}]{sluse2015}
{Sluse} D.,  {Hutsem{\'e}kers} D.,  {Anguita} T.,  {Braibant} L.,   {Riaud} P.,
   2015, \mn@doi [\aap] {10.1051/0004-6361/201526832}, \href
  {https://ui.adsabs.harvard.edu/abs/2015A&A...582A.109S} {582, A109}

\bibitem[\protect\citeauthoryear{{Smith}, {Lucey}  \& {Collier}}{{Smith}
  et~al.}{2018}]{smith2018}
{Smith} R.~J.,  {Lucey} J.~R.,   {Collier} W.~P.,  2018, \mn@doi [\mnras]
  {10.1093/mnras/sty2328}, \href
  {https://ui.adsabs.harvard.edu/abs/2018MNRAS.481.2115S} {481, 2115}

\bibitem[\protect\citeauthoryear{{Sonnenfeld} et~al.,}{{Sonnenfeld}
  et~al.}{2020}]{sonnenfeld2020}
{Sonnenfeld} A.,  et~al., 2020, \mn@doi [\aap] {10.1051/0004-6361/202038067},
  \href {https://ui.adsabs.harvard.edu/abs/2020A&A...642A.148S} {642, A148}

\bibitem[\protect\citeauthoryear{{Sorini}, {Dav{\'e}}  \&
  {Angl{\'e}s-Alc{\'a}zar}}{{Sorini} et~al.}{2020}]{sorini2020}
{Sorini} D.,  {Dav{\'e}} R.,   {Angl{\'e}s-Alc{\'a}zar} D.,  2020, \mn@doi
  [\mnras] {10.1093/mnras/staa2937}, \href
  {https://ui.adsabs.harvard.edu/abs/2020MNRAS.499.2760S} {499, 2760}

\bibitem[\protect\citeauthoryear{{Spiniello} et~al.,}{{Spiniello}
  et~al.}{2018}]{spiniello2018}
{Spiniello} C.,  et~al., 2018, \mn@doi [\mnras] {10.1093/mnras/sty1923}, \href
  {https://ui.adsabs.harvard.edu/abs/2018MNRAS.480.1163S} {480, 1163}

\bibitem[\protect\citeauthoryear{{Spiniello}, {Agnello}, {Sergeyev}, {Anguita},
  {Rodr{\'\i}guez}, {Napolitano}  \& {Tortora}}{{Spiniello}
  et~al.}{2019}]{spiniello2019}
{Spiniello} C.,  {Agnello} A.,  {Sergeyev} A.~V.,  {Anguita} T.,
  {Rodr{\'\i}guez} {\'O}.,  {Napolitano} N.~R.,   {Tortora} C.,  2019, \mn@doi
  [\mnras] {10.1093/mnras/sty3407}, \href
  {https://ui.adsabs.harvard.edu/abs/2019MNRAS.483.3888S} {483, 3888}

\bibitem[\protect\citeauthoryear{{Stern} et~al.,}{{Stern}
  et~al.}{2012}]{stern2012}
{Stern} D.,  et~al., 2012, \mn@doi [\apj] {10.1088/0004-637X/753/1/30}, \href
  {https://ui.adsabs.harvard.edu/abs/2012ApJ...753...30S} {753, 30}

\bibitem[\protect\citeauthoryear{{Stern} et~al.,}{{Stern}
  et~al.}{2021}]{stern2021}
{Stern} D.,  et~al., 2021, \mn@doi [\apj] {10.3847/1538-4357/ac0f04}, \href
  {https://ui.adsabs.harvard.edu/abs/2021ApJ...921...42S} {921, 42}

\bibitem[\protect\citeauthoryear{{Tang} et~al.,}{{Tang}
  et~al.}{2021}]{tang2021}
{Tang} S.,  et~al., 2021, \mn@doi [\apj] {10.3847/1538-4357/ac1ff0}, \href
  {https://ui.adsabs.harvard.edu/abs/2021ApJ...922...83T} {922, 83}

\bibitem[\protect\citeauthoryear{{Torrey} et~al.,}{{Torrey}
  et~al.}{2015}]{torrey2015}
{Torrey} P.,  et~al., 2015, \mn@doi [\mnras] {10.1093/mnras/stv1986}, \href
  {https://ui.adsabs.harvard.edu/abs/2015MNRAS.454.2770T} {454, 2770}

\bibitem[\protect\citeauthoryear{{Treu} et~al.,}{{Treu}
  et~al.}{2018}]{treu2018}
{Treu} T.,  et~al., 2018, \mn@doi [\mnras] {10.1093/mnras/sty2329}, \href
  {https://ui.adsabs.harvard.edu/abs/2018MNRAS.481.1041T} {481, 1041}

\bibitem[\protect\citeauthoryear{{Wang}, {Liu}, {Qiu}, {Bai}, {Yang}, {Guo}  \&
  {Zhang}}{{Wang} et~al.}{2016}]{wang2016}
{Wang} S.,  {Liu} J.,  {Qiu} Y.,  {Bai} Y.,  {Yang} H.,  {Guo} J.,   {Zhang}
  P.,  2016, \mn@doi [\apjs] {10.3847/0067-0049/224/2/40}, \href
  {https://ui.adsabs.harvard.edu/abs/2016ApJS..224...40W} {224, 40}

\bibitem[\protect\citeauthoryear{{Webb} et~al.,}{{Webb}
  et~al.}{2020}]{webb2020}
{Webb} N.~A.,  et~al., 2020, \mn@doi [\aap] {10.1051/0004-6361/201937353},
  \href {https://ui.adsabs.harvard.edu/abs/2020A&A...641A.136W} {641, A136}

\bibitem[\protect\citeauthoryear{{Yue}, {Fan}, {Yang}  \& {Wang}}{{Yue}
  et~al.}{2022}]{yue2022}
{Yue} M.,  {Fan} X.,  {Yang} J.,   {Wang} F.,  2022, arXiv e-prints, \href
  {https://ui.adsabs.harvard.edu/abs/2022arXiv220106761Y} {p. arXiv:2201.06761}

\bibitem[\protect\citeauthoryear{{Zhou} et~al.,}{{Zhou}
  et~al.}{2021}]{zhou2021}
{Zhou} R.,  et~al., 2021, \mn@doi [\mnras] {10.1093/mnras/staa3764}, \href
  {https://ui.adsabs.harvard.edu/abs/2021MNRAS.501.3309Z} {501, 3309}

\makeatother
\end{thebibliography}






\bsp    
\label{lastpage}
\end{document}